\documentstyle[amsmath,amssymb,bm,graphicx,epsf]{elsart}
\def\be{\begin{equation}}
\def\ee{\end{equation}}
\def\bea{\begin{eqnarray}}
\def\eea{\end{eqnarray}}
\def\ep{\epsilon}

\def\sh{{\sigma}}

\def\Va{{V_\alpha}}

\def\wt{\widetilde}
\def\p{\partial}

\def\L{{\pounds}}
\def\R{{{\cal{R}}}}
\def\Q{{\cal{Q}}}
\def\C{{\cal{C}}}

\def\S{{\cal{S}}}
\def\H{{\cal H}}
\def\cs2{c_{\rm{s}}^2}
\def\Mpl{{M_{\rm{Pl}}}}
\def\U0{{\bar U_0}}
\def\wt{\widetilde}
\def\dT{{\delta{\bf T}_1}}
\def\dTT{{\delta{\bf T}_2}}
\def\drho{{\delta\rho_1}}
\def\drhorho{{\delta\rho_2}}

\def\dij{\delta_{ij}}
\newcommand\eq[1]{Eq.~(\ref{#1})}
\newcommand\eqs[1]{Eqs.~(\ref{#1})}
\def\bi{\begin{itemize}}
\def\ei{\end{itemize}}
\def\vp{{\varphi}}
\newcommand\dvp[1]{\delta\varphi_{{#1}}}

\newcommand\dvpK[1]{{\delta\varphi_{{#1}K}}}

\newcommand\dU[1]{{\delta U_{#1}}}
\def\dvpdvpKll{\delta\vp_{1K,l}\delta\vp_{1K,}^{~~~~l}}
\def\hp{{\cal{P}}}
\newcommand\gami[1]{{\gamma_{{#1}}^{~i}}}
\newcommand\gamk[1]{{\gamma_{{#1}}^{~k}}}
\newcommand{\vect}[1]{\bm{\mathrm{{#1}}}}

\newcommand\A[1]{{\phi_{{#1}}}}

\def\X{{\cal{X}}}
\def\Xv{{{\cal{X}}_{\rm{v}}}}
\def\XB{{{\cal{X}}_{\rm{B}}}}

\def\U{{\cal{U}}}
\def\Uivec {{\U_{\rm{vec}}^i}}

\def\bkone{\mathbf k_1}
\def\bktwo{\mathbf k_2}
\def\bkthree{\mathbf k_3}
\def\bkfour{\mathbf k_4}
\def\fNL{{f_{NL}}}
\def\Mp{m_{\rm{Pl}}}

\def\tom{{\rm{tom}}}
\def\syn{{\rm{syn}}}
\def\com{{\rm{com}}}
\def\fg{{\rm{flat}}}

\def\udg{{\delta\rho}}

\renewcommand{\theequation}{\arabic{section}.\arabic{equation}}

\begin{document}
\begin{frontmatter}
\title{Cosmological perturbations}
\author[ad1]{Karim A.~Malik} and
\author[ad2]{David Wands}
\address[ad1]{Astronomy Unit, School of Mathematical Sciences,
Queen Mary, University of London, Mile End Road, \cty London E1 4NS,
\cny United Kingdom}
\address[ad2]{Institute of Cosmology and Gravitation, University of Portsmouth,
\cty Portsmouth~PO1~2EG, \cny United Kingdom}

\begin{abstract}

We review the study of inhomogeneous perturbations about a
homogeneous and isotropic background cosmology. We adopt a
coordinate based approach, but give geometrical interpretations of
metric perturbations in terms of the expansion, shear and curvature
of constant-time hypersurfaces and the orthogonal timelike vector
field. We give the gauge transformation rules for metric and matter
variables at first and second order. We show how gauge invariant
variables are constructed by identifying geometric or matter
variables in physically-defined coordinate systems, and give the
relations between many commonly used gauge-invariant variables. In
particular we show how the Einstein equations or energy-momentum
conservation can be used to obtain simple evolution equations at
linear order, and discuss extensions to non-linear order. We present
evolution equations for systems with multiple interacting fluids and
scalar fields, identifying adiabatic and entropy perturbations. As
an application we consider the origin of primordial curvature and
isocurvature perturbations from field perturbations during inflation
in the very early universe.

\end{abstract}

\end{frontmatter}
%
\newpage
\pagenumbering{roman}
\normalsize
\tableofcontents

\newpage
\pagenumbering{arabic}
%
\section{Introduction}
\setcounter{equation}{0}

The standard model of hot big bang cosmology is based upon the
spatially homogeneous and isotropic Friedmann-Robertson-Walker (FRW)
model. This successfully describes the average expansion of the
universe on large scales according to Einstein's theory of general
relativity, and the evolution from a hot, dense initial state
dominated by radiation to the cool, low density state dominated by
non-relativistic matter and, apparently, vacuum energy at the present
day. This standard model is described by just a handful of numbers
specifying the expansion rate, the temperature of the present
microwave background radiation, the density of visible matter, dark
matter and dark vacuum energy.

But an homogeneous model cannot describe the complexity of the actual
distribution of matter and energy in our observed universe where stars
and galaxies form clusters and superclusters of galaxies across a wide
range of scales. For this we need to be able to describe spatial
inhomogeneity and anisotropy. But there are few exact solutions of
general relativity that incorporate spatially inhomogeneous and
anisotropic matter and hence geometry. Therefore we use a perturbative
approach starting from the spatially homogeneous and isotropic FRW
model as a background solution with simple properties within which we
can study the increasing complexity of inhomogeneous perturbations
order by order.

The introduction of a homogeneous background model to describe the
inhomogeneous universe leads to an ambiguity in the choice of
coordinates. In the FRW model the homogeneous three-dimensional
hypersurfaces provide a natural time slicing of four-dimensional
spacetime. For instance, hypersurfaces of a uniform density coincide
with hypersurfaces with uniform spatial curvature. However in the
real, inhomogeneous universe spatial hypersurfaces of uniform density
do not in general have uniform spatial curvature, and hypersurfaces of
uniform curvature do not have uniform density. In general relativity
there is, a priori, no preferred choice of coordinates. Choosing a set
of coordinates in the inhomogeneous universe which will then be
described by an FRW model plus perturbations amounts to assigning a
mapping between spacetime points in the inhomogeneous universe and the
homogeneous background model. The freedom in this choice is the gauge
freedom, or gauge problem, in general relativistic perturbation theory.
It can lead to apparently different descriptions of the same
physical solution simply due to the choice of coordinates. This
freedom can be a powerful tool as it allows one to work in terms of
variables best suited to the problem in hand, however it is
important to work with gauge-invariant definitions of the
perturbations variables if the results are to be easily assimilated
and cross-referenced to the work of others.

In this review we will focus on how one can construct a variety of
gauge-invariant variables to deal with perturbations in different
cosmological models at first order and beyond. We will emphasise the
geometrical meaning of metric and matter perturbations and their
gauge-invariant definitions.

Most work to date has been done only to linear order where the
perturbations obey linear field equations. Even then we must
consistently solve the linear evolution equations, subject to the
constraint equations of general relativity which relate the matter
variables to the geometry. Beyond first order, the non-linearity of
Einstein's equations becomes evident, making progress much more
difficult. But in limiting cases, notably the large scale limit, it is
possible to extend some of the simple results of linear theory to
higher orders.

Previous reviews on the topic of linear perturbation theory in
cosmology include, for example,
Refs.~\cite{KS,MFB,Durrer1993,Bertschinger1995,Riotto:2002yw,Hu_review},
and the relevant chapters in
Refs.~\cite{weinberg,Peebles80,Landau:1987gn,LLBook}. For
second-order perturbation theory see
Refs.~\cite{Noh2004,Bartolo:2004if,Nakamura:2004rm}.
For reviews on inflation in particular see, e.g.,
Refs.~\cite{Lindebook,KolbTurner,Lidsey:1995np,LRreview,LLBook,Langlois:2004de,Bassett:2005xm,Wands:2007bd}
and for cosmic microwave background anisotropies see, e.g.,
Refs.~\cite{Kodama:1986fg,Lewis:1999bs,cmbeasy2,Giovannini:2004rj}
Finally, for reviews on perturbation theory in the context of the
higher-dimensional brane-world scenario see, e.g., Refs.
~\cite{Kodama:2000fa,Bridgman:2001mc,Riazuelo:2002mi,Maartens:2003tw,Brax:2004xh}
and for perturbation theory in the context of the low-energy string
effective action see, e.g., Refs.~\cite{Lidsey:1999mc,Cartier01}.

For simplicity we work with a flat background spatial metric which
is compatible with current observations. For generalisation to
spatially hyperbolic or spherical FRW models see other reviews,
e.g.,~Ref.\cite{KS}.
We use a prime to denote derivatives with respect to conformal time,
and we use a comma to denote partial derivatives with respect to
comoving spatial coordinates, i.e.,
\be
T_{,i} \equiv \frac{\partial
T}{\partial x^i} \,.
\ee
For further definitions and notation see Appendix
\ref{defandnot_sect}.

\section{Perturbations in cosmology}
\label{cosmopertsect}

\setcounter{equation}{0}
%
%

Throughout this review we will assume that our observable Universe can
approximately be described by a homogeneous and isotropic
Friedmann-Robertson-Walker (FRW) spacetime. Thus we assume that the
physical quantities can usefully be decomposed into a homogeneous
background, where quantities depend solely on cosmic time, and
inhomogeneous perturbations. The perturbations thus ``live'' on the
background spacetime and it is this background spacetime which is used
to split four-dimensional spacetime into spatial three-hypersurfaces,
using a (3+1) decomposition.
In addition we work with a spatially flat FRW background, though
results can be readily extended to non-zero spatial curvature.

We start this section by defining arbitrary perturbations of
tensorial quantities and then proceed by decomposing vectors and
tensors into ``time'' and ``space'' parts on the spatial
hypersurfaces.

\subsection{Defining perturbations}

Any tensorial quantities can then be split into a homogeneous
background and an inhomogeneous perturbation
\be
\label{tensor_split1}
{\bf T}(\eta,x^i) = {\bf T}_0(\eta) + \delta {\bf T}(\eta,x^i)\,.
\ee
The background part is a time-dependent quantity, ${\bf T}_0\equiv
{\bf T}_0(\eta)$, whereas the perturbations depend on time and space
coordinates $x^\mu=[\eta,x^i]$.
The perturbation can be further expanded as a power series,
\be
\label{tensor_split2}
\delta {\bf T}(\eta,x^i)=
\sum_{n=1}^\infty\frac{\ep^n}{n!}\delta {\bf{T}}_n(\eta,x^i)\,,
\ee
where the subscript $n$ denotes the order of the perturbations, and
we explicitly include here the small parameter $\ep$.
In linear perturbation theory, for example, we only consider
first-order terms, $\ep^1$, and can neglect terms resulting from the
product of two perturbations, which would necessarily be of order
$\ep^2$ or higher, which considerably simplifies the resulting
equations.
In the following sections we shall omit the small parameter $\ep$
whenever possible, as is usually done to avoid the equations getting
too cluttered.

\subsection{Decomposing tensorial quantities}
\label{decomposingsection}

It is convenient to slice the spacetime manifold into a one-parameter
family of spatial hypersurfaces of constant time, which is the
standard (3+1) split of spacetime.
This foliation was introduced by Darmois already in 1927 (see
Ref.~\cite{Gourgoulhon:2007ue}) and popularised by Arnowitt, Deser and
Misner \cite{ADM} (for conditions on the existence of the foliation
see e.g.~Ref.~\cite{Wald84}). We refer to the foliation of spacetime
by spatial hypersurfaces of given conformal time as the time-slicing,
and the identification of spatial coordinates on each hypersurface as
the threading.

Note that the (3+1) split of spacetime precedes the decomposition of
3-dimensional quantities into scalars and vectors, or scalar,
vectors, and tensors below.

\subsubsection{Vectors}
\label{decomposingvectors}

We can split any 4 vector $V^\mu$ into a temporal and spatial part,
\be
\label{vectorsplit}
\U^\mu=\left[\U^0, \U^i\right]\,.
\ee
Note that $\U^0$ is a {\em scalar} on spatial hypersurfaces. The
spatial part $\U^i$ can then be further decomposed into a further
{\em scalar} part $\U$ and a {\em vector} part $\Uivec$,
\be 
\label{vectordecomp} 
\U^i \equiv \delta^{ij} \U_{,j} + \Uivec\,,
\ee
where $\p \Uivec/\p x^i=0$.
The denominations {\em scalar} and {\em vector} part go back to
Bardeen \cite{Bardeen80} and are due to the transformation behaviour
of $\U$ and $\Uivec$ on spatial hypersurfaces.
The decomposition of a vector field into potential (or curl-free)
part and a divergence-free part in Euclidean space is known as
Helmholtz's theorem. The curl-free and divergence-free parts are
also called longitudinal and solenoidal parts, respectively.

In our isotropic (FRW) background, there can be no spatial vector
part at zeroth order (as this would correspond to a preferred
direction), but there can be a non-zero temporal part,
\be
\label{backgroundvector}
\U^0_{(0)}\neq 0\,,\qquad \U^i_{(0)}=0\,.
\ee
A non-trivial vector, with a non-zero spatial part, can appear only
at first order. We give examples of vector fields in a FRW
background including perturbations in Section \ref{vectorfieldsec},
where we discuss the unit vector field normal to constant-time
hypersurfaces and the fluid 4-velocity.

Note that ``divergence-free'', etc, is defined with respect to the
flat-space metric, rather than using covariant derivatives, since
perturbations are defined with respect to the spatially flat
background.

\subsubsection{Tensors}
\label{decomposingtensors}

As for vector fields, we can decompose a rank-2 tensor into a time
part and spatial part, but now have also mixed time-space parts.

Consider the metric tensor $g_{\mu\nu}$ which we require to be
symmetric,
\be
g_{\mu\nu} \equiv g_{\nu\mu}\,.
\ee
The metric tensor has therefore only 10 independent components in 4
dimensions.
We first split the metric tensor into a background and a perturbed
part, using \eq{tensor_split1}.
It then turns out to be useful to split the metric perturbation into
different parts labelled {\em scalar}, {\em vector} or {\em tensor}
according to their transformation properties on spatial
hypersurfaces \cite{Bardeen80,Stewart1990}, which are themselves
expanded into first and higher order parts using \eq{tensor_split2}.

Our background spacetime is described by a spatially flat FRW
background metric
\be
\label{def_bg_metric}
ds^2 = a^2 \left[ -d\eta^2 + \delta_{ij} dx^i dx^j \right] \,,
\ee
where $\eta$ is conformal time and $a=a(\eta)$ is the scale factor.
The cosmic time, measured by observers at fixed comoving spatial
coordinates, $x^i$, is given by $t=\int a(\eta)\, d\eta$.

The perturbed part of the metric tensor can be written as
\begin{eqnarray}
 \label{defA}
\delta g_{00} &=& - 2a^2 \A{} \, ,\\
\delta g_{0i} &=&   a^2 B_{i} \, ,\\
\delta g_{ij} &=&  2a^2  C_{ij}  \, .
\end{eqnarray}
The $0-i$ and the $i-j$-components of the metric tensor can be
further decomposed into scalar, vector and tensor parts
\bea
\label{decompBi}
B_i  &=& B_{,i} -S_i\, ,\\
\label{decompCij} C_{ij} &=& -\psi\;\delta_{ij} + E_{,ij}+ F_{(i,j)}
+ \frac12 h_{ij}\,.
 \eea
where $\A{}$, $B$, $\psi$ and $E$ are {\em scalar} metric
perturbations, $S_i$ and $F_i$ are {\em vector} metric
perturbations, and $h_{ij}$ is a {\em tensor} metric perturbation,
which we will now define.

{\em Scalar} perturbations can always be constructed from a scalar,
or its derivatives, and background quantities.
Any 3-vector, such as $B_{,i}$, constructed from a scalar is
necessarily curl-free, i.e., $B_{,[ij]}=0$.

{\em Vector} perturbations are divergence-free. For instance one can
distinguish an intrinsically {\em vector} part of the metric
perturbation $\delta g_{0i}$, which we denote by $-S_i$, which gives
a non-vanishing contribution to $\delta g_{0[i,j]}$. Similarly we
define the vector contribution to $\delta g_{ij}$ constructed from
the (symmetric) derivative of a vector $F_{(i,j)}$.

Finally there is a {\em tensor} contribution to $\delta g_{ij}=a^2
h_{ij}$ which is both transverse, $h_{ij,}^{~~~j}=0$ (i.e.,
divergence-free), and trace-free ($h_{i}^i=0$) which therefore
cannot be constructed from inhomogeneous scalar or vector
perturbations.

Note that when raising and lowering spatial indices of vector and
tensor perturbations we use the comoving background spatial metric,
$\delta_{ij}$, so that, for instance, $h_i^j\equiv
\delta^{jk}h_{ik}$.

The reason for splitting the metric perturbation into these three
types is that the governing equations decouple at linear order, and
hence we can solve each perturbation type separately.
At higher order this is no longer true: we find for example at
second order that although the ``true'' second order perturbations,
$\delta g_{2\mu\nu}$, still decouple, but their governing equations
have source terms quadratic in the first order variables, $\delta
g_{1\mu\nu}$, mixing the different types \cite{Nakamura:2006rk}.
Indeed at all higher orders, $n>1$, the different types of
perturbations of order $n$ decouple, but are sourced by terms
comprising perturbations of lower order.

We have introduced four scalar functions, two spatial vector valued
functions with three components each, and a symmetric spatial tensor
with six components. But these functions are subject to several
constraints: $h_{ij}$ is transverse and traceless, which contributes
four constraints, $F_i$ and $S_i$ are divergence-free, one
constraint each. We are therefore finally left 10 new degrees of
freedom, the same number as the independent components of the
perturbed metric.

The choice of variables to describe the perturbed metric is not
unique. Already at first order there are different conventions for
the split of the spatial part of the metric. We follow the notation
of Mukhanov et al~\cite{MFB} so that the metric perturbation,
$\psi$, can be identified directly with the intrinsic scalar
curvature of spatial hypersurfaces at first order, see later.
Sometimes it is useful to work in terms of the trace of the
perturbed spatial metric
 \be
 C = C_i^i = -3\psi + \nabla^2 E
 \ee
At first order this coincides with the perturbation of the
determinant of the spatial metric. Including terms up to second order
we have
 \bea
 \label{determinant}
 \det \left( \delta_{ij}+2C_{ij} \right)
 &=& 1 + 2C + 2 \left( C^2-C_{ij}C^{ij} \right)
  \,
 \nonumber\\
 &=& 1 -6\psi +2\nabla^2 E
 \nonumber\\
&& + 12\psi^2 - 8\psi\nabla^2E + 2(\nabla^2E)^2 -
 2E_{,ij}E_{,}^{~ij}
 - 2F_{i,j}F_{~,}^{i~j}
 \nonumber\\
&&
  - \frac12 h_{ij}h^{ij} - 2E_{,ij}h^{ij} - 2F_{i,j}h^{ij} \,.
 \eea
where we have used the general result $\det(e^\gamma)=e^{{\rm Tr}(\gamma)}$.
There are further choices for the way the spatial metric is split
into the different perturbation variables at second (and higher)
order in the perturbations.

Note that our metric perturbations in
Eqs.~(\ref{defA}--\ref{decompCij}) include all orders. If we write
out the complete metric tensor, up to and including second-order in
the perturbations we have
\bea
\label{metric1}
g_{00}&=&-a^2\left(1+2\A1+\A2\right) \,,\nonumber \\
g_{0i}&=&a^2\left(B_{1i}+\frac{1}{2}B_{2i}\right) \,,\nonumber \\
g_{ij}&=&a^2\left[\delta_{ij}+2C_{1ij}+C_{2ij}\right]\,,
\eea
where the first and second order perturbations $B_{1i}$ and
$C_{1ij}$, and $B_{2i}$ and $C_{2ij}$, can be further split
according to \eqs{decompBi} and (\ref{decompCij}).

The contravariant metric tensor follows from the constraint (to the
required order),
\be
g_{\mu\nu} ~ g^{\nu\lambda}=\delta_{\mu}^{~\lambda}\,,
\ee
which up to second order gives
\bea
\label{metric2}
g^{00}&=&-a^{-2}\left[1-2\A1-\A2+4\A1^2-
 B_{1k}B_{1}^{~k}\right] \,,\nonumber \\
g^{0i}&=&a^{-2}\left[B_{1}^{~i}+\frac{1}{2}B_{2}^{~i}-2\A1 B_1^i
-2B_{1k}C_{1}^{~ki}\right]\,, \nonumber \\
g^{ij}&=&a^{-2}\left[
\delta^{ij}-2C_{1}^{~ij}-C_{2}^{~ij}+4C_{1}^{~ik}C_{1k}^{~~j}
-B_{1}^{~i}B_{1}^{~j}\right]\,.
\eea
%

\section{Geometry of spatial hypersurfaces}
\label{geo_sec}
\setcounter{equation}{0}

\subsection{Timelike vector fields}
\label{vectorfieldsec}

The perturbed metric given in Section \ref{decomposingtensors}
implicitly defines a unit time-like vector field orthogonal to
constant-$\eta$ hypersurfaces,
\be
n_\mu \propto \frac{\p \eta}{\p x^\mu}\,, \ee
subject to the constraint
\be
n^\mu n_\mu=-1\,.
\ee
In the FRW background this coincides
with the 4-velocity of matter and the expansion of the velocity
field $\theta=3H$, where $H$ is the Hubble expansion rate. We define
the conformal Hubble parameter
\be
\H \equiv aH\,.
\ee
In this section we calculate corresponding geometrical quantities
for $n^\nu$, and thus the space-time, defined by the perturbed
metric tensor.
Note however that in the perturbed spacetime the vector field
$n^\mu$, need no longer coincide with the 4-velocity of matter
fields at first order and beyond.

Up to and including to second order, the covariant vector field is
\be
\label{defnmu}
n_\mu= -a\left[1+\A1+\frac{1}{2}\A2
+\frac{1}{2}\left(B_{1k}B_1^{k}-\phi_1^2\right),\vect{0}
\right]\,,
\ee
and the contravariant vector field is
\bea
\label{defnmuup}
n^0 &=& \frac{1}{a}\left[
1-\phi_1-\frac{1}{2}\phi_2+\frac{3}{2}\phi_1^2
-\frac{1}{2}B_{1k}B_{1}^{k}\right]\,,
 \nonumber
\\
n^i &=& \frac{1}{a}\left[-\left(B_1^i+\frac{1}{2}B_{2}^{i}\right)
+2B_{1k}C_1^{ki}+\A1 B_1^i\right]\,. 
\eea
Observers moving along the hypersurface-orthogonal vector field,
$n^\mu$, have a vanishing 3-velocity with respect to the spatial
coordinates $x^i$ when the shift vector $B^i$ is zero. We will refer
to these as orthogonal coordinate systems.

\subsection{Geometrical quantities}
\label{geo_quant_sec}

The covariant derivative of any time-like unit vector field
$n_{\mu}$ can be decomposed uniquely as follows~\cite{Wald84}:
\be
n_{\mu;\nu}= \frac{1}{3}\theta \,\hp_{\mu\nu} + \sigma_{\mu\nu} +
\omega_{\mu\nu} - a_{\mu}n_{\nu}
\, ,
\ee
where the spatial projection tensor $\hp_{\mu\nu}$, orthogonal to
$n^\mu$, is given by
\be
\label{defPmunu}
\hp_{\mu\nu}=g_{\mu\nu}+n_{\mu}n_{\nu}.
\ee

The overall expansion rate is given by
\be
\label{theta}
\theta=n^{\mu}_{~;\mu} \,,
\ee
the (trace-free and symmetric) shear is
\be
\label{defshear}
\sigma_{\mu\nu}=
\frac{1}{2} \hp^{~\alpha}_{\mu}\hp^{~\beta}_{\nu}
\left( n_{\alpha; \beta} + n_{\beta; \alpha} \right)
-\frac{1}{3} \theta \hp_{\mu\nu} \,,
\ee
the (antisymmetric) vorticity is
\be
\label{defomega}
\omega_{\mu\nu}=\frac{1}{2} \hp^{~\alpha}_{\mu}
\hp^{~\beta}_{\nu} \left( n_{\alpha; \beta} - n_{\beta; \alpha}
\right) \,,
\ee
and the acceleration is
\be
 \label{defacc} a_{\mu}=n_{\mu;\nu} n^{\nu} \,.
 \ee
On spatial hypersurfaces the expansion, shear, vorticity and
acceleration coincide with their Newtonian counterparts in fluid
dynamics \cite{Hawking1973,Stephani2004}.
In this subsection we focus on the unit normal vector field $n^\mu$,
but the expansion, shear, vorticity and acceleration defined in this
way can readily be applied to any other 4-vector field, such as the
4-velocity $u^\mu$.
One can easily verify that the vorticity (\ref{defomega}) is
automatically zero for the hypersurface orthogonal vector field,
$n_\mu$ defined in Eq.~(\ref{defnmu}).
Note however that the perturbed fluid velocity can have vorticity
and this is described by the vector (divergence-free) part of the
fluid 3-velocity which we will define in
section~\ref{Tmunufluid_sec}.

The projection tensor $\hp_{\mu\nu}$ is the induced 3-metric on the
spatial hypersurfaces, and the Lie derivative of $\hp_{\mu\nu}$
along the vector field $n^\mu$ is the extrinsic curvature of the
hypersurface embedded in the higher-dimensional spacetime
\cite{Wald84,Deruelle:1995kd}.
The extrinsic curvature of the spatial hypersurfaces defined by
$n_{\mu}$ is thus given by
\be
 K_{\mu\nu} \equiv \frac12 \pounds_n \hp_{\mu\nu} =
 \hp_{\nu}^{~\lambda} n_{\mu;\lambda} =
\frac{1}{3}\theta \,\hp_{\mu\nu} + \sigma_{\mu\nu} \,. \ee

At first order we can easily identify the metric perturbations with
geometrical perturbations of the spatial hypersurfaces or the
associated vector field, $n_\mu$.

The intrinsic curvature of spatial hypersurfaces up to first order
is given by
\be
{}^{(3)}R_1 = \frac{4}{a^2} \nabla^2 \psi_1 \,.
\ee

The scalar part of the shear (\ref{defshear}) up to first order is
given by
\be \sigma_{1ij} =
\left(\p_i\p_j-\frac{1}{3}\nabla^2\dij\right)a\sh_1\,,
\ee
where we define the shear potential
\be
\label{shearscalar1} \sh_1 \equiv E_1'-B_1 \,.
\ee
The vector and tensor parts are given by, respectively,
\bea
\sigma_{1{\rm{V}}ij}= a\left(F_{1(i,j)}'-B_{1(i,j)}
\right)\,,\qquad
\sigma_{1{\rm{T}}ij}=\frac{a}{2}h_{1ij}'\,.
\eea

The acceleration up to first order is
\be
a_i = \A{}_{,i} \,.
\ee

The expansion rate up to first order is given by
\be
\theta = \frac{3}{a} \left[ \H - \H\A{} - \psi' + \frac13
\nabla^2\sigma \right]
\,.
\ee

The intrinsic spatial curvature, shear and acceleration of $n_\mu$
are given up to second order in Appendix~\ref{App_n_components}.

The overall expansion, up to second order is given by
\bea
\label{expansion_comp}
\theta=\frac{1}{a}\Bigg[&&
3\frac{a'}{a}-3\frac{a'}{a}\A1+{C_{1k}^{~~k}}'-B_{1k,}^{~~k}\nonumber\\
&&-\frac{3}{2}\frac{a'}{a}\left(\A2-3\A1^2\right)
+\frac{1}{2}\left({C_{2k}^{~~k}}'-B_{2k,}^{~~k}\right)
+\phi_1\left(B_{1k,}^{~~k}-{C_{1k}^{~~k}}'\right)\nonumber\\
&&-\frac{3}{2}\frac{a'}{a}B_{1k}B_{1}^k
-2C_1^{kl}C_{1kl}'+2C_1^{kl}B_{1l,k}
+2B_1^lC_{1lk,}^{~~~k}-B_1^k C^l_{1~l,k}
\Bigg]\,.
\eea

Focusing for the moment only on scalar perturbations (neglecting
first order vectors and tensors) on large scales (neglecting spatial
derivatives) the perturbed part of the expansion simplifies to
\bea
\delta\theta_1&\simeq&-\frac{3}{a}\left(
\frac{a'}{a}\phi_1+\psi_1'\right)\,, \\
\delta\theta_2&\simeq&-\frac{3}{a}\left[
\frac{a'}{a}\phi_2
+\psi_2'+2\psi_1'\left(2\psi_1-\phi_1\right)
-3\frac{a'}{a}\phi_1^2
\right] \,.
\eea
Note that the expansion rate, $\theta$ in
Eq.~(\ref{expansion_comp}), is defined with respect to proper time
(comoving with $n^\mu$).
The expansion rate with respect to conformal time on large scales is
\be
\theta_{\rm{coord}}
\simeq3\left[
\frac{a'}{a}-\psi_1'-\frac{1}{2}\psi_2'
-2\psi_1'\psi_1
\right]\,.
\ee
We see that on large scales (and considering only scalar
perturbations) spatially flat hypersurfaces ($\psi=0$) are also
uniform coordinate expansion hypersurfaces (on which the perturbed
expansion vanishes).

In the following we will often refer to the perturbed logarithmic
expansion, or ``number of e-foldings''
 \be
 N \equiv \frac13 \int \theta d\tau = \frac13 \int \theta_{\rm{coord}} d\eta
 = \ln a - \psi_1 - \frac12\psi_2 - \psi_1^2 \,.
 \ee
This quantity, and its perturbation $\delta N$ becomes a
particularly useful quantity to describe the primordial scalar
perturbation beyond linear order, as we will discuss in
Section~\ref{non_linear_sec}.

\section{Energy-momentum tensor for fluids}
\label{Tmunufluid_sec} 
\setcounter{equation}{0}

We are interested in how the spacetime geometry, described by the
metric tensor, is affected by the perturbed matter content,
described by the energy-momentum tensor.

The four-velocity of matter, $u^\mu$, is defined by
\be
u^\mu=\frac{dx^\mu}{d\tau}\,,
\ee
where $\tau$ is the proper time comoving with the fluid,
subject to the constraint
\begin{equation}
u_{\mu}u^{\mu}=-1\,.
\end{equation}

The components of the 4-velocity up to second order are then given
by
\bea
\label{defumu}
 u_0 &=& -a\left[1+\A1+\frac{1}{2}\A2
-\frac{1}{2}\A1^2+\frac{1}{2}v_{1k}v_1^k\right]\,, \nonumber \\
u_i &=& a\left[v_{1i}+B_{1i}+\frac{1}{2}\left(v_{2i}+B_{2i}\right)
-\phi_1B_{1i}+2C_{1ik}v_1^k
\right] \,, \\
\label{defumuup} u^0 &=& a^{-1}\left[1-\A1-\frac{1}{2}\A2
+\frac{3}{2}\A1^2+\frac{1}{2}v_{1k}v_1^k+v_{1k}B_1^k
\right]\,, \nonumber \\
u^i &=& a^{-1}\left(v_1^i+\frac{1}{2}v_2^i\right)\,, \eea

The spatial part of the velocity can be split into a scalar
(potential) part and a vector (solenoidal) part, order by order,
following \eq{vectordecomp} as
\be \label{defv}
 v^i \equiv \delta^{ij}v_{,j}+v_{{\rm vec}}^i\,,
\ee
for each order $n$ where we refer to $v_{(n)}$ as the scalar
velocity potential, and to $v_{(n)vec}^i$ as the vector part.

Note that $v^i$ is the 3-velocity of matter defined with respect to
the spatial coordinates, $x^i$, and so is not the velocity with
respect to the hypersurface-orthogonal vector field $n^i$, except in
orthogonal coordinate systems for which $B^i=0$. In comoving
orthogonal coordinates, which we will discuss later, $v^i=0$ and
$B^i=0$.

\subsection{Single fluid}
\label{tmunu_single_sec}

The energy-momentum tensor of a fluid with density $\rho$, isotropic
pressure $P$ and 4-velocity $u^{\mu}$, defined above in \eq{defumu},
is defined as~\cite{KS,Stephani2004,vanElst,weinberg}
\be
\label{defTmunugeneral}
T_{\mu\nu}
= \left( \rho +P\right)
u_{\mu}u_{\nu}+P g_{\mu\nu} +\pi_{\mu\nu} \,.
\ee
The anisotropic stress tensor $\pi_{\mu\nu}$ is split into first and
second order parts in the usual way,
\be
\pi_{\mu\nu}\equiv \pi_{(1)\mu\nu}+\frac{1}{2}\pi_{(2)\mu\nu}\,,
\ee
and is subject to the constraints
\be
\label{pi_constraint}
\pi_{\mu\nu} u^\nu=0\,, \qquad \pi^{\mu}_{~\mu}=0\,.
\ee
The anisotropic stress vanishes for a perfect fluid or minimally
coupled scalar fields, but may be non-zero in the presence of
free-streaming neutrinos or a non-minimally coupled scalar field.

We follow Kodama and Sasaki \cite{KS} in defining the proper energy
density as the eigenvalue of the energy-momentum tensor, and the
four velocity $u^\mu$ as the corresponding eigenvector
\bea
\label{eigen1}
T^{\mu}_{~\nu}u^\nu=-\rho u^\mu\,.
\eea

The anisotropic stress tensor decomposes into a trace-free scalar
part, $\Pi$, a vector part,
$\Pi_i$, and a tensor part, $\Pi_{ij}$,
at each order according to
\be
\label{def:Pi_ij}
\pi_{ij}=a^2\left[ \Pi_{,ij}- \frac{1}{3} \nabla^2\Pi \delta_{ij} +
\frac{1}{2}\left(\Pi_{i,j} +\Pi_{j,i}\right) + \Pi_{ij} \right]\, .
\ee

We get for the components of the stress energy tensor in the
background
\bea T^0_{~0} = -\rho_0 \,, \qquad
T^0_{~i} = 0\,,  \qquad
T^i_{~j} = \delta^i_{~j} P_0 \,, \eea
at first order,
\bea
{}^{(1)}\delta T^0_{~0} &=& -\delta\rho_1 \, , \\
\label{defTmtm}
 {}^{(1)}\delta T^0_{~i}
&=& \left(\rho_0+P_0\right)\left(v_{1i}+B_{1i}\right) \,, \\
\label{perfT_ij} {}^{(1)}\delta T^i_{~j} &=& \delta
P_1\delta^i_{~j}+a^{-2}\pi_{(1)~j}^{~i}
 \,,
\eea
and at second order
\bea
{}^{(2)}\delta T^0_{~0} &=& -\delta\rho_2
-2\left(\rho_0+P_0\right)v_{1k}\left(v_1^{~k}+B_1^{~k}\right) \, , \\
{}^{(2)}\delta T^0_{~i} &=&
\left(\rho_0+P_0\right)\Big[v_{2i}+B_{2i}+
4C_{1ik}v_1^{~k}-2\phi_1\left(v_{1i}+2B_{1i}\right)\Big]\nonumber\\
&&\qquad +2\left(\delta\rho_1+\delta
P_1\right)\left(v_{1i}+B_{1i}\right) +\frac{2}{a^2}\left(B^k_1 +
v_1^k \right)\pi_{1ik}
\,, \\
\label{perfT_2ij}
{}^{(2)}\delta T^i_{~j} &=& \delta
P_2\,\delta^i_{~j}+a^{-2} \pi^{~i}_{2~j} -\frac{4}{a^2}C^{ik}_1
\pi_{1jk}
+2\left(\rho_0+P_0\right)v_1^i\left(v_{1j}+B_{1j}\right)\,.
\eea
Note, that for compactness of presentation we have not split
perturbations into their constituent scalar, vector and tensor parts
in the above expressions. The decompositions are given for $v_i$ and
$\pi_{ij}$ in \eqs{defv} and (\ref{def:Pi_ij}), and given for $B_i$
and $C_{ij}$ in Section \ref{decomposingtensors}, in \eqs{decompBi}
and (\ref{decompCij}).

Note that contracting the $i-j$ part of the energy-momentum tensor,
\eq{perfT_2ij}, including the constraints for the anisotropic
stress, \eq{pi_constraint}, guarantees that the anisotropic stress
cancels on the trace,
\bea
{}^{(1)}\delta T^k_{~k} &=& 3 \delta P_1\,,\\
{}^{(2)}\delta T^k_{~k} &=&
3\delta P_2
+2\left(\rho_0+P_0\right)v_1^k\left(v_{1k}+B_{1k}\right)
\,.
\eea
This cancellation is true at all orders.

Coordinate transformations affect the split between spatial and
temporal components of the matter fields and so quantities like the
density, pressure and 3-velocity are gauge-dependent, as described
in Section \ref{gaugesec}. Density and pressure are 4-scalar
quantities which transform as given in Eq.~(\ref{rhotransform1}) in
the following section, but the 4-velocity is a 4-vector which
transforms described in Section~\ref{gauge_vector_sec}.
The anisotropic stress is gauge-invariant at first order, but
becomes gauge-dependent at second order.

\subsection{Multiple fluids}

In the multiple fluid case the total energy-momentum tensor is the
sum of the energy-momentum tensors of the individual fluids
\be
 \label{def:sumT}
T^{\mu\nu}=\sum_\alpha T^{\mu\nu}_{(\alpha)}\,. \ee
For each fluid we define the local energy-momentum transfer 4-vector
$Q^\nu_{(\alpha)}$ through the relation
\be \label{nablaTalpha} \nabla_\mu
T^{\mu\nu}_{(\alpha)}=Q^\nu_{(\alpha)}\,, \ee
where the energy-momentum tensor, $T^{\mu\nu}_{(\alpha)}$, is
locally conserved only for non-interacting fluids, for which
$Q^\nu_{(\alpha)}=0$. Equations~(\ref{nablaTmunu})
and~(\ref{nablaTalpha}) imply the constraint
\be \label{Qconstraint} \sum_\alpha Q^\nu_{(\alpha)}=0 \,. \ee

Following Refs.~\cite{KS,MW2004} we split the energy-momentum
transfer 4-vector using the total fluid velocity $u^\mu$ as
\be \label{defQmu} Q^\mu_{(\alpha)}=Q_{(\alpha)}
u^\mu+f^\mu_{(\alpha)}\,, \ee
where $Q_{(\alpha)}$ is the energy transfer rate and
$f^\mu_{(\alpha)}$ the momentum transfer rate, subject to the
constraint
\be \label{fmucon} u_\mu f^\mu_{(\alpha)}=0\,, \ee
{}From \eq{fmucon} we find for the temporal component of the momentum
transfer rate vector $f^\mu_{(\alpha)}$
\be
f^0_{1(\alpha)}=0\,, \qquad f^0_{2(\alpha)}
=2 f^k_{1(\alpha)}\left(v_{1k}+B_{1k}\right)\,,
\ee
We then find for the temporal components of the energy transfer
4-vector to be at zeroth, first and second order, respectively
\bea
Q^0_{(\alpha)}&=&\frac{1}{a}Q_{0\alpha}\,,\\
Q^0_{(\alpha)}&=&\frac{1}{a}\left(
\delta Q_{1\alpha}-\A1 Q_{0\alpha} \right)\,,\\
Q^0_{(\alpha)}&=&\frac{1}{2a}\Big[\delta Q_{2\alpha}
+Q_{0\alpha}\left(3\A1^2-\A2\right)
-2\A1 \delta Q_{1\alpha}\nonumber\\
&&\qquad\qquad +\left(v_{1k}+B_{1k}\right)
\left(
\frac{2}{a} f^k_{1(\alpha)}+Q_{0\alpha}v_1^k\right)\Big]\,,
\eea
where $Q_{0\alpha}$, $\delta Q_{1\alpha}$, and $\delta Q_{2\alpha}$
are the energy transfer to the $\alpha$-fluid in the background, at
first and at second order, respectively.

For the spatial components of the energy transfer 4-vector, the
momentum part, we get at first and second order, respectively
\bea
Q^i_{(\alpha)}&=&\frac{1}{a} Q_{0\alpha} v_1^i
+\frac{1}{a^2}f^i_{1(\alpha)}\,,\\
Q^i_{(\alpha)}&=&\frac{1}{2a}\Big[\frac{1}{a} f^i_{2(\alpha)} +\delta
Q_{1\alpha}v_1^i + Q_{0\alpha}\left(v_2^i+2\A1
B_1^i-4C_{1k}^{~i}v_1^k\right) \Big]\,, \eea
where $f^i_{1(\alpha)}$ and $f^i_{2(\alpha)}$ are the spatial parts
of the momentum transfer rates at first and second order.

Note that the homogeneous and isotropic FRW background excludes a
zeroth order momentum transfer.

Using \eq{vectordecomp} the spatial momentum transfer vector of order
$n$, $f^i_{n(\alpha)}$, can be further decomposed into a scalar and a
vector part, according to
\be
 f^i_{n(\alpha)}\equiv \delta^{ij} f_{n(\alpha),j} + \hat
f^i_{n(\alpha)}\,. 
\ee
%


\section{Energy-momentum tensor for scalar fields}
\label{Tmunu_scal_sec}
\setcounter{equation}{0}

\subsection{Single field}
\label{Tmunu_scal_sec_single}

A minimally coupled scalar field is specified by the Lagrangian
density
\be
\label{Lscal}
{\cal L}=-\frac{1}{2} g^{\mu\nu}
\varphi_{,\mu}\varphi_{,\nu} - U(\varphi) \,,
\ee
where the scalar field kinetic energy is then non-negative for our
choice of metric signature.

The energy momentum tensor is defined as
\be
T_{\mu\nu}=-2\frac{\partial {\cal L}}{\partial g^{\mu\nu}}
+g_{\mu\nu}{\cal L} \,,
\ee
and we therefore get for a scalar field $\varphi$
\be
 T^{\mu}_{~\nu} =
 g^{\mu\alpha}\varphi_{,\alpha}\varphi_{,\nu}
-\delta^{\mu}_{~\nu}\left( U(\varphi) +\frac{1}{2}g^{\kappa\lambda}
\varphi_{,\kappa}\varphi_{,\lambda} \right)
 \,.
 \ee
Comparing to the energy-momentum tensor for a perfect fluids,
Eq.~(\ref{defTmunugeneral}), we can identify the non-linear
4-velocity, density and pressure of the scalar field \cite{Taub73}
\bea
 u_\mu &=&
 \frac{\varphi_{,\mu}}{|g^{\lambda\kappa}\varphi_{,\lambda}\varphi_{,\kappa}|}
 \,, \qquad
 \rho = - g^{\lambda\kappa}\varphi_{,\lambda}\varphi_{,\kappa} + U
 \,, \nonumber\\
 P &=& - g^{\lambda\kappa}\varphi_{,\lambda}\varphi_{,\kappa} - U
 \,.
\eea
Note that the anisotropic stress, $\pi_{\mu\nu}$, is identically
zero for minimally coupled scalar fields. In fact we can subdivide
the energy-momentum tensor for a single field into that of a stiff
kinetic fluid with $\rho_{(\varphi)}=P_{(\varphi)}=-
g^{\lambda\kappa}\varphi_{,\lambda}\varphi_{,\kappa}$ and a vacuum
energy $\rho_{(U)}=-P_{(U)}=U(\varphi)$, which exchange
energy-momentum
$Q^\mu_{(\varphi)}=-Q^\mu_{(U)}=(dU/d\varphi)\nabla^\mu \varphi$.

Splitting the scalar field into a homogeneous background field and a
perturbation,
\be \varphi(\eta,x^i)=\varphi_0(\eta)+\delta\varphi_1(\eta,x^i) \, ,
\ee
and using the definitions above we find for the components of the
energy momentum tensor of a perturbed scalar field at linear order
without specifying a gauge yet
\begin{eqnarray}
\label{scalT_00} T^0_{~0} &=& -\frac12 a^{-2} \varphi'^{~2}_0 - U_0
+ a^{-2} \varphi_0' \left(\phi_1~\varphi_0'-\delta\varphi_1'\right)
-U_{,\varphi} \delta\varphi_1 \, ,  \\
\label{scalT_0i} T^0_{~i} &=& -a^{-2} \left(\varphi_0^{\prime}
\delta\varphi_{1,i} \right) \, , \\
\label{scalT_ij} T^i_{~j} &=& \left[ \frac{1}{2}a^{-2}
\varphi'^{~2}_0 - U_0 - U_{,\varphi} \delta\varphi_1 +a^{-2}
\varphi_0' \left(\delta\varphi_1'
-\phi_1~\varphi_0'\right)\right]~\delta^i_{~j} \,,
\end{eqnarray}
where $U_{,\varphi}\equiv dU/d\varphi$ and $U_0=U(\varphi_0)$.
By comparing Eq.~(\ref{scalT_ij}) with Eq.~(\ref{perfT_ij}), we see
that scalar fields do not support vector or tensor perturbations to
first order.

\subsection{Multiple fields}

For $N$ minimally coupled scalar fields the Lagrangian density is
given by
\be
{\cal L} =-\frac{1}{2} \sum_I \left(
g^{\mu\alpha}\varphi_{I,\alpha}\varphi_{I,\mu}\right) -
U(\vp_1,\ldots,\vp_N) \,.
\ee
The energy-momentum tensor is
\be \label{multiTmunu} T_{\mu\nu}=\sum_{K=1}^N\left[
\vp_{K,\mu}\vp_{K,\nu}
-\frac{1}{2}g_{\mu\nu}g^{\alpha\beta}\vp_{K,\alpha}\vp_{K,\beta}\right]
-g_{\mu\nu}U \,, \ee
where $\vp_K$ is the $K$th scalar field and $U$ the scalar field
potential and $\vp_{K,\mu}\equiv\frac{\p\vp_{K}}{\p x^\mu}$.

Analogous to the energy-momentum tensor for a single field, we can
identify the non-linear 4-velocity, density and pressure of each of
the scalar fields \cite{MW2004}
\bea
 u_{(I)\mu} &=&
 \frac{\varphi_{I,\mu}}{|g^{\lambda\kappa}\varphi_{I,\lambda}\varphi_{I,\kappa}|}
 \,, \nonumber\\
 \rho_{(I)} &=& P_{(I)} = - g^{\lambda\kappa}\varphi_{I,\lambda}\varphi_{I,\kappa}
 \,.
\eea
The energy-momentum transfer to each fluid is $Q_{(I)\mu}=(\partial
U/\partial \vp_I)\vp_{I,\mu}$.

The total energy-momentum tensor (\ref{multiTmunu}) is the sum over
$N$ stiff fluids plus the vacuum energy
 \be
  T_{\mu\nu} = \sum_{I=1}^N \rho_{(I)} \left[ 2 u_{(I)\mu}
  u_{(I)\nu} + g_{\mu\nu} \right] - U g_{\mu\nu} \,.
  \ee
The anisotropic stress, $\pi_{\mu\nu}$, is identically zero for any
number of minimally coupled scalar fields.
The total energy-momentum tensor is only equivalent to that of a
single scalar field in the special case where all the 4-velocities
of the fields, $u_{(I)\mu}$, are identical. This is true in the
homogeneous FRW cosmology, but in general breaks down when
inhomogeneous perturbations are considered.

We split the scalar fields $\vp_I$ into a background and
perturbations up to and including second order according to
\eqs{tensor_split1} and (\ref{tensor_split2}),
\be
\vp_I(x^\mu)=\vp_{0I}(\eta)+\dvp{1I}(x^\mu)+\frac{1}{2}\dvp{2I}(x^\mu)\,.
\ee
The potential $U\equiv U(\vp_I)$ can be split similarly according to
\be U(\vp_I)=U_0+\dU1+\frac{1}{2}\dU2\,, \ee
where
\bea \label{defdU1}
\dU1&=&\sum_K U_{K}  \dvp{1K}\,,\\
\label{defdU2} \dU2&=&\sum_{K,L}U_{KL}\dvp{1K}\dvp{1L} +\sum_K
U_{K}\dvp{2K}\,. \eea
and we use the shorthand
 $U_{K}\equiv \p U/\p\vp_K$.
The energy-momentum tensor, \eq{multiTmunu}, expanded up to second
order in the perturbations for the metric tensor \eq{metric1} is
given in Appendix \ref{scalarTmunu_sect}.

\section{Gauge transformations}
\label{gaugesec}
\setcounter{equation}{0}

We now review how tensorial quantities change under coordinate
transformations \cite{Bardeen80,Stewart1990,Mukhanov96,Bruni96,MM2008}
(see Ref.~\cite{sachs,weinberg,Stewart74} for earlier work on this
subject).
While the order of the perturbation is indicated by a subscript, we
also keep the small parameter $\epsilon$ in the following equations
whenever appropriate.

A problem which arises in cosmological perturbation theory is the
presence of spurious coordinate artefacts or gauge modes in the
calculations. Although the gauge modes had been dealt with on a ``case
by case'' basis before, the gauge issue was resolved in a systematic
way by Bardeen~\cite{Bardeen80}.
The gauge issue will arise in any approach to GR that splits
quantities into a background and a perturbation. Although GR is
covariant, i.e.~manifestly coordinate choice independent, splitting
variables into a background part and a perturbation is not a covariant
procedure, and therefore introduces a coordinate or gauge
dependence. By construction this only affects the perturbations; the
background quantities remain the same in the different coordinate
systems. Here we assume the difference between the coordinate systems
is small, of $O(\epsilon)$, however the gauge problem would persist
also for finite transformations. Note that the ``covariant approach''
\cite{Ellis1989} also corresponds to a choice of gauge, namely the
comoving one, which is made explicit by the inclusion of the velocity
field \cite{Challinor:1998xk,Langlois:2005ii}.

In order to restore covariance as far as possible, we usually wish
to eliminate the gauge degrees of freedom. We will show in Section
\ref{gi_var_sec} how, by constructing variables corresponding to
perturbations in physically defined coordinate systems, the gauge
dependencies can be made to cancel out (the quantities so
constructed will not change under a gauge transformation).

\subsection{Active and passive approaches to gauge transformations}

There are two approaches to calculate how perturbations change under a
small coordinate or gauge transformation.
For the \emph{active view} we study how perturbations change under a
mapping, where the map directly induces the transformation of the
perturbed quantities. In the \emph{passive view} the relation between
two coordinate systems is specified, and we calculate how the
perturbations are changed under this coordinate transformation.
In the passive approach the transformation is taken at the same physical point,
whereas in the active approach the transformation of the perturbed
quantities is evaluated at the same coordinate point.  We
will discuss both approaches briefly in the following, but shall use
the active approach to calculate the transformation behaviour of the
first and second order variables. For a mathematically more rigorous
discussion see e.g.~Ref.~\cite{MM2008}.

\subsubsection{Active approach}

The starting point in the active approach is the exponential map, that
allows us to immediately write down how a tensor ${\bf T}$ transforms
up to second order,
once the generator of the gauge transformation, $\xi^\mu$, has been
specified. The exponential map is
\be
\label{Ttransgeneral}
\widetilde{\bf T}=e^{\pounds_{\xi}}{\bf T}\,,
\ee
where $\pounds_{\xi}$ denotes the Lie derivative with respect
to $\xi^\lambda$.  The vector field generating the transformation,
$\xi^\lambda$, is up to second order
\be
\xi^\mu\equiv \epsilon \xi_1^{\mu}
+\frac{1}{2}\epsilon^2\xi_2^{\mu} +O(\epsilon^3)\,.
\ee
The exponential map can be readily expanded
\be
\label{exp_expand}
\exp(\pounds_{\xi})=1+\epsilon\pounds_{\xi_1}
+\frac{1}{2}\epsilon^2\pounds_{\xi_1}^2
+\frac{1}{2}\epsilon^2\pounds_{\xi_2}
+\ldots
\ee
where we kept terms up to $O(\epsilon^2)$.
Splitting the tensor ${\bf T}$ up to second order, as given in
\eq{tensor_split2}, and collecting terms of like order in $\ep$ we
find that tensorial quantities transform at zeroth, first and second order,
respectively, as \cite{Mukhanov96,Bruni96}
\bea
\widetilde {{\bf T}_0} &=& {\bf T}_0 \,,\\
\label{Ttrans1}
\ep\widetilde \dT
&=& \ep\dT + \ep\L_{\xi_1} {\bf T}_0  \,,\\
\ep^2\widetilde \dTT
\label{Ttrans2}
&=& \ep^2\left(\dTT +\L_{\xi_2} {\bf T}_0 +\L^2_{\xi_1}
{\bf T}_0 + 2\L_{\xi_1} \dT\right)\,.
\eea
Note that the background quantities are not affected by the mapping.
We will apply Eqs.~(\ref{Ttrans1}) and (\ref{Ttrans2}) to scalars,
vectors, and tensors after discussing the passive approach next.

Applying the map (\ref{Ttransgeneral}) to the coordinate functions
$x^\mu$ we get a relation for the coordinates of a point $q$ in and
a point $p$ as
\be
\label{defcoordtrans}
{x^\mu}( {{q}})
= e^{\xi^\lambda \frac{\p}{\p x^\lambda}\big|_{{p}}} \
x^\mu( {{p}})\,,
\ee
where we have used the fact that when acting on scalars $\pounds_{\xi}
= \xi^{\mu} (\partial/\partial x^{\mu})$ and the partial
derivatives are evaluated at $p$.
The left-hand-side and the right-hand-side of \eq{defcoordtrans} are
evaluated at different points.
Equation (\ref{defcoordtrans}) can then be expanded up to second-order
as
\bea
\label{coordtrans2}
{x^\mu}(q) = x^\mu(p)+\epsilon\xi_1^{\mu}(p)
+\frac{1}{2}\epsilon^2\left(\xi^{\mu}_{1,\nu}(p)\xi_1^{~\nu}(p)
+ \xi_2^{\mu}(p)
\right)   \,.
\eea
Note that we do not need \eq{coordtrans2} to calculate how
perturbations change under a gauge transformation in the active
approach, it simply tells us how the coordinates of the points $p$ and
$q$ are related in this approach.

\subsubsection{Passive approach}

In the passive approach we specify the relation between two coordinate
systems directly, and then calculate the change in the metric and
matter variables when changing from one system to the other. As long
as the two coordinate systems are related through a small
perturbation, the functional form relating them is quite arbitrary.

However, in order to make contact with the active approach,
discussed above, we take \eq{coordtrans2} as our starting point.
Note, that all quantities in the passive approach are evaluated at
the same physical point.  Equation (\ref{coordtrans2}) can be
rewritten to give a relation between the ``old'' (untilded) and the
``new'' (tilde) coordinate systems \cite{Bruni96,MM2008},
\be
\label{passive_coord}
\widetilde{x^\mu}(q)=x^\mu(q)-\epsilon\xi_1^{\mu}(q)
+\epsilon^2\frac{1}{2}\left({\xi^{\mu}_1(q)}_{,\nu}\xi_1^{~\nu}(q)
-\xi_2^{\mu}(q)\right)\,,
\ee
evaluated at the same physical point $q$.

The passive point of view is very popular at first order, see
e.g.~the original paper by Bardeen \cite{Bardeen80} and the widely
used reviews by Kodama and Sasaki \cite{KS}, and Mukhanov, Feldman,
and Brandenberger \cite{MFB}.

The starting point in the {passive approach} is to identify an
invariant quantity, that allows us to relate quantities to be evaluated
in the two coordinate systems. We denote the two coordinate systems by
$\tilde x^\mu$ and $x^\mu$ system, and their relation is given by
\eq{passive_coord}.
We choose as an example the energy density, $\rho$, which as a four
scalar will not change under a coordinate transformation. However,
once it has been split into the background quantity and perturbation
at different orders, these variables will change.

The two coordinate systems are related by \eq{passive_coord}, which we
can use to expand the right-hand-side of \eq{tilderho} in a Taylor
expansion up to second order.
To first order, the two coordinate systems are simply related, using the
linear part of \eq{coordtrans2}, by
\be
\tilde x^\mu=x^\mu-\xi_1^\mu\,.
\ee

We get the transformation behaviour of the density perturbation,
$\delta\rho$, from the requirement that the total density,
$\rho=\rho_0+\delta\rho$, has to be invariant under a change of
coordinate system and therefore has to be the same in the $\tilde
x^\mu$ and the $x^\mu$ system, that is
\be
\label{tilderho}
\tilde\rho(\tilde x^\mu)=\rho ( x^\mu)\,.
\ee

Expanding both sides of \eq{tilderho} using \eq{tensor_split1}, we
get
\bea
\rho(x^\mu) &=& \rho_0(x^0) + \epsilon \delta\rho_1(x^\mu)
+ O(\epsilon^2)\,,\\
\wt\rho(\tilde x^\mu)&=&\rho_0\left(\wt{x^0}\right)
+\epsilon \wt{\delta\rho_1}\left(\wt{x^\mu}\right)
+ O(\epsilon^2)\,,\nonumber\\
&=&\rho_0(x^0)+ \epsilon \left( - \rho_0'(x^0) \xi_1^0(x^\mu)
+ \wt{\delta\rho_1}(x^\mu) \right)
+O(\epsilon^2)\,.
\eea
Note that we use the same background solution $\rho_0(\eta)$ in both
expressions.
Thus we obtain the transformation rule at first order
\be
\label{rho_trans_passive}
\wt{\delta\rho_1} = \delta\rho_1+\rho_0'\xi_1^0 \,.
\ee

Another invariant is the line element $ds^2$, which allows us to
deduce the transformation properties of the metric tensor by
exploiting the invariance of $ds^2$, i.e.,
\be
ds^2=\tilde g_{\mu\nu}d \tilde x^\mu d\tilde x^\nu
= g_{\mu\nu}dx^\mu d x^\nu\,,
\ee
We here will not follow this approach further here, but see
e.g.~\cite{KS,M2001}.

\subsection{Four-scalars}
\label{gauge_scalar_sec}

We now return to the active approach by studying the simplest
tensorial quantity, the four-scalar.  Examples of four-scalar are the
energy density, $\rho$, and the scalar field $\vp$, and we shall use
the former below.

{}From Eqs.~(\ref{tensor_split1}) and (\ref{tensor_split2}) we
immediately get the perturbed four-scalar up to second order
\be
\rho=\rho_0+\delta\rho_1+\frac12\delta\rho_2\,,
\ee
using the energy density as an example.

\subsubsection{First order}

Before we can study the transformation behaviour of the
perturbations at first order, we split the generating vector
$\xi_1^\mu$ into a scalar temporal part $\alpha_1$ and a spatial
scalar and vector part, $\beta_{1}$ and $\gami1$, according to
\be
\label{def_xi1}
\xi_1^\mu=\left(\alpha_1,\beta_{1,}^{~~i}+\gami1\right)\,,
\ee
where the vector part is divergence-free $\p_k\gamk1=0$.

Under a first-order transformation a four scalar, here the energy
density, $\rho$, then transforms from \eqs{Ttrans1} and
(\ref{lie_scalar}) as,
\be
\label{rhotransform1}
\widetilde\drho = \drho + \rho_0'\alpha_1 \,.
\ee
The first-order density perturbation is fully specified by
prescribing the first order temporal gauge or time slicing,
$\alpha_1$.

\subsubsection{Second order}

At second order, as at first order, we split the generating vector
$\xi_2^\mu$ into a scalar time and scalar and vector spatial part,
similarly as at first order, as
\be
\label{def_xi2}
\xi_2^\mu=\left(\alpha_2,\beta_{2,}^{~~i}+\gami2\right)\,,
\ee
where the vector part is divergence-free $\p_k\gamk2=0$.
We then find from Eqs.~(\ref{Ttrans2}) and (\ref{lie_scalar}) that a
four scalar transforms as
\bea
\label{rhotransform2}
\widetilde\drhorho = \drhorho
+\rho_0'\alpha_2&+&\alpha_1\left(
\rho_0''\alpha_1+\rho_0'{\alpha_1}'+2\drho'\right)\nonumber\\
&+&\left(2\drho+\rho_0'{\alpha_1}\right)_{,k}
(\beta_{1,}^{~~k}+\gamk1)
\,.
\eea
We see here already the coupling between vector and scalar
perturbations in the last term through the gradient and $\gami1$.  The
gauge is only specified once the scalar temporal gauge perturbations
at first and second order, $\alpha_1$ and $\alpha_2$, and the first
order spatial gauge perturbations, $\beta_{1}$ and $\gami1$, are
specified.

\subsection{Four-vectors}
\label{gauge_vector_sec}

We now turn to four-vectors and their transformation properties.
Of particular interest in cosmology is the unit four-velocity $u^\mu$,
which we defined in Section \ref{Tmunufluid_sec} above \footnote{Note
  that under a gauge transformation the vector field normal to the
  constant-$\eta$ hypersurface, $n^\mu$ defined in section
  \ref{vectorfieldsec}, will be replaced by a new vector field normal
  to the constant-$\tilde\eta$ hypersurfaces, therefore it is not
  particularly helpful to study the transformation of $n^\mu$. We will
  study the gauge transformation of the metric tensor in section
  \ref{gauge_tensor_sec}, from which we can derive the transformation
  rules for the hypersurface-orthogonal field and its expansion,
  shear, etc.}.

\subsubsection{First order}

A four-vector transforms at first order, using \eqs{Ttrans1} and
(\ref{lie_vector}), as
\bea
\label{transu1}
\wt{\delta\U_{1\mu}}=\delta\U_{1\mu}+ \U_{(0)\mu}'\alpha_1
+\U_{(0)\lambda} \xi^\lambda_{1,\mu}\,,
\eea
where we used the fact that in a FRW spacetime background quantities
are time dependent only.

For the specific example of the four-velocity, defined in \eq{defumu},
we find,
\bea
\wt{v_{1i}}+\wt{B_{1i}}={v_{1i}}+{B_{1i}}-\alpha_{1,i}\,.
\eea
Using the transformation of the metric perturbation $B_{1i}$,
given below in \eq{B1itrans}, and using the decompositions given in
Section \ref{decomposingvectors} above, we get the transformations for
the scalar and vector parts, respectively, at first order
\bea
\label{transv1}
\wt {v_{1}} &=& v_{1}-\beta_1'\,,\\
\label{transv1vec}
\wt {v_{\rm{vec}(1)}^i} &=& v_{\rm{vec}(1)}^i-\gami1'\,,
\eea
and the perturbed temporal part of $u^\mu$ does indeed transform as a scalar.

\subsubsection{Second order}

At second order we find that a four-vector transforms, using \eqs{Ttrans2} and
(\ref{lie_vector}), as
\bea
\label{transu2}
\wt{\delta\U_{2\mu}}
&=&\delta\U_{2\mu}+\U_{(0)\mu}'\alpha_2+\U_{(0)0}\alpha_{2,\mu}
+\U_{(0)\mu}''\alpha_1^2+\U_{(0)\mu}'\alpha_{1,\lambda}\xi^\lambda_1\\
&&+2\U_{(0)0}'\alpha_1\alpha_{1,\mu}
+\U_{(0)0}\left(\xi^\lambda_1\alpha_{1,\mu\lambda}
+\alpha_{1,\lambda}\xi^\lambda_{1,\mu}\right)
+2\left(\delta\U_{1\mu,\lambda}\xi^\lambda_1
+\delta\U_{1\lambda}\xi^\lambda_{1,\mu}
\right)\,,\nonumber
\eea
where as before we used for the background
$\U_{(0)\mu}\equiv\U_{(0)\mu}(\eta)$ and $\U_{(0)i}=\vect{0}$.

Focusing again on the four-velocity, \eq{defumu}, and following a
similar procedure as at first order, we find that the second order combined
scalar and vector spatial part transforms as
\bea
\label{transv2i}
\wt {v_{2i}} &=& v_{2i}-\xi'_{2i}+\Xv_i\,,
\eea
where $\Xv_i$ contains the terms quadratic in the first order
perturbations and is given by
\bea
\label{defXvi}
\Xv_i
\equiv&&
\xi'_{1i}\left(2\A1+\alpha_1'+2\H\alpha_1\right)-\alpha_1\xi''_{1i}
\nonumber\\
&&
-\xi_1^k\xi'_{1i,k}+\xi_1^{k\prime}\xi_{1i,k}
-2\alpha_1\left(v_{1i}'+\H v_{1i}\right)
+2v_{1i,k}\xi_1^k-2v_1^k\xi_{1i,k}\,,
\eea
and we already substituted for the transformation of the metric
perturbation $B_{2i}$, given below in \eq{B2itrans}.
Decomposing then the second order velocity transformation, \eq{transv2i},
into scalar and vector parts, we get the transformations as
\bea
\label{transv2}
\wt {v_{2}} &=& v_{2}-\beta_2'+\nabla^{-2}\Xv^k_{~,k}\,,\\
\label{transv2vec}
\wt {v_{\rm{vec}(2)}^i} &=& v_{\rm{vec}(2)}^i-\gami2'
+\Xv_i-\nabla^{-2}\Xv^k_{~,ki}
\,.
\eea

\subsection{Tensors}
\label{gauge_tensor_sec}

\subsubsection{First order coordinate transformation}

We can now calculate how the first order metric perturbations change
under a gauge transformation.
We get the change of the $\delta^{(1)} g_{00}$ and hence the lapse
function $\A1$ immediately from \eqs{Ttrans1} and (\ref{lie_tensor}),
since this component of the metric is scalar in nature.
The change of the $\delta^{(1)} g_{0i}$ slightly more involved, since
this component contains scalar and vector perturbations. We therefore
have to compute the overall transformation of this metric component
using \eqs{Ttrans1} and (\ref{lie_tensor}), and then split the result
for $B_{1i}$ into its scalar part, $B_1$, and its divergence-free part
$-S_i$.
We therefore get for the combined part $B_{1i}$,
\bea
\label{B1itrans}
\widetilde B_{1i}&=&B_{1i}+\xi_{1i}'-\alpha_{1,i}\,,
\eea
and taking the divergence gives for the scalar part,
\bea
\label{B1idiv}
\nabla^2 \widetilde B_{1}&=&\nabla^2 B_{1}
+\nabla^2 \beta_{1}'-\nabla^2 \alpha_{1}\,,
\eea
which, after ``removal'' of the Laplacian gives the transformation
behaviour of $B_1$. We can then subtract the scalar part from
\eq{B1idiv} and are left with the vector part. The results are given
below.

To get the change of the metric functions in the spatial part of
the metric under a gauge transformation, we again first use
\eqs{Ttrans1} and (\ref{lie_tensor}) to get transformation of the
spatial part of the metric $\delta^{(1)} g_{ij}$, or $C_{1ij}$,
\bea
\label{Cij1trans}
2\wt C_{1ij}=2 C_{1ij} +2\H\alpha_1\delta_{ij}+\xi_{1i,j}+\xi_{1j,i}\,,
\eea
where we reproduce \eq{decompCij} above for convenience,
\[
2 C_{1ij}=-2\psi_1\delta_{ij}+2E_{1,ij}+2 F_{1(i,j)}+h_{1ij}\,.
\]
Taking the trace of \eq{Cij1trans} we get
\bea
\label{Cij1trace}
-3\wt\psi_1+\nabla^2\wt E_1
=-3\psi_1+\nabla^2 E_1+3\H\alpha_1+\nabla^2\beta_{1}\,.
\eea
Now applying the operator $\p^i\p^j$ to \eq{Cij1trans} we get a second
equation relating the scalar perturbation $\psi_1$ and $E_1$,
\bea
\label{Cij1divdiv}
-3\wt\nabla^2\psi_1+\nabla^2\nabla^2\wt E_1
=-3\nabla^2\psi_1+\nabla^2\nabla^2 E_1+3\H\nabla^2\alpha_1
+\nabla^2\nabla^2\beta_1\,.
\eea
Taking the divergence of \eq{Cij1trans} we get
\bea
2\wt C_{1ij,}^{~~~j}=2 C_{1ij,}^{~~~j} +2\H\alpha_{1,i}
+\nabla^2\xi_{1i}+\nabla^2\beta_{1,i}\,.
\eea
Substituting in our results for $\wt\psi_1$ and $\wt E_1$ we
arrive at
\be
\nabla^2 \wt F_{1i}= \nabla^2 F_{1i}+\nabla^2 \gami1\,.
\ee

We can sum up the transformations of the first order metric
perturbations we have from the above, first for the scalars as
\bea
\label{transphi1}
\widetilde {\A1} &=& \A1 +\H\alpha_1+\alpha_1'\,,\\
\label{transpsi1}
\widetilde \psi_1 &=& \psi_1-\H\alpha_1 \,,\\
\label{transB1}
\widetilde B_1 &=& B_1-\alpha_1+\beta_1'\,,\\
\label{transE1}
\widetilde E_1 &=& E_1+\beta_1\,,
\eea
and for the vector perturbations as
\bea
\label{transS1}
\widetilde {S_{1}^{~i}} &=& S_{1}^{~i}-\gami1'\,, \\
\label{transF1}
\widetilde {F_{1}^{~i}} &=& F_{1}^{~i}+\gami1\,.
\eea

The first order tensor perturbation is found to be gauge-invariant,
\be
\widetilde h_{1ij} = h_{1ij}\,.
\ee
by substituting \eqs{transphi1} to (\ref{transF1}) into
\eq{Cij1trans}.

For later use, we note that the scalar shear potential,
$\sh_1=E_1'-B_1$, defined in \eq{shearscalar1} above, and the
combination $v_1+B_1$ corresponding to the momentum scalar,
transform as
\bea
\wt{\sh_1}&=&\sh_1+\alpha_1\,,\\
\wt{v_1}+\wt{B_1}&=&v_1+B_1-\alpha_1\,.
\eea

\subsubsection{Second order gauge transformations}

The metric tensor transforms at second order, from \eqs{Ttrans2} and
(\ref{lie_tensor}), as
\bea
\label{general_gmunu2}
\wt{\delta g^{(2)}_{\mu\nu}}&=&\delta g^{(2)}_{\mu\nu}
+g^{(0)}_{\mu\nu,\lambda}\xi^\lambda_2
+g^{(0)}_{\mu\lambda}\xi^\lambda_{2~,\nu}
+g^{(0)}_{\lambda\nu}\xi^\lambda_{2~,\mu}
+2\Big[
\delta g^{(1)}_{\mu\nu,\lambda}\xi^\lambda_1
+\delta g^{(1)}_{\mu\lambda}\xi^\lambda_{1~,\nu}
+\delta g^{(1)}_{\lambda\nu}\xi^\lambda_{1~,\mu}
\Big]\nonumber \\
&&+g^{(0)}_{\mu\nu,\lambda\alpha}\xi^\lambda_1\xi^\alpha_1
+g^{(0)}_{\mu\nu,\lambda}\xi^\lambda_{1~,\alpha}\xi^\alpha_1
+2\Big[
g^{(0)}_{\mu\lambda,\alpha} \xi^\alpha_1\xi^\lambda_{1~,\nu}
+g^{(0)}_{\lambda\nu,\alpha} \xi^\alpha_1\xi^\lambda_{1~,\mu}
+g^{(0)}_{\lambda\alpha}  \xi^\lambda_{1~,\mu} \xi^\alpha_{1~,\nu}
\Big]
\nonumber \\
&&+g^{(0)}_{\mu\lambda}\left(
\xi^\lambda_{1~,\nu\alpha}\xi^\alpha_1
+\xi^\lambda_{1~,\alpha}\xi^\alpha_{1,~\nu}
\right)
+g^{(0)}_{\lambda\nu}\left(
\xi^\lambda_{1~,\mu\alpha}\xi^\alpha_1
+\xi^\lambda_{1~,\alpha}\xi^\alpha_{1,~\mu}
\right)\,.
\eea

As at first order, in the previous subsection, we get the
transformation behaviour for the second order lapse function $\A2$
straight from the $0-0$ component of \eq{general_gmunu2}, which
gives
\bea
\label{transphi2}
\widetilde {\A2} &=& \A2+\H\alpha_2+{\alpha_2}'
+\alpha_1\left[{\alpha_1}''+5\H{\alpha_1}' +\left(\H'+2\H^2
\right)\alpha_1 +4\H\phi_1+2\phi_1'\right]\nonumber \\
&&+2{\alpha_1}'\left({\alpha_1}'+2\phi_1\right)
+\xi_{1k}
\left({\alpha_1}'+\H{\alpha_1}+2\phi_1\right)_{,}^{~k}\nonumber \\
&&+\xi_{1k}'\left[\alpha_{1,}^{~k}-2B_{1k}-{\xi_1^k}'\right]\,.
\eea

The combined scalar and vector $0-i$ metric part transforms from
\eqs{Ttrans2} and (\ref{lie_tensor}) as
\bea
\label{B2itrans}
\widetilde B_{2i}&=&B_{2i}+\xi_{2i}'-\alpha_{2,i}
+\XB_i\,,
\eea
where $B_{2i}$ is similarly vector and scalar combined, and
we defined $\XB_i$ to contain the terms quadratic in the first order
perturbations, as
\bea
\label{defXBi}
\XB_i
\equiv&&
2\Big[
\left(2\H B_{1i}+B_{1i}'\right)\alpha_1
+B_{1i,k}\xi_1^k-2\phi_1\alpha_{1,i}+B_{1k}\xi_{1,~i}^k
+B_{1i}\alpha_1'+2 C_{1ik}{\xi_{1}^k}'
 \Big]\nonumber\\
&&+4\H\alpha_1\left(\xi_{1i}'-\alpha_{1,i}\right)
+\alpha_1'\left(\xi_{1i}'-3\alpha_{1,i}\right)
+\alpha_1\left(\xi_{1i}''-\alpha_{1,i}'\right)\nonumber\\
&&+{\xi_{1}^k}'\left(\xi_{1i,k}+2\xi_{1k,i}\right)
+\xi_{1}^k\left(\xi_{1i,k}'-\alpha_{1,ik}\right)
-\alpha_{1,k}\xi_{1,~i}^k\,.
\eea

To get the transformation behaviour of the vector and the scalar part
separately, we take the divergence of \eq{B2itrans} and find after
applying the inverse Laplacian, the transformation scalar part $\wt
B_2$,
\bea
\label{transB2}
\widetilde B_{2} &=& B_{2}-\alpha_2+\beta_2' +\nabla^{-2} \XB^k_{~,k}
\,,
\eea
or explicitly
\bea
\label{transB2detail}
\widetilde B_{2} &=& B_{2}-\alpha_2+\beta_2' \nonumber\\
&+& \nabla^{-2}\Bigg\{ 2\Big[ \nabla^2\left(2\H
B_{1}+B_{1}'\right)\alpha_1 +\left(2\H
B_{1k}+B_{1k}'\right)\alpha_{1,}^{~k} +\nabla^2
B_{1,k}\xi_1^{k}+B_{1~,k}^{~l}\xi_{1~,l}^{~k}
\nonumber\\
&&\qquad\qquad -2\phi_1\nabla^2\alpha_1-2\phi_{1,k}\alpha_{1,}^{~k}
+ B_{1k}\nabla^2\xi_1^{k}+B_{1k,}^{~~l}\xi_{1~,l}^{~k}
\nabla^2 B_{1}\alpha_1'+B_{1k}{\alpha_{1,}^{~k}}'\nonumber\\
&&\qquad\qquad +2C_{1~k}^{l}{\xi_{1~,l}^{~k}}'
+2C_{1~k,l}^{l}{\xi_{1}^{~k}}' \Big]
+4\H\Big[\alpha_1\nabla^2\left(\beta_1'-\alpha_1\right)
+\alpha_{1,k}\left({\xi_{1}^{~k}}'-\alpha_{1,}^{~k}\right)\Big]
\nonumber\\
&&\qquad +\alpha_1'\nabla^2\left(\beta_1'-3\alpha_1\right)
+\alpha_{1,k}\left({\xi_{1}^{~k}}'-3\alpha_{1,}^{~k}\right)
+\alpha_1\nabla^2\left(\beta_1''-\alpha_1'\right)
+\alpha_{1,k}\left({\xi_{1}^{~k}}''-{\alpha_{1,}^{~k}}'\right)\nonumber\\
&&\qquad +{\xi_{1}^{~k}}'\nabla^2\left(\beta_{1,k}+2\xi_{1k}\right)
+{\xi_{1~~l}^{~k}}'\left({\xi_{1~~k}^{~l}}+2\xi_{1k,}^{~~l}\right)
+{\xi_{1}^{~k}}\nabla^2\left(\beta_{1,k}'-\alpha_{1,k}\right)
\nonumber\\
&&\qquad
+{\xi_{1~~l}^{~k}}\left({\xi_{1~~k}^{~l}}'-\alpha_{1,~~k}^{~l}
\right) -\alpha_{1,k}\nabla^2\xi_{1}^{~k}
-\alpha_{1,k}^{~~l}\xi_{1~~,l}^{~k}
 \Bigg\}\,.
\eea
The vector part is then simply found by subtracting the scalar part
from \eq{B2itrans}, and is given by
\bea
\label{transS2}
\widetilde S_{2i}&=&S_{2i}-\gami2'-\XB_i+\nabla^{-2}\XB^k_{~,ki}
\,.
\eea

We now turn to the transformation behaviour of the perturbations in
the spatial part of the metric tensor. We can follow here along
similar lines as in the linear case. However, the task is made more
complicated not only by the size of the expressions but more
importantly by the fact that now we will have to let inverse
gradients operate on products of first order quantities.

The perturbed spatial part of the metric, $C_{2ij}$, transforms at
second order as
\bea
\label{Cij2trans}
2\widetilde C_{2ij}&=&2C_{2ij}+2\H\alpha_2 \delta_{ij}
+\xi_{2i,j}+\xi_{2j,i}+\X_{ij}\,,
\eea
where we defined $\X_{ij}$ to contain the terms quadratic in the first
order perturbations as
\bea
\label{Xijdef}
\X_{ij}&\equiv&
2\Big[\left(\H^2+\frac{a''}{a}\right)\alpha_1^2
+\H\left(\alpha_1\alpha_1'+\alpha_{1,k}\xi_{1}^{~k}
\right)\Big] \delta_{ij}\nonumber\\
&&
+4\Big[\alpha_1\left(C_{1ij}'+2\H C_{1ij}\right)
+C_{1ij,k}\xi_{1}^{~k}+C_{1ik}\xi_{1~~,j}^{~k}
+C_{1kj}\xi_{1~~,i}^{~k}\Big]
+2\left(B_{1i}\alpha_{1,j}+B_{1j}\alpha_{1,i}\right)
\nonumber\\
&&
+4\H\alpha_1\left( \xi_{1i,j}+\xi_{1j,i}\right)
-2\alpha_{1,i}\alpha_{1,j}+2\xi_{1k,i}\xi_{1~~,j}^{~k}
+\alpha_1\left( \xi_{1i,j}'+\xi_{1j,i}' \right)
+\left(\xi_{1i,jk}+\xi_{1j,ik}\right)\xi_{1}^{~k}
\nonumber\\
&&+\xi_{1i,k}\xi_{1~~,j}^{~k}+\xi_{1j,k}\xi_{1~~,i}^{~k}
+\xi_{1i}'\alpha_{1,j}+\xi_{1j}'\alpha_{1,i}
\,.
\eea
The perturbed spatial part of the metric, $C_{2ij}$, is decomposed in
\eq{decompCij} above into scalar, vector, and tensor part, which we
reproduce here at second order,
\[
2 C_{2ij}=-2\psi_2\delta_{ij}+2E_{2,ij}+2 F_{2(i,j)}+h_{2ij}\,.
\]
Taking the trace of \eq{Cij2trans} we get
\bea
\label{Cij2trace}
-3\wt{\psi_2}+\nabla^2\wt{E_2}
&=&-3{\psi_2}+\nabla^2 {E_2}
+3\H\alpha_2 +\nabla^2\beta_2
+\frac{1}{2}\X^k_{~k}\,,
\eea
where we find $\X^k_{~k}$ to be
\bea
\label{Xtrace}
\frac{1}{2}\X^k_{~k}&=&
3\left(\H^2+\frac{a''}{a}\right)\alpha_1^2
+3\H\left(\alpha_1\alpha_1'+\alpha_{1,k}\xi_{1}^{~k}
\right)\nonumber\\
&& +2\Big[\alpha_1\left( {C_{1~~k}^{~k}}'+2\H{C_{1~~k}^{~k}} \right)
+ {C_{1~~k,l}^{~k}}\xi_{1}^{~l} +2C_{1}^{~kl}\xi_{1l,k}\Big]
\nonumber\\
&& +2B_{1k}\alpha_{1,}^{~~k} -\alpha_{1,k}\alpha_{1,}^{~~k}
 +\xi_{1k,}^{~~~l}\xi_{1~,l}^{~k}
 +\xi_{1k,}^{~~~l}\xi_{1l,}^{~~k}
 \nonumber\\
&& +\alpha_1\nabla^2\left(\beta_1'+4\H\beta_1\right)
 +\nabla^2\beta_{1,k}\xi_{1}^{~k}
 +\xi_{1k}'\alpha_{1,}^{~~k} \,. \eea
Now applying the operator $\p^i\p^j$ to \eq{Cij2trans} we get a second
equation relating the scalar perturbations $\psi_2$ and $E_2$,
\be
\label{Cij2divdiv}
-\wt\nabla^2\psi_2+\nabla^2\nabla^2\wt E_2
=-\nabla^2\psi_2+\nabla^2\nabla^2 E_2+\H\nabla^2\alpha_2
+\nabla^2\nabla^2\beta_2 +\frac{1}{2} \X^{ij}_{~~,ij}  \,,
\ee
This gives the second-order scalar metric perturbations
\be
\label{transpsi2}
\wt\psi_2=\psi_2-\H\alpha_2-\frac{1}{4}\X^k_{~k}
+\frac{1}{4}\nabla^{-2} \X^{ij}_{~~,ij}\,,
\ee
and
\be
\label{transE2}
\wt E_2=E_2+\beta_2+\frac{3}{4}\nabla^{-2}\nabla^{-2}\X^{ij}_{~~,ij}
-\frac{1}{4}\nabla^{-2}\X^k_{~k}\,.
\ee
Taking the divergence of \eq{Cij2trans} we get
\be
2\wt C_{2ij,}^{~~~j}=2 C_{2ij,}^{~~~j} +2\H\alpha_{2,i}
+\nabla^2\xi_{2i}+\nabla^2\beta_{2,i}+\X_{ik,}^{~~k}\,.
\ee
Substituting in our results for $\wt\psi_2$ and $\wt E_2$ we then
arrive at
\be
\nabla^2 \wt F_{2i}= \nabla^2 F_{2i}+\nabla^2 \gamma_{2i}
+\X_{ik,}^{~~~k}-\nabla^{-2}\X^{kl}_{~~,kli}
\,.
\ee
Finally we obtain the second-order vector metric perturbation
\be
\label{transFi2}
\wt F_{2i}= F_{2i}+\gamma_{2i}
+\nabla^{-2}\X_{ik,}^{~~~k}-\nabla^{-2}\nabla^{-2}\X^{kl}_{~~,kli}
\,.
\ee

We can now turn to the tensor perturbation at second order.
Substituting our previous results for $\psi_2$, $E_2$, and $F_{2i}$
into \eq{Cij2trans} we get
\bea
\label{transhij2}
\wt h_{2ij}&=& h_{2ij}+\X_{ij}
+\frac{1}{2}\left(\nabla^{-2}\X^{kl}_{~~,kl}-\X^k_{~k}
\right)\delta_{ij}
+\frac{1}{2}\nabla^{-2}\nabla^{-2}\X^{kl}_{~~,klij}\nonumber\\
&&+\frac{1}{2}\nabla^{-2}\X^k_{~k,ij}
-\nabla^{-2}\left(\X_{ik,~~~j}^{~~~k}+\X_{jk,~~~i}^{~~~k}
\right)
\,.
\eea
Although the second-order tensor transformation $h_{2ij}$ is not
dependent on the second-order part of the gauge transformation,
$\xi_{2}^\mu$, it does depend on first order quantities
quadratically. The tensor metric perturbations are no longer
gauge-invariant at second and higher order.

\subsubsection{The large scale or small $k$ limit}

{}From Eqs.~(\ref{rhotransform2}), (\ref{transpsi2}), and
(\ref{transphi2}) we see that on large scales, where gradient terms
can be neglected, the definition of the second order perturbations
in the ``new'' coordinate is independent of the spatial coordinate
choice (the ``threading'') at second order in the gradients.
It is therefore sufficient on large scales (at $O(k^2)$) to
specify the time slicing by prescribing $\alpha_1$ and
$\alpha_2$, in order to define gauge-invariant variables
\cite{MW2003,LMS,Malik:2006pm}.
The procedure to neglect the gradient terms, is explained in
detail in Ref.~\cite{LMS}.
For the approximation to hold one assumes that each quantity can be
treated as smooth on some sufficiently large scale. Formally one
multiplies each spatial gradient $\partial_i$ by a fictitious parameter
$k$, and expands the exact equations as a power series in
$k$, keeping only the zero- and first-order terms, finally
setting $k=1$.
%

\section{Gauge-invariant variables}
\label{gi_var_sec} \setcounter{equation}{0}

The notion of invariance under coordinate reparametrisation is
central to Einstein's theory of general relativity. This is both a
blessing and a curse in the study of cosmological perturbations. We
are free to pick coordinate systems best adapted to the problem at
hand, but we also obtain apparently different results depending upon
this arbitrary choice of coordinates. This is the gauge problem.

Ultimately physical observables are not dependent on the choice of
coordinate, though observables may be different for different
observers. All one can do is to specify quantities unambiguously,
such that they have a {\em gauge-invariant} definition. This is not
the same as {\em gauge independence}. A quantity like the tensor
metric perturbation, $h_{1ij}$, is truly gauge independent at first
order in that the tensor part of the metric perturbation is the same
in all gauges. The scalar curvature perturbation, $\psi_1$, on the
other hand is intrinsically gauge-dependent. It is different under
different time slicings. (Indeed in the spatially flat slicing the
curvature perturbation is zero by construction.) One can construct
gauge-invariant combinations, which may be referred to as the
gauge-invariant curvature perturbation, but they only correspond
with the curvature perturbation in one particular gauge. As a result
one can find in the literature many different gauge-invariant
curvature perturbations corresponding to the many different choices
of gauge, such as $\Psi$, $\zeta$ and $\R$, corresponding to the
curvature perturbation in the longitudinal, the uniform density, and
the comoving gauge, respectively, to name just three.

In this section we shall show how different gauge invariant
combinations of otherwise gauge dependent quantities can be
constructed by fixing the otherwise arbitrary coordinate
transformations at first order and beyond, yielding gauge invariant
definitions of the physical perturbations in specified gauges.
Residual gauge degrees of freedom only remain in cases where the
coordinate choice is not unambiguously fixed, as in the synchronous
gauge.

At first order the tensor metric perturbation, $h_{1ij}$, is not
affected by the mapping, or by the change of coordinate system, and
hence is gauge-invariant. Thus we only have to construct
gauge-invariant scalar and vector perturbations at first order.
However at second order the tensor part of the metric perturbation
also becomes gauge-dependent.

At first order we can define scalar and vector type gauge-invariant
variables independently of each other, but matters are more
complicated at second order. Whereas we can still specify the
``proper'' second order scalar and vector slicings and threadings
independently, we now also have to specify the first order gauge
functions of both types simultaneously.

We could specify different gauges at first and second order, but we
would be losing the physical interpretation of the quantity thus
constructed.  We therefore choose the same gauge at first and second
order, and at first order the same physical gauge condition for the
scalar and vector gauge functions.
It is however sometimes necessary to combine different temporal and
spatial gauge conditions. For example imposing the uniform density
condition only specifies the slicing, and we are free to combine it
with a flat threading.

\subsection{Longitudinal gauge}
\label{longitudinal1}

\subsubsection{First order}

The gauge-dependence of the metric perturbations lead
Bardeen~\cite{Bardeen80} to propose that only quantities that are
explicitly invariant under gauge transformations should be
considered. By studying the transformation
Eqs.~(\ref{transphi1}--\ref{transE1}), Bardeen\footnote{In Bardeen's
notation these gauge-invariant perturbations are given as
$\Phi\equiv\Phi_A Q^{(0)}$ and $\Psi\equiv-\Phi_H Q^{(0)}$.}
constructed two such quantities~\cite{Bardeen80}
\begin{eqnarray}
\label{defPhi}
\Phi &\equiv& \A1 + \H(B_1-E_1') + (B_1-E_1')'  \,,\\
\label{defPsi}
\Psi &\equiv& \psi_1 - \H \left( B_1-E_1' \right) \,.
\end{eqnarray}
These turn out to coincide with the scalar metric perturbations in a
particular gauge, called variously the orthogonal
zero-shear~\cite{Bardeen80,KS}, conformal
Newtonian~\cite{Bertschinger1995,Ma1995} or longitudinal
gauge~\cite{MFB}.
It may therefore appear that this gauge is somehow preferred over
other gauge choices. However any unambiguous choice of time-slicing
and threading can be used to define explicitly gauge-invariant
perturbations. The longitudinal gauge of Ref.~\cite{MFB} provides but
one example.

The two scalar gauge functions, $\alpha$ and $\beta$ defined in
Eq.~(\ref{def_xi1}), which represent different choices of
time-slicings (choice of spatial hypersurfaces) and threading
(choice of spatial coordinates on these hypersurfaces) respectively,
allow two of the scalar metric perturbations to be eliminated,
implying that the two remaining gauge-invariant combinations should
then be gauge-invariant.
If we choose to work on spatial hypersurfaces with vanishing shear,
we find from Eqs.~(\ref{transB1}),(\ref{transE1}) and
(\ref{shearscalar1}) that the shear scalar transforms as
$\wt{\sigma_1}=\sigma_1+\alpha_1$ and this implies that to obtain
perturbations in the longitudinal gauge starting from arbitrary
coordinates we should perform a transformation
\begin{equation}
 \label{alpha1l}
\alpha_{1\ell} = -\sh_1= B_1-E_1' \,.
\end{equation}
This is sufficient to determine the geometrical perturbations
$\phi_1$, $\psi_1$, $\sigma_1$ or other scalar quantities on these
hypersurfaces. In addition, the longitudinal gauge is completely
determined by the spatial gauge choice $\wt{E_{1\ell}}=0$ [and hence
from Eq.~(\ref{alpha1l}) $\wt{B_{1\ell}}=0$] which requires from
Eq.~(\ref{transE1})
\begin{equation}
 \label{beta1l}
\beta_{1\ell} = -E_1 \,.
\end{equation}
The remaining scalar metric perturbations, $\phi_1$ and $\psi_1$,
are given from Eqs.~(\ref{transphi1}) and (\ref{transpsi1})
as
\begin{eqnarray}
\label{defphi1l}
\wt{\A{}_{1\ell}} &=& \A1 + \H(B_1-E_1') + (B_1-E_1')' \,, \\
\label{defpsi1l} \wt{\psi_{1\ell}} &=& \psi_1 - \H \left( B_1-E_1'
\right) \,,
\end{eqnarray}
Note that $\wt{\A{}_{1\ell}}$ and $\wt{\psi_{1\ell}}$ are then
identical to $\Phi$ and $\Psi$ defined in Eqs.~(\ref{defPhi})
and~(\ref{defPsi}).

The fluid density perturbation, $\delta\rho_1$, and scalar velocity,
$v_1$, are given from (\ref{rhotransform1}) and (\ref{transv1})
\begin{eqnarray}
\label{defrho1l}
 \wt{\delta\rho_{1\ell}} &=& \delta \rho_1 + \rho_0'
\left(B_1-E_1'\right) \,,\\
\label{defv1l}
 \wt{v_{1\ell}} &=& v_1 + E_1' \,.
\end{eqnarray}
These gauge-invariant quantities are simply a gauge-invariant
definition of the perturbations in the longitudinal gauge.

This gauge is widely used, for example, throughout Ref.~\cite{MFB}.
It has also proven useful for calculations on small scales, since it
gives evolution equations closest to the Newtonian ones,
e.g.~Ref.~\cite{Green:2005fa}.
Recently it has also become popular in backreaction studies,
e.g.~\cite{Behrend:2007mf,Kolb:2008bn,VanAcoleyen:2008cy}.
After imposing the gauge-conditions the metric tensor is diagonal,
which simplifies many calculations, for example the derivation of the
governing equations of the Boltzmann-hierarchy.
Moreover we shall show in Section~\ref{sect:dynamics} that in many
cases of physical interest (in the absence of anisotropic stress)
one finds $\Phi=\Psi$ and there is only one variable required to
describe all scalar metric perturbations.

However, it can be difficult to define quantities in this gauge in the
super-horizon limit, since in the super-horizon limit the shear
vanishes and hence numerical instabilities can occur on large scales
in the longitudinal gauge, see
e.g.~Refs.~\cite{Ma1995,Hu:1998gy,Lewis:1999bs}.

The extension to include vector and tensor metric perturbations is
called the Poisson gauge \cite{Bertschinger1995,Bruni96}. Tensor
metric perturbations are automatically gauge independent at first
order (and hence gauge invariant). Eliminating the spatial part of
the contravariant vector field $n^\mu$ in Eq.~(\ref{defnmuup})
requires both $\wt{B_{\ell}}_,^{~i}$ and $\wt{S_\ell}^i=0$ which
from Eq.~(\ref{transS1}) fixes the vector part of the spatial gauge
transformation~(\ref{def_xi1})
 \be
 \label{gamma1l}
 \gamma_{1\ell}^i = \int S_1^i d\eta + \hat\C_1^i(x^j) \,,
 \ee
up to an arbitrary constant 3-vector $\hat\C_1^i$ which depends on
the choice of spatial coordinates on an initial hypersurface. The
remaining vector metric perturbation is
\be
\label{def:Sgi}
\wt{F_{1\ell}}^i = F_1^{i\prime}+\int S_1^i d\eta + \hat\C_1^i(x^j)
\,. 
\ee

\subsubsection{Second order}

It is possible to extend the longitudinal, or Poisson, gauge to
higher orders. The principle for constructing gauge invariant
variables remains the same. We use a physical choice of gauge to
specify the vector field $\xi^\mu$ generating the transformation
(\ref{defcoordtrans}) from an arbitrary gauge
\cite{MW2003,Nakamura:2003wk}.
Requiring first that
$\wt{E_2}_\ell=0$ fixes the scalar part of the spatial gauge using
Eq.~(\ref{transE2}), which gives
\be
\beta_{2\ell}=- E_2 - \frac{3}{4}\nabla^{-2}\nabla^{-2}\X^{ij}_{~~,ij}
+\frac{1}{4}\nabla^{-2}\X^k_{~k}\,.
\ee
Note that having already imposed the Poisson gauge at first order,
$\alpha_1$, $\beta_1$ and $\gamma_{1i}$ are fixed by \eqs{alpha1l},
(\ref{beta1l}) and~(\ref{gamma1l}), and thus so is $\X_{ij}$, given
in \eq{Xijdef}.

Requiring that the scalar part of the perturbed shift function
vanishes, $\wt{B_2}_\ell=0$, then sets the temporal gauge,
$\alpha_{2\ell}$ using Eq.~(\ref{transB2}), while requiring that the
vector part vanishes, $F_2^i={\bf 0}$, can be used along with
Eq.~(\ref{transFi2}) to fix the vector part of the spatial gauge,
$\gamma_{2\ell}^i$, up to a constant of integration as at first
order, Eq.~(\ref{gamma1l}).

We then obtain gauge invariant definition of
$\Phi$, $\Psi$ and other perturbations at second order by
substituting these specific gauge transformations into
Eq.~(\ref{transphi2}) and (\ref{transpsi2}) to obtain
\bea
\label{transphi2l}
\widetilde {\A{2\ell}} &=& \A2+\H\alpha_{2\ell}+{\alpha_{2\ell}}'
+\alpha_{1\ell}\left[{\alpha_{1\ell}}''+5\H{\alpha_{1\ell}}' +\left(\H'+2\H^2
\right)\alpha_{1\ell} +4\H\phi_1+2\phi_1'\right]\nonumber \\
&&+2{\alpha_{1\ell}}'\left({\alpha_{1\ell}}'+2\phi_1\right)
+\xi_{1\ell k}
\left({\alpha_{1\ell}}'+\H{\alpha_{1\ell}}+2\phi_1\right)_{,}^{~k}\nonumber \\
&&+\xi_{1\ell k}'\left[\alpha_{1\ell,}^{~k}-2B_{1k}-{\xi_{1\ell}^k}'\right]\,.\\
\label{transpsi2l}
\wt{\psi_{2\ell}} &=& \psi_2-\H\alpha_{2\ell}-\frac{1}{4}\X^{~k}_{\ell k}
+\frac{1}{4}\nabla^{-2} \X^{~ij}_{\ell~~,ij}\,,
\eea
where $\X_{\ell ij}$ is denotes the quadratic first order terms in
Eq.~(\ref{Xijdef}) using the longitudinal gauge transforms
$\alpha_{1\ell}$ and $\xi_{\ell i}$.

The tensor (transverse, tracefree) part of the metric perturbation at
second order, $h_{ij}$ in Eq.~(\ref{transhij2}), is not affected by
the second order gauge transformations $\alpha_2$ and $\xi_{2i}$, but
it does depend on the choice of gauge at first order, $\alpha_1$ and
$\xi_{1i}$. Thus we need to include the corresponding first-order
gauge definition to obtain a gauge invariant definition of the tensor
metric perturbations at second order.

In particular, recent work on the generation of gravitational waves at
second order
\cite{Mollerach:2003nq,Nakamura:2004rm,Ananda:2006af,Baumann:2007zm,Bartolo2007}
has calculated the resulting tensor mode in the Poisson gauge. To give
a gauge invariant definition of the tensor perturbation in the Poisson
gauge one needs to explicitly include the transverse and tracefree
(tensor) part of the second order gauge transformation from an
arbitrary gauge.  {}From Eqs.~(\ref{transhij2}) we obtain the gauge
invariant definition of the tensor metric perturbation in the Poisson
gauge:
\bea
\label{transhij2l}
\wt h_{2\ell ij}&=& h_{2ij}+\X_{\ell ij}
+\frac{1}{2}\left(\nabla^{-2}\X^{\ell kl}_{~~,kl}-\X^{~k}_{\ell k}
\right)\delta_{ij}
+\frac{1}{2}\nabla^{-2}\nabla^{-2}\X^{\ell kl}_{~~~,klij}\nonumber\\
&&+\frac{1}{2}\nabla^{-2}\X^{~k}_{\ell k,ij}
-\nabla^{-2}\left(\X_{\ell ik,~~~j}^{~~~~k}+\X_{\ell jk,~~~i}^{~~~~k}
\right)
\,.
\eea
%

\subsection{Spatially flat gauge}

\subsubsection{First order}

An alternative gauge choice, defined purely by local metric
quantities is the spatially flat or uniform curvature
gauge~\cite{KS,Hwang1993,Hwang1994,Hwang1996,Stewart1993}, also
called the off-diagonal gauge~\cite{Brustein1994}. In
this gauge one selects spatial hypersurfaces on which the induced
3-metric on spatial hypersurfaces is left unperturbed by scalar or
vector perturbations, which requires $\wt{\psi_1}=\wt{E_1}=0$ and
$\wt{F_{1}}_i={\bf 0}$. Using Eqs.~(\ref{transpsi1}),
(\ref{transE1}) and~(\ref{transF1}) this corresponds to a gauge
transformation (\ref{def_xi1}) where
\begin{equation}
\label{deffg1}
\alpha_{1\fg} = \frac{\psi_1}{\H} \,, \qquad
\beta_{1\fg} = -E_1 \,, \qquad
\gamma_{1\fg}^i = -F_1^i \,.
\end{equation}
The gauge-invariant definitions of the remaining scalar metric
degrees of freedom are then from Eqs.~(\ref{transphi1})
and~(\ref{transB1})
\begin{eqnarray}
\wt{\A{}_{1\fg}} &=& \A1 + \psi_1 + \left( \frac{\psi_1}{\H}
\right)^{\prime} \,, \\
\wt{B_{1\fg}} &=& B_1-E_1'-\frac{\psi_1}{\H}  \, .
\end{eqnarray}
These gauge-invariant combinations were denoted ${\cal A}$ and ${\cal
B}$ by Kodama and Sasaki~\cite{KS}.
The gauge-invariant definition of the remaining vector metric
perturbation is the time derivative of the vector metric
perturbation in the Poisson gauge (\ref{def:Sgi}):
\be
\wt{S_{1\fg}}^i = S_1^i+F_1^{i\prime}
= \wt{F_{1\ell}}^{i\prime}
\,.
\ee

Perturbations of scalar quantities in this gauge, such as the
density perturbation, have gauge invariant definitions from
Eq.~(\ref{rhotransform1}):
\begin{equation}
\label{defrho1psi} \wt{\delta\rho_{1\fg}} = \delta\rho_1 +
\rho_0'\frac{\psi_1}{\H}  \,,
\end{equation}
and the velocity potential (\ref{transv1}) is given from
Eq.~(\ref{transv1}):
\be
\wt{v_{1\fg}} = v_1+E_1'\,.
\ee

The shear perturbation in the spatially flat gauge is given by
$\wt{\sigma_{1\fg}}=-\wt{B_{1\fg}}$. This is closely related to the
curvature perturbation in the zero-shear (longitudinal) gauge,
$\wt{\psi_\ell}=\Psi$, given in \eqs{defPsi} or (\ref{defpsi1l}),
\be
\wt{B_{1\fg}} = -\frac{\wt{\psi_{1\ell}}}{\H} =
-\frac{\Psi}{\H} \,.
\ee
Gauge-invariant quantities, such as $\wt{B_{1\fg}}$ or
$\wt{\psi_{1\ell}}$ are proportional to the displacement between two
different choices of spatial hypersurface,
\be
\wt{B_{1\fg}} = -\frac{\wt{\psi_{1\ell}}}{\H}
= \alpha_{1\fg} - \alpha_{1\ell} \,,
\ee
which would vanish for a homogeneous cosmology.

In some circumstances it is more convenient to use the spatially-flat
gauge-invariant variables instead of those in the longitudinal
gauge. For instance, when calculating the evolution of perturbations
during a collapsing ``pre Big Bang'' era the perturbations
$\wt{\A{}_{1\fg}}$ and $\wt{B_{1\fg}}$ may remain small even when
$\Phi$ and $\Psi$ become large~\cite{Brustein1994,Copeland1997}.
On the other hand the metric perturbation $\wt{B_{1\fg}}$ grows on
large scales in radiation or matter dominated eras,
$\wt{B_{1\fg}}\propto\H^{-1}\propto \eta$, when the longitudinal gauge
metric perturbation $\Psi$ remains constant.

Note that the scalar field perturbation on spatially flat
hypersurfaces,
\be
\label{defvpflat1}
\wt{\delta\vp_{1\fg}}\equiv\dvp1+\varphi_0'\frac{\psi_1}{\H}
\,,
\ee
is the gauge-invariant Sasaki-Mukhanov
variable~\cite{Sasaki1986,Mukhanov1988}, often denoted $\Q$.

\subsubsection{Second order}

At second order we get from the gauge condition $\wt{\psi_2}=0$ using
\eq{transpsi2}
\bea
 \label{deffg2}
\alpha_{2\fg}=\frac{\psi_2}{\H}+\frac{1}{4\H}\left[
\nabla^{-2}\X^{ij}_{\fg,ij}-\X^k_{\fg k}\right]\,, \eea
where we get $\X_{\fg ij}$ from \eq{Xijdef} using the first order gauge
generators given above, as
\bea
\label{Xijflat}
\X_{\fg ij}&=&2\left[
\psi_1\left(\frac{\psi_1'}{\H}+2\psi_1\right)+\psi_{1,k}\xi_{1\fg}^k
\right]\delta_{ij}
+\frac{4}{\H}\psi_1\left(C_{1ij}'+2\H C_{1ij}\right)\nonumber\\
&& +4 C_{1ij,k}\xi_{1\fg}^k + \left(4 C_{1ik}+\xi_{1\fg
i,k}\right)\xi_{1\fg,j}^k
+ \left(4 C_{1jk}+\xi_{1\fg j,k}\right)\xi_{1\fg,i}^k\nonumber\\
&&
+\frac{1}{\H}\Big[
\psi_{1,i}\left(2B_{1j}+\xi_{1\fg j}'\right)
+\psi_{1,j}\left(2B_{1i}+\xi_{1\fg i}'\right)
\Big]
-\frac{2}{\H^2}\psi_{1,i}\psi_{1,j}\nonumber\\
&&
+\frac{2}{\H}\psi_1\left(
\xi_{1\fg (i,j)}'+4\H  \xi_{1\fg (i,j)}\right)
+2\xi_{1\fg}^k\xi_{1\fg (i,j)k}
+2\xi_{1\fg k,i}\xi_{1\fg,j}^k\,,\nonumber\\
\eea
where we define
\be
\xi_{1\fg i}=-\left(E_{1,i}+F_{1i}\right)\,.
\ee
The trace of \eq{Xijflat} is then
\bea
\X^k_{\fg k}&=&6\left[
\psi_1\left(\frac{\psi_1'}{\H}+2\psi_1\right)+\psi_{1,k}\xi_{1\fg}^k
\right]
+\frac{4}{\H}\psi_1\left(C_{1~k}^{k\prime}+2\H C_{1~k}^{k} \right)\nonumber\\
&& +4 C_{1~k,l}^{k}\xi_{1\fg}^l + 4 \left( 2 C_{1}^{kl}+\xi_{1\fg
,}^{k~~~~l}\right)\xi_{1\fg (k,l)} -2\nabla^2E_{1,k}\xi_{1\fg}^k
\\
&&
+\frac{2}{\H}\left(
2B_{1k}+\xi_{1\fg k}'-\frac{1}{\H}\psi_{1,k}\right)\psi_{1,}^{~k}
-\frac{2}{\H}\left(\psi_1\nabla^2E_1'+4\H\nabla^2E_1\right)
\,.\nonumber
\eea

As an example of a second-order scalar metric perturbation we give
the lapse function in the flat gauge
\bea
\wt{\A{}_{2\fg}}&=&
\A2+\frac{1}{\H}\left[
\psi_2'+\left(\H-\frac{\H'}{\H}\right)\psi_2\right]
\\
&&
+\frac{1}{4\H}\left[
\nabla^{-2}\X^{kl\prime}_{\fg,kl}-\X^{k\prime}_{\fg k}
+\left(\H-\frac{\H'}{\H}\right)\left(
\nabla^{-2}\X^{kl}_{\fg,kl}-\X^k_{\fg k}
\right)
\right]\nonumber\\
&&
+\frac{1}{\H^2}\left(\psi_1''\psi_1+2\psi_1^{\prime 2}\right)
+\left(2-\frac{\H''}{\H^3}\right)\psi_1^2
+\frac{1}{\H}\left(5-6\frac{\H'}{\H^2}\right)\psi_1\psi_1'
+\frac{2}{\H}\A1'\psi_1\nonumber\\
&&
+\frac{4}{\H}\A1 \left[
\psi_1'+\left(\H-\frac{\H'}{\H}\right)\psi_1\right]
+\frac{1}{\H}\left[
\psi_1'+\left(\H-\frac{\H'}{\H}\right)\psi_1
+2\H \A1\right]_{,k} \xi_{1\fg}^{k}
\nonumber\\
&&
+\frac{1}{\H}\left[
\left(\psi_1'+\left(\H-\frac{\H'}{\H}\right)\psi_1\right)_{,k}
-2\H B_{ik}\right]\xi_{1\fg}^{k\prime}\,.
\nonumber
\eea
The second order tensor perturbation is in the flat gauge
\bea
\label{transhij2flat}
\wt h_{2\fg ij}&=& h_{2ij}+\X_{\fg ij}
+\frac{1}{2}\left(\nabla^{-2}\X^{kl}_{\fg,kl}-\X^k_{\fg k}
\right)\delta_{ij}
+\frac{1}{2}\nabla^{-2}\nabla^{-2}\X^{kl}_{\fg ,klij}\nonumber\\
&&+\frac{1}{2}\nabla^{-2}\X^k_{\fg k,ij}
-\nabla^{-2}\left(\X_{\fg ik,j}^{~~~k}+\X_{\fg jk,i}^{~~~k}
\right)
\,.
\eea
As an example of a matter variable we choose the energy density, which
in the flat gauge is
\bea
\label{rhotransform2flat}
\widetilde{\delta\rho_{2\fg}} &=& \drhorho
+\frac{\rho_0'}{\H}\psi_2
+\frac{\rho_0'}{4\H}\left(
\nabla^{-2}\X^{ij}_{\fg,ij}-\X^k_{\fg k}\right)
\\
&&+\frac{\psi_1}{\H^2}\Big[\rho_0'' {\psi_1}
+\rho_0'\left(\psi_1'-\frac{\H'}{\H}\psi_1\right)+2\H\delta\rho_1'\Big]
+\left(2\drho+\frac{\rho_0'}{\H}\psi_1\right)_{,k}\xi_{1\fg}^k
\,.\nonumber
\eea

To show the relation between gauge-invariant perturbations defined
in the flat gauge and those previously defined in the longitudinal
gauge we note that in the longitudinal gauge we have
\be
 2\wt{C_{1\ell}}_{ij} = -2\Psi\delta_{ij} + 2 \wt{F_{1\ell}}_{(i,j)} +
 h_{1ij} \,,
 \ee
and the first-order gauge shifts to the flat gauge,
Eqs.~(\ref{deffg1}), become
\begin{equation}
\label{deffg1ell}
\alpha_{1\fg}|_\ell = \frac{\Psi}{\H} \,, \qquad
\beta_{1\fg}|_\ell = 0 \,, \qquad
\gamma_{1\fg}^i = -\wt{F_{1\ell}}^i \,.
\end{equation}
Substituting these into Eq.(\ref{Xijflat}) we obtain
\bea
 \label{Xijflatell}
 \X_{\fg ij}|_\ell &=&\frac{2}{\H}\left[ - \Psi\left(\Psi'+2\H\Psi\right)
- \Psi_{,k} \wt{F_{1\ell}}^k
\right]\delta_{ij}
+\frac{2}{\H}\Psi \left( h_{1ij}'+2\H h_{1ij}\right)\nonumber\\
&& +2 \left( 2\Psi_{,k} \delta_{ij} - 2 \wt{F_{1\ell}}_{(i,j)k} -
h_{ij,k} \right) \wt{F_{1\ell}}^k
 - 4 h_{k(i} \wt{F_{1\ell}}_{,j)}^k
 \nonumber\\
&& - \frac{2}{\H} \Psi_{,(i} \wt{F_{1\ell}}_{j)}
 -\frac{2}{\H^2}\Psi_{,i}\Psi_{,j}\nonumber\\
&& - \frac{2}{\H}\Psi \wt{F_{1\ell}}_{(i,j)}'
 + 2 \wt{F_{1\ell}}^k\wt{F_{1\ell}}_{(i,j)k}
 + 2 \wt{F_{1\ell}}_{k,i}\wt{F_{1\ell}}^k_{,j}
 \,.
\eea
If we can neglect the first-order vector and tensor perturbations,
$\wt{F_{1\ell}}_i$ and $h_{1ij}$, then we have
\be
 \label{Xijflatellspecial}
 \X_{\fg ij}|_\ell = - \frac{2}{\H} \Psi\left(\Psi'+2\H\Psi\right) \delta_{ij}
 -\frac{2}{\H^2}\Psi_{,i}\Psi_{,j}
 \,.
\ee

Similarly, neglecting first vector and tensor perturbations,
i.e.~setting $F_{1i}=S_{1i}=h_{1ij}=0$ and hence only considering
scalar perturbations, the energy density at second order,
\eq{rhotransform2flat}, simplifies to
\bea
\label{rhotransform2flatscal}
&&\widetilde{\delta\rho_{2\fg}} = \delta\rho_{2}
+\frac{\rho_0'}{\H}\psi_2
+\frac{\psi_1}{\H^2}\Big[\left(\rho_0''+2\H\rho_0'\right) {\psi_1}
+\rho_0'\left(2\psi_1'-\frac{\H'}{\H}\psi_1\right)+2\H\delta\rho_1'\Big]
\nonumber\\
&&
-2\left(\drho+\frac{\rho_0'}{\H}\psi_1\right)_{,k}E_{1,}^{~k}
\nonumber\\
&&
+\frac{\rho_0'}{2\H}\Bigg\{
E_{1,kl}E_{1,}^{~kl}+\nabla^2E_{1,k}E_{1,}^{~k}
-2\psi_1\nabla^2\left(\frac{E_1'}{\H}+2E_1\right)
-\frac{\psi_{1,k}}{\H}\left(
2B_1+E_1'-\frac{\psi_{1}}{\H}\right)_{,k}\nonumber\\
&&+\nabla^{-2}\Big[2\psi_1\left(
\frac{E_{1,}^{\prime~ ij}}{\H}+2E_{1,}^{~ij}\right)
-E_{1,}^{~ijk}E_{1,k}-E_{1,}^{~ik}E_{1,k}^{~~j}
+\frac{\psi_{1,}^{~(i}}{\H}\left(
2B_1+E_1'-\frac{\psi_{1}}{\H}\right)_{,}^{~j)}
\Big]_{,ij}
\Bigg\}
\,.\nonumber\\
\eea
This expression was first derived in Ref.~\cite{MW2003}, however with
a different, incorrect sub-horizon part~\cite{Pitrou}.
On super-horizon scales, where gradient terms can be neglected, we
recover the expressions given in Ref.~\cite{M2005}.

\subsection{Synchronous gauge}

The synchronous gauge is defined by
$\wt{\A{}}=\wt{B}_i=0$, so that the proper time for observers at
fixed spatial coordinates coincides with cosmic time in the FRW
background, i.e., $d\tau = ad\eta$. This simplifies dynamical
equations as the time derivatives can be directly related to proper
time derivatives.
This gauge is very popular for numerical studies, and used in many
Boltzmann solvers such as CMBFAST \cite{cmbfast}. It is also popular
in the older literature \cite{weinberg,Peebles80}.

The gauge condition at first order is $\wt{\A1}=\wt{B_{1}}_i=0$,
which from Eqs.~(\ref{transphi1}) and (\ref{transB1}) gives
\bea \alpha_{1\syn}
&=&-\frac{1}{a}\left(\int a\phi_1 d\eta-\C_1(x^i)\right)\,,\\
\beta_{1\syn}&=&\int\left( \alpha_{1\syn}-B_1\right) d\eta
+\hat\C_1(x^i)\,,\\
 \gamma_{1\syn}^i &=& \int S_1^i d\eta + \hat\C_1^i(x^i) \,.
 \eea
This does not determine the time-slicing unambiguously and we are
left with two arbitrary scalar functions of the spatial coordinates,
$\C_1$ and $\hat\C_1$. Note that $\hat\C_{1,i}+\hat\C_{1i}$ affects
only the labelling of the coordinates on the initial spatial
hypersurface, but $\C_1$ affects scalar perturbations on spatial
hypersurfaces.
We are left with two non-zero geometrical scalar perturbations,
\bea
\wt{\psi_{1\syn}}
&=&\psi_1+\frac{\H}{a}\left(\int a\phi_1 d\eta-\C(x^i)\right)\,,\\
\wt\sh_{1\syn} &=& \sh_1 + \alpha_{1\syn}-B_1 \,,
 \eea
and the matter variables are
\bea
\wt{\delta\rho_{1\syn}} &=&\delta\rho_1
-\frac{\rho_0'}{a}\left(\int a\phi_1 d\eta-\C(x^i)\right)\,,\\
\wt{v_{1\syn}}&=&v_1+B_1- \alpha_{1\syn}\,.
\eea
Thus it is not possible to define gauge-invariant quantities in
general using this gauge condition~\cite{Martin:1997zd}. This gauge
was originally used by Lifshitz in his pioneering work on
perturbations in a FRW spacetime \cite{Lifshitz1945} (see also
Ref.~\cite{Landau:1987gn}). He dealt with the residual gauge freedom
by eliminating the unphysical gauge modes through symmetry arguments.

To remove the ambiguity, we can follow Ref.~\cite{Bucher:1999re} and
choose the initial velocity of cold dark matter to be zero,
$\wt{v_{1{\rm cdm}}}\equiv 0$, which fixes the residual gauge
freedom
 \be
 \C_1(x) = a (v_{1{\rm cdm}} + B_1) \,.
 \ee
Note that for pressureless matter, momentum conservation equation
ensures that $a(v_{1{\rm cdm}}+B_1)$ is a constant (see
Section~\ref{sect:dynamics}).

\subsection{Comoving orthogonal gauge}

The comoving gauge is defined by choosing spatial coordinates such
that the 3-velocity of the fluid vanishes, $\wt{v}_i=0$.
Orthogonality of the constant-$\eta$ hypersurfaces to the
4-velocity, $u^\mu$, then requires $\wt{v}_i+\wt{B}_i=0$, which
shows that the momentum vanishes as well. From Eqs.~(\ref{transB1})
and~(\ref{transv1}) this implies
\begin{eqnarray}
\label{alpha1com}
\alpha_{1\com} &=& v_1+B_1 \, , \nonumber \\
\beta_{1\com}&=& \int v_1 d\eta + \hat\C(x^i) \, ,
\end{eqnarray}
where $\hat\C(x^i)$ represents a residual gauge freedom,
corresponding to a constant shift of the spatial coordinates.  All
the 3-scalars like curvature, expansion, acceleration and shear are
independent of $\hat\C(x^i)$.  Applying the above transformation
from arbitrary coordinates, the scalar perturbations in the comoving
orthogonal gauge can be written as
\begin{eqnarray}
\label{defphi1com} \wt{{\A{}}_{1\com}} &=&
 \A1+\H\left(v_1+B_1\right)+\left(v_1'+B_1'\right)
 \, , \\
\label{defpsi1com}
\label{defR}
\R \equiv \wt{\psi_{1\com}}
&=& \psi_1 - \H \left( v_1+B_1 \right) \, , \\
\wt{\sh_{1\com}} &=& v_1 + E_1' \,.
\end{eqnarray}
Defined in this way, these combinations are gauge-invariant under
transformations of their component parts in exactly the same way as,
for instance, $\Phi$ and $\Psi$ defined in Eqs.~(\ref{defPhi})
and~(\ref{defPsi}).

Note that the curvature perturbation in the comoving gauge given
above, Eq.~(\ref{defpsi1com}) was used (with a constant pre-factor)
by Lukash \cite{Lukash80}. It was later employed by Lyth and denoted
$\cal{R}$ in his seminal paper, \cite{Lyth85}, and in many
subsequent works, e.g.~\cite{Liddle1993} and \cite{Lidsey:1995np}.

The density perturbation on the comoving orthogonal hypersurfaces is
given by Eqs.~(\ref{rhotransform1}) and~(\ref{alpha1com}) in
gauge-invariant form as
\begin{equation}
 \label{defrho1com}
 \wt{\delta\rho_{1\com}} = \delta\rho_1+\rho_0'\left(v_1+B_1\right) \,,
\end{equation}
and corresponds to the gauge-invariant density perturbation
$\epsilon_mE_0Q^{(0)}$ in the notation of Bardeen~\cite{Bardeen80}.
The gauge-invariant scalar density perturbation $\Delta$ introduced
in Refs.~\cite{Bruni1992,Ellis1989} corresponds to
$\delta\tilde\rho_{1\com,i}^{~~~i}/\rho_0$.

If we wish to write these gauge-invariant quantities in terms of the
metric perturbations rather than the velocity potential then we can
use the Einstein equations, presented in
section~\ref{sect:dynamics}, to obtain
\begin{equation}
v_1+B_1 = \frac{\H \phi_1 + \psi_1'}{\H' - \H^2} \,.
\end{equation}
In particular we note that we can write the comoving curvature
perturbation, given in Eq.~(\ref{defR}), in terms of the longitudinal
gauge-invariant quantities as
\begin{equation}
\R = \Psi - \frac{\H(\H\Phi+\Psi')}{\H'-\H^2} \, ,
\end{equation}
which coincides the quantity denoted $\zeta$ by Mukhanov, Feldman
and Brandenberger in Ref.~\cite{MFB}.

By comparing the definitions of the energy-momentum tensor for a
single fluid and a single scalar field in Sections
\ref{tmunu_single_sec} and \ref{Tmunu_scal_sec_single} we can relate
the $v_1+B_1$ to $\dvp1$, which allows one to rewrite the definition
of the comoving curvature perturbations, \eq{defR}, as
\be
\label{defRfield1}
\R=\psi_1+\frac{\H}{\vp_0'}\dvp1\,.
\ee
{}From the definition above we immediately that the comoving
curvature perturbation is related to the field fluctuation on flat
slices, defined in \eq{defvpflat1}, by
\be
\widetilde{{\delta\varphi}_{1\fg}}
= \frac{\varphi_0'}{\H} \R \,.
\ee
For extensions to the multi-field case see Section \ref{multi_field_sec}.

\subsection{Total matter gauge}
\label{totmat}

This gauge is also known as the velocity orthogonal isotropic gauge
\cite{KS} but here we follow the terminology of Ref.~\cite{LLBook}.
It is closely related to comoving orthogonal and longitudinal
gauges.

To fix the temporal gauge we require the total momentum potential on
spatial hypersurfaces to vanish
\be
\wt{v_1}+\wt{B_1}=0\,.
\ee
In addition we require $\wt{E_1}=0$ and $\wt{F_1}_i=0$, which fixes
the spatial gauge.
These require
\bea
 \alpha_{1\tom}=v_1+B_1\,,
 \qquad \beta_{1\tom} =-E_1\,, \qquad \gamma_{1\tom}^i=-F_1^i\,.
\eea
We therefore get the metric perturbations in the total matter gauge
related to the comoving orthogonal and longitudinal gauge
perturbations:
\bea
\wt{{\A{}}_{1\tom}}&=&
\wt{{\A{}}_{1\com}}\,,\\
\wt{\psi_{1\tom}}&=&
\wt{\psi_{1\com}}\,,\\
\wt{B_{1\tom}}&=&
 - \wt{v_{1\ell}}\,,
\eea
and for the matter quantities we get in the total matter gauge
\bea
\wt{\delta\rho_{1\tom}}&=&
\wt{\delta\rho_{1\com}}
\,,\\
\wt{v_{1\tom}}&=&
\wt{v_{1\ell}}
\,. \eea
Note that in the total matter gauge velocity potential is not
identically zero (unlike in the comoving orthogonal gauge), but
equal to the shear potential,
$\wt{v_{1\tom}}=\wt{\sh_{1\tom}}=-\wt{B_{1\tom}}$, which also
coincides with velocity potential in longitudinal gauge.

\subsection{Uniform density gauge}
\label{udg_sec}

Alternatively we can use the matter to pick out a foliation of
uniform density hypersurfaces on which to define perturbed
quantities.

\subsubsection{First order}

Using Eq.~(\ref{rhotransform1}) we see that $\wt{\delta\rho_1}=0$
implies a temporal gauge transformation
\begin{equation}
\label{defunid1}
\alpha_{1\udg} = -\frac{\delta\rho_1}{\rho_0'} \,.
\end{equation}
On these hypersurfaces the gauge-invariant curvature perturbation
is~\cite{Deruelle:1995kd,Martin:1997zd}
\begin{equation}
\label{defzeta}
-\zeta_1 \equiv \wt{\psi_{1\udg}}
= \psi_1 + \H \frac{\delta\rho_1}{\rho_0'} \, .
\end{equation}
The sign is chosen to coincide with $\zeta$ defined in
Refs.~\cite{Bardeen83,Bardeen88}\footnote{Note, that $\zeta_1$ defined
in \eq{defzeta} is related to the curvature
perturbation $\zeta_{\rm{SBB}}$ defined in
Ref.~\cite{Salopek:1988qh} by $\zeta_{\rm{SBB}}\equiv 3\zeta_1$.}.
There is still the freedom to choose the spatial gauge. In
particular we can choose either $\tilde B$, $\tilde E$ or $\tilde v$
to be zero and thus fix $\beta$.

Note that $\zeta_1$ is simply related to the density perturbation in
the flat gauge, \eq{defrho1psi} by
\be
-\zeta_1=\frac{\H}{\rho_0'}\delta\rho_{1\fg}\,.
\ee

The curvature perturbation in the uniform-density gauge is also
closely related to the comoving curvature perturbation (\ref{defR}).
At first order we have
\begin{equation}
\zeta_1 = -\R_1 - \frac{\H}{\rho_0'} \delta\rho_{1\com}\,,
\end{equation}
where $\delta\rho_{1\com}$ is the comoving density perturbation
(\ref{defrho1com}). In Section \ref{sect:dynamics} we shall use the
Einstein equations to show that $\zeta_1$ and $\R_1$ differ only by an
overall minus sign in the large scale limit where the comoving density
perturbation vanishes.

\subsubsection{Second order}

The transformation behaviour of scalars at second order,
\eq{rhotransform2}, allows us to define the temporal gauge
corresponding to uniform density hypersurfaces as
\be \label{interdefunid2}
\alpha_{2\udg}=-\frac{\delta\rho_2}{\rho_0'}
-\frac{\alpha_1}{\rho_0'}\left(
\rho_0''\alpha_{1}+\rho_0'\alpha_{1}'+2\delta\rho_1'\right)
-\frac{1}{\rho_0'}\left(2\delta\rho_1+\rho_0'\alpha_{1}\right)_{,k}
\left(\beta_{1,}^{~~k}+\gamk1\right)\,. \ee
Using then the definition of uniform density hypersurfaces at first
order, \eq{defunid1}, and choosing a spatially flat threading by using
\eq{deffg1}, we finally get
\be
\label{defunid2}
\alpha_{2\udg}=\frac{1}{\rho_0'}\left[
-\delta\rho_2 +\frac{\delta\rho_1'}{\rho_0'}\delta\rho_1
+\left(E_{1,}^{~~k}+F_1^{~k}\right)\delta\rho_{1,k}\right]\,.
\ee

Using \eq{Xijdef} we find
\bea
\label{Xijud2}
&&\X_{ij\udg}\equiv -2\frac{\H}{\rho_0'}\Big[
\H\left(1+3\cs2\right)\left(\frac{\delta\rho_{1}^2}{\rho_0'}\right)
-\frac{\delta\rho_{1}'}{\rho_0'}\delta\rho_{1}
+{\delta\rho_{1,k}\xi_{1}^{~k}}
\Big] \delta_{ij}\\
&& +4\Big[-\frac{\delta\rho_{1}}{\rho_0'}\left(C_{1ij}'+2\H
C_{1ij}\right) +C_{1ij,k}\xi_{1}^{~k}\big]
+\left(4 C_{1ik}+\xi_{1\fg i,k}\right)\xi_{1\fg,j}^k\nonumber\\
&&
+\left(4 C_{1jk}+\xi_{1\fg j,k}\right)\xi_{1\fg,i}^k
-\frac{1}{\rho_0'}\Big[
\delta\rho_{1,i}\left(2B_{1j}+\xi_{1\fg j}'\right)
+\delta\rho_{1,j}\left(2B_{1i}+\xi_{1\fg i}'\right)
\Big]\nonumber\\
&&
-2\frac{\delta\rho_{1}}{\rho_0'}\left(
\xi_{1\fg (i,j)}'+4\H  \xi_{1\fg (i,j)}\right)
-\frac{2}{\rho_0^{\prime 2}}\delta\rho_{1,i}\delta\rho_{1,j}
+2\xi_{1\fg}^k\xi_{1\fg (i,j)k}
+2\xi_{1\fg k,i}\xi_{1\fg,j}^k\,,\nonumber
\eea
where we choose a flat threading by defining,
\be
\xi_{1\fg}^i\equiv-\left(E_{1,}^{~~i}+F_1^i\right)\,.
\ee
{}From \eq{Xijud2} we get for the trace in the uniform density gauge
\bea
\label{Xtrace_udg}
\X^k_{~k\udg}&=& -6\frac{\H}{\rho_0'}\Big[
\H\left(1+3\cs2\right)\left(\frac{\delta\rho_{1}^2}{\rho_0'}\right)
-\frac{\delta\rho_{1}'}{\rho_0'}\delta\rho_{1}
+{\delta\rho_{1,k}\xi_{1}^{~k}}
\Big] \\
&& +4\Big[
C_{1~k,l}^{k}\xi_{1\fg}^l
-\frac{\delta\rho_{1}}{\rho_0'}
\left(C_{1~k}^{k\prime}+2\H C_{1~k}^{k} \right)
\Big]\nonumber\\
&&
+4\left(2 C_{1}^{kl}+\xi_{1\fg ,}^{k~~~~l}\right)\xi_{1\fg (k,l)}
-\frac{2}{\rho_0^{\prime}}
\left(2B_1^k+\xi_{1\fg k}'\right)\delta\rho_{1,k}\nonumber\\
&&
+2\frac{\delta\rho_{1}}{\rho_0'}\nabla^2\left( E_1'+4\H\nabla^2E_1\right)
-\frac{2}{\rho_0^{\prime 2}}\delta\rho_{1,}\delta\rho_{1,}^k
-2\nabla^2E_{1,k}\xi_{1\fg}^k
\,.\nonumber
\eea

Finally, from \eq{transpsi2} we get for $\zeta_2$, the curvature
perturbation on uniform density hypersurfaces,
\bea
\label{zeta2}
-\zeta_2=\psi_2+\frac{\H}{\rho_0'}
\left[\delta\rho_2-\frac{\delta\rho_1'}{\rho_0'}\delta\rho_1
+\xi_{1\fg}^k\delta\rho_{1,k}\right]
-\frac{1}{4}\X^k_{\udg\ k}
+\frac{1}{4}\nabla^{-2}\X^{ij}_{\udg\ ,ij}\,.
\eea

The second order tensor perturbation is in the uniform density gauge
\bea
\label{transhij2udg}
\wt h_{2\udg ij}&=& h_{2ij}+\X_{\udg ij}
+\frac{1}{2}\left(\nabla^{-2}\X^{kl}_{\udg,kl}-\X^k_{\udg k}
\right)\delta_{ij}
+\frac{1}{2}\nabla^{-2}\nabla^{-2}\X^{kl}_{\udg ,klij}\nonumber\\
&&+\frac{1}{2}\nabla^{-2}\X^k_{\udg k,ij}
-\nabla^{-2}\left(\X_{\udg ik,~~~j}^{~~~k}+\X_{\udg jk,~~~i}^{~~~k}
\right)
\,.
\eea

The curvature perturbation on uniform density hypersurfaces, defined
in \eq{zeta2}, simplifies considerably if we neglect first order
vector and tensor perturbations, i.e.~setting
$F_{1i}=S_{1i}=h_{1ij}=0$ and hence only considering scalar
perturbations,
\bea
\label{zeta2scal}
&&-\zeta_2=\psi_2+\frac{\H}{\rho_0'}\delta\rho_2
-\frac{\H}{\rho_0'}\left[
2\frac{\delta\rho_1'}{\rho_0'}\delta\rho_1
-\H\left(1+3\cs2\right)\frac{\delta\rho_1^2}{\rho_0'}
+2\delta\rho_1\left(\frac{\psi_1'}{\H}+2\psi_1\right)
\right]\nonumber\\
&&-\left(\frac{5}{2}\frac{\H}{\rho_0'}\delta\rho_{1}
+\psi_1\right)_{,k}E_{1,}^{~k}
+\psi_1\nabla^2E_1-E_{1,kl}E_{1,}^{~kl}
\\
&&
-\frac{1}{2}\Big[
\frac{\delta\rho_1}{\rho_0'}\nabla^2E_1'
+\nabla^2E_{1,k}E_{1,}^{~k}
+\frac{\delta\rho_{1,}^{~k}}{\rho_0'}
\left(2B_1-E_1'\frac{\delta\rho_{1}}{\rho_0'}\right)_{,k}
\Big]
\nonumber\\
&&
+\nabla^{-2}\Big\{2E_{1,}^{~ij}\left(
4\psi_1-\frac{\delta\rho_{1}}{\rho_0'}\right)
-2E_{1,}^{~ijk}E_{1,k}-4E_{1,}^{~ik}E_{1,k}^{~~j}
-\frac{2\delta\rho_{1,}^{~(i}}{\rho_0'}\left(
2B_1-E_1'\frac{\delta\rho_{1}}{\rho_0'}
\right)_{1,}^{~j)}
\Big\}_{,ij}\,.\nonumber
\eea
This expression was first derived in Ref.~\cite{MW2003}, however with
a different, incorrect sub-horizon part~\cite{Pitrou}.
On super-horizon scales, where gradient terms can be neglected, we
recover the expressions given in Ref.~\cite{M2005}.
%

\section{Dynamics}
\label{sect:dynamics}
\setcounter{equation}{0}

In this section we give the Einstein equations governing the
evolution of the FRW background and perturbations in general
relativity. This will allow us to derive some key properties of the
perturbation variables, such as the conservation of the curvature
perturbation $\zeta$ on super-horizon scales in the adiabatic case.

In general relativity the Einstein equations relate the local
spacetime curvature to the local energy-momentum:
\begin{equation}
\label{Einstein}
G_{\mu\nu} = 8\pi G T_{\mu\nu} \,.
\end{equation}
In more general theories of gravity we can still equate the local
spacetime curvature, $G_{\mu\nu}$, with an effective energy-momentum,
though this may not be simply related to the energy-momentum tensor
derived, say, from the matter Lagrangian. Moreover, many modified
gravity theories, including Brans-Dicke gravity or higher-order
theories, may be rewritten in terms of general relativity plus
non-minimally coupled matter fields through a conformal rescaling of
coordinates \cite{Maeda:1988ab,Wands:1993uu}. In this review we will
restrict our analysis to general relativity.

We can project the tensor equation (\ref{Einstein}) into components
tangent to and orthogonal to the timelike 4-vector field, $n^\mu$
defined in Eq.~(\ref{defnmu}), which defines the coordinate system
(see section \ref{geo_sec}). This gives two constraint equations for
the metric perturbations, which we will refer to as the energy and
momentum constraint equations.
We also have two evolution equations driven by the trace and
trace-free parts of the pressure.
Through the Bianchi identities, $\nabla_\mu G^\mu_\nu=0$, the field
equations (\ref{Einstein}) imply the local conservation the total
energy and momentum,
\be
\label{nablaTmunu}
\nabla_\mu T^{\mu\nu}=0\,.
\ee
which can similarly be split into energy and momentum conservation
equations with respect to a given coordinate system.

In the case of multiple matter components the total energy-momentum
tensor is the sum of the energy-momentum tensors of the individual
fluids, $T^{\mu\nu}_{(\alpha)}$, given in \eq{def:sumT}.
For each fluid the local energy-momentum ``conservation'' equation
(\ref{nablaTalpha}), has an energy-momentum transfer 4-vector,
$Q^\nu_{(\alpha)}$ on the right-hand-side, which is zero only for
non-interacting fluids. However local conservation of the total
energy-momentum imposes the constraint \eq{Qconstraint}.

We also have at our disposal the equations of motion for specific
matter fields, such as the Klein-Gordon equation for canonical scalar
fields, $\varphi_I$, with interaction potential energy $U$:
\begin{equation}
\Box \varphi_I = \frac{dU}{d\varphi_I} \,.
\end{equation}

In the following we equate terms order by order in a perturbative
expansion about a homogeneous background spacetime.


\subsection{Background}
\label{background}

The Einstein equations (\ref{Einstein}) give the
Friedmann constraint and evolution equation for
the background FRW universe
\bea
\label{Friedmann}
\H^2&=& \frac{8\pi G}{3} a^2 \rho \,,\\
\label{Hdot}
\H' &=& -\frac{4\pi G}{3} a^2 \left( \rho+3P\right) \,, \eea
and energy-momentum conservation, Eq.~(\ref{nablaTmunu}), gives the
continuity equation
\be
\label{continuity}
\rho'=-3\H\left( \rho+P\right)\,,
\ee
where $\rho$ and $P$ are the total energy density and the total
pressure, a prime denotes a derivative with respect to conformal time,
$\eta$, the scale factor is $a$, and $\H\equiv a'/a$ is the conformal
Hubble parameter.

The total density and the total pressure are related to the density
and pressure of the component fluids by
\be
\sum_\alpha \rho_{\alpha} =\rho \,, \qquad
\sum_\alpha P_{\alpha} =P \,.
\ee
The continuity equation (\ref{nablaTalpha}) for each individual
fluid in the background is \cite{KS}
\be
\label{cont_alpha}
\rho_{\alpha}'=-3\H\left(\rho_{\alpha}+P_{\alpha}\right) +aQ_{\alpha}\,,
\ee
where the energy transfer to the $\alpha$-fluid is given by the time
component of the energy-momentum transfer vector
\begin{equation}
 Q_{\alpha} \equiv - u_\mu Q^\mu_{(\alpha)} \,.
\end{equation}
Equation~(\ref{Qconstraint}) implies that the energy
transfer obeys the constraint
\be
\label{backconstraint}
\sum_\alpha Q_{\alpha}=0 \,.
\ee

Homogeneous scalar fields in the FRW metric obey the Klein-Gordon
equation
\begin{equation}
\label{homogKGI}
\varphi_I'' + 2\H \varphi_I' + a^2 \frac{dU}{d\varphi_I} = 0 \,.
\end{equation}
It is sometimes useful to identify the kinetic energy density and
(isotropic) pressure of each field as
\begin{equation}
\rho_I = P_I =
\frac12 a^{-2} \varphi_I^{\prime2}
\end{equation}
The Klein-Gordon equation (\ref{homogKGI}) then implies an energy
transfer of the form given by Eq.~(\ref{cont_alpha})
\begin{equation}
a Q_I = - \varphi' \frac{dU}{d\varphi_I} \,.
\end{equation}
where this energy is transferred to the potential energy
\begin{equation}
\rho_U = -P_U = U \,,
\end{equation}
and overall energy conservation (\ref{backconstraint}) implies
\begin{equation}
Q_U = - \sum_I Q_I =
 a^{-1} U' \,.
\end{equation}


\subsection{First order scalar perturbations}

In the following we discuss the linear constraint and evolution
equations for inhomogeneous perturbations at first order. We omit the
subscript ``1'' denoting the order of the perturbations to avoid
unnecessary clutter.

\subsubsection{Einstein equations}

The scalar metric perturbations in an arbitrary gauge are related to
matter perturbations via the first-order energy and momentum
constraints
\cite{KS,MFB}
\begin{eqnarray}
\label{eq:densitycon}
3\H\left(\psi'+\H\A{}\right) - \nabla^2\left[\psi+\H\sh\right]
 &=& -4\pi G a^2 \delta\rho, \\
 \psi'+\H\A{} &=& -4\pi G a^2 (\rho+P) V
 \label{eq:mtmcon}
\end{eqnarray}
where the total covariant velocity perturbation is given by
\be \label{defV} V \equiv v+B\,, \ee
and $v$ is the total scalar velocity potential (\ref{defv}).

In a specific gauge, such as the spatially flat gauge these can be
written in terms of the corresponding gauge-invariant quantities.
For instance, in the spatially flat gauge we have
\begin{eqnarray}
\label{eq:densitycon:flat}
3\H^2\A{}_\fg - \H\nabla^2\sh_\fg
 &=& -4\pi G a^2 \delta\rho_\fg, \\
 \H\A{}_\fg &=& -4\pi G a^2 (\rho+P)V_\fg \,,
 \label{eq:mtmcon:flat}
\end{eqnarray}
which makes it straightforward to eliminate the metric variables
$\A{}_\fg$ and $\sh_\fg=-B_\fg$ in favour of the energy and
momentum in the flat gauge.

Alternatively in the longitudinal gauge the shear terms are absent
and we obtain first order differential equations for the curvature
perturbation
\begin{eqnarray}
 \label{eq:densitycon:long}
 3\H\left(\Psi'+\H\Phi\right) - \nabla^2\Psi
 &=& -4\pi G a^2 \delta\rho_\ell, \\
 \Psi'+\H\Phi &=& -4\pi G a^2 (\rho+P)v_\ell \,.
 \label{eq:mtmcon:long}
\end{eqnarray}
Typically one then uses these equations to eliminate the density and
velocity perturbations, $\delta\rho_\ell$ and $v_\ell$, in terms of
the metric perturbations in the longitudinal gauge.

The same energy and momentum constraints can be re-written in terms
of gauge invariant variables to give expressions for the curvature
perturbation in the uniform-density gauge (\ref{defzeta}) and the
comoving curvature perturbation (\ref{defR}), respectively, in terms
of the longitudinal gauge metric perturbations (\ref{defPhi}) and
(\ref{defPsi}):
\begin{eqnarray}
\label{gi:densitycon}
 \Psi' +\H\Phi - \frac{\H'-\H^2}{\H} \Psi -
\frac{1}{3\H} \nabla^2\Psi = \frac{\H'-\H^2}{\H} \zeta
 \,,\\
 \label{gi:mtmcon}
 \Psi' +\H\Phi - \frac{\H'-\H^2}{\H} \Psi = - \frac{\H'-\H^2}{\H} \R
 \,.
\end{eqnarray}

These can be combined to give the gauge-invariant generalisation of
the Newtonian Poisson equation
\begin{equation}
 \label{eq:rhomcon}
\nabla^2 \Psi = - 3 \left( \H'-\H^2 \right) (\zeta+\R) = 4\pi G a^2
\delta\rho_{\rm com} \,,
\end{equation}
relating the longitudinal gauge curvature perturbation
(\ref{defPsi}) to the comoving density perturbation
(\ref{defrho1com}). We see that the comoving density perturbation is
suppressed relative to the metric perturbation $\Psi$ on large
scales, and that $\zeta$ and $-\R$ coincide in the large scale limit
so long as $\Psi$ is finite in this limit\footnote{Note that
Eq.~(\ref{gi:mtmcon}) shows that the variable denoted $\zeta$ in the
review by Mukhanov, Feldman and Brandenberger:
\begin{equation}
\zeta_{\rm MFB} \equiv \Phi - \frac{\H(\Phi'+\H\Phi)}{\H'-\H^2} \,,
\end{equation}
coincides with the comoving curvature perturbation when
$\Psi=\Phi$.}.

The perturbed Einstein equations at first order also yield two
evolution equations for the scalar metric perturbations
\begin{eqnarray}
 \label{eq:psievol}
\psi''+2\H\psi'+\H\A{}'
+ \left( 2\H'+\H^2 \right) \A{}
=4\pi G a^2 \left(\delta P+\frac{2}{3}\nabla^2\Pi\right)\,,\\
\label{eq:aniso}
\sh'+2\H\sh+\psi-\A{}=8\pi G a^2 \Pi\,,
\end{eqnarray}
where $\Pi$ is the scalar part of the (tracefree) anisotropic
stress, defined in Eq.~(\ref{def:Pi_ij}).

Equation~(\ref{eq:aniso}) in a general gauge can be interpreted as
the evolution equation for the scalar shear, but in the longitudinal
gauge it becomes a constraint equation for the gauge-invariant
perturbations $\Phi$ and $\Psi$, defined in Eqs.~(\ref{defPhi})
and~(\ref{defPsi}),
\begin{equation}
\Psi - \Phi =
8\pi G a^2 \Pi \,,
\end{equation}
and hence we have $\Psi=\Phi$ in the absence of anisotropic
stresses.

Equation~(\ref{eq:psievol}) then provides a second-order evolution
equation for the metric perturbation in the longitudinal gauge
driven by isotropic pressure:
\begin{equation}
 \label{eq:Psievol}
\Psi''+3\H\Psi' + \left( 2\H'+\H^2 \right) \Psi =4\pi G a^2 \delta P
\,.
\end{equation}
For adiabatic perturbations we can relate the pressure to the
density, $\delta P=c_s^2\delta\rho$ where $c_s^2$ is the adiabatic
sound speed, in which case (\ref{eq:Psievol}) and
(\ref{eq:densitycon}) yield a closed second-order differential
equation~\cite{MFB}
\begin{equation}
\Psi''+3(1+c_s^2)\H\Psi' + [2\H'+(1+3c_s^2)\H^2-c_s^2\nabla^2]\Psi =
 0 \,.
\end{equation}

\subsubsection{Energy and momentum conservation}

Energy-momentum conservation gives evolution equations for the
perturbed energy and momentum
\bea
 \label{pertcontinuity}
 \delta\rho' +3\H \left( \delta\rho +\delta P \right)
-3 \left( \rho+P \right) \psi'
  + (\rho+P) \nabla^2 \left( V+\sh \right) = 0
 \,,\\
V'+(1-3c^2_{\rm s})\H V + \phi +\frac{1}{\rho+P}\left(\delta
P+\frac{2}{3}\nabla^2\Pi\right)=0\,, \eea
where  $c^2_{\rm s}$ is the adiabatic speed of sound, defined as
\be
\label{c2s}
c^2_{\rm s}\equiv
\frac{P'}{\rho'} \,.
\ee

{}From the momentum conservation equation
in the total matter gauge, such that $V=0$, we see that the
acceleration is proportional to the pressure perturbation:
$(\rho+P)\phi=\delta P + (2/3)\nabla^2\Pi$. Alternatively, for
pressureless, non-interacting dust we have
$(a{V})'+a\A{}=0$ and hence the scalar velocity potential redshifts
as $V\propto 1/a$ in a synchronous gauge.

Re-writing the energy conservation equation (\ref{pertcontinuity})
in terms of the curvature perturbation on uniform-density
hypersurfaces, $\zeta$ in (\ref{defzeta}), we obtain the important
result
\begin{equation}
\label{eq:dotzeta}
\zeta' = -\H \frac{\delta P_{\rm nad}}{\rho+P} - {\Sigma_V} \,,
\end{equation}
where $\delta P_{\rm nad}$ is the non-adiabatic pressure
perturbation, defined in (\ref{defPnad}), and $\Sigma$
describes the divergence of the velocity in the longitudinal gauge,
\eq{defv1l} or, equivalently, the scalar shear along comoving
worldlines~\cite{Lyth:2003im}
\begin{equation}
 \label{SigmaVH}
{\Sigma_V} \equiv \frac{1}{3} \nabla^2 \left( V + \sh \right)
= \frac{1}{3} \nabla^2 \wt{v_\ell} \,.
\end{equation}
Thus the curvature perturbation on uniform-density hypersurfaces is
constant for adiabatic perturbations on large scales when the shear
of comoving worldlines becomes negligible. This follows directly
from local energy conservation and holds independently of the
gravitational field
equations~\cite{Wands00,Starobinsky:2001xq,Bertschinger:2006aw,Cardoso:2008gz}.

Using the definition of $\zeta$ and $\Psi$ in Eqs.~(\ref{defzeta})
and (\ref{defPsi}) and the constraint equation for the comoving
density perturbation (\ref{eq:rhomcon}), we have
\begin{equation}
 \label{constrainSigmaVH}
\frac{\Sigma_V}{\H} =
 \frac{\nabla^2}{3\H^2} \left( \zeta + \Psi \right)
 + \frac{2\rho}{3(\rho+P)}
     \left( \frac{\nabla^2}{\H^2} \right)^2 \Psi \,.
\end{equation}
Thus we see that $\zeta$ is constant for adiabatic perturbations
($\delta P_{\rm nad}=0$) on super-Hubble scales ($k/\H\ll1$), so
long as $\Psi$ remains finite.
This makes $\zeta$ a convenient variable to characterise the
primordial density perturbation on super-Hubble scales, either
during a period of inflation in the very early universe, or in the
subsequent radiation dominated era. This is an excellent
approximation throughout reheating at the end of inflation and the
subsequent radiation era on scales relevant for observations of
degree-scale anisotropies in the cosmic microwave background and
large-scale galaxy surveys \cite{LLBook}.

Conversely, local variations in the pressure leading to a
non-adiabatic pressure perturbation, $\delta P_{\rm nad}$, will lead
to a change in the curvature perturbation $\zeta$ on super-Hubble
scales \cite{Mollerach:1989hu,Garcia-Bellido:1995qq,Linde:1996gt}.
This mechanism is at the heart of the curvaton scenario for the
origin of large scale structure in the universe
\cite{Enqvist02,Lyth02,Moroi01}.

\subsubsection{Multiple fluids}

The perturbations in the total energy-momentum can be related to
the perturbations of individual fluids by
\be
\label{sum_delta_rho}
\sum_\alpha \delta\rho_{\alpha} =\delta\rho \,, \qquad
\sum_\alpha \delta P_{\alpha} = \delta P \,, \qquad
\sum_\alpha \Pi_\alpha =\Pi\,,
\ee
and
\be
\label{sum_V}
V=\sum_\gamma\frac{\rho_\gamma+P_\gamma}{\rho+P}V_\gamma\,,
\ee
where $\delta\rho_{\alpha}$ and $\delta P_{\alpha}$ are the perturbed
energy density and the perturbed pressure of the $\alpha$-fluid,
respectively, and $\Va$ is the covariant
velocity perturbation of the $\alpha$-fluid defined as
\be
 \label{defVa} \Va\equiv v_\alpha+B\,,
 \ee
where $v_{\alpha}$ is the scalar velocity potential of the
$\alpha$-fluid.

The perturbed energy transfer 4-vector, Eq.~(\ref{nablaTalpha}), for
individual fluids including terms up to first order, is written
as~\cite{KS,MWU}
\bea
\label{pert_q_vector}
Q_{(\alpha)0} &=& -aQ_{\alpha}(1+\phi)-a \delta Q_\alpha\,,\nonumber \\
Q_{(\alpha)i} &=& \left( f_\alpha + aQ_\alpha V \right)_{,i} \,,
\eea
and Eq.~(\ref{Qconstraint}) implies that the perturbed energy and
momentum transfer obey the constraints
\be
\label{pertconstraint}
\sum_\alpha \delta Q_{\alpha} = 0 \,,
\quad
\sum_\alpha f_\alpha =0\,.
\ee
Note that the momentum transfer, $f_\alpha$, is by convention
\cite{KS,MW2004} defined with respect to the total momentum, $V$, so
is non-zero only if the momentum transfer vanishes in the total matter
frame ($V=0$).

The perturbed energy conservation equation for a particular fluid,
including energy transfer, is then obtained by the first-order part
of the time-component of the perturbed continuity equation
(\ref{nablaTalpha}) to give \cite{KS,M2001}
\be
\label{pertenergyexact}
\delta\rho_{\alpha}' +3\H(\delta\rho_{\alpha}+\delta P_{\alpha})
- 3\left(\rho_{\alpha}+P_{\alpha}\right)\psi'
+ (\rho_{\alpha}+P_{\alpha}) \nabla^2\left(\Va+\sh\right) =
aQ_{\alpha}\phi + a\delta Q_{\alpha} \,,
\ee
The momentum conservation equation of the $\alpha$-fluid is
\bea
\label{dotVa}
{V}_\alpha'
&+&\left[\frac{aQ_\alpha}{\rho_\alpha+P_\alpha}(1+c^2_\alpha) +(1-3
c^2_\alpha)\H \right]\Va + \phi
 \nonumber\\
 &+&\frac{1}{\rho_\alpha+P_\alpha}\left[ \delta
P_\alpha+\frac{2}{3}\nabla^2\Pi_\alpha - aQ_\alpha V
-f_\alpha\right] =0\,,
 \eea
where $c^2_{\alpha}\equiv P_{\alpha}'/ \rho_{\alpha}'$ is the
adiabatic sound speed of the $\alpha$-fluid and
$a^{2}[\Pi_{\alpha,ij}-(1/3)\delta_{ij}\nabla^2\Pi_\alpha]$ is the
scalar anisotropic stress of that fluid.
The total adiabatic sound speed, \eq{c2s}, is the weighted sum of
the adiabatic sound speeds of the individual fluids,
\be
c^2_{\rm{s}} = \sum_\alpha \frac{\rho_\alpha'}{\rho'} c^2_\alpha
\,.
\ee

We recover the evolution equation for the total density perturbation
(\ref{pertcontinuity}) from Eq.~(\ref{pertenergyexact}) by summing
over all fluids, using Eq.~(\ref{sum_delta_rho}) and the constraint
(\ref{pertconstraint}).

Analogous to the curvature perturbation on uniform-total-density
hypersurfaces, $\zeta$ defined in Eq.~(\ref{defzeta}), we can define
a gauge-invariant perturbation on the uniform-$\alpha$-density
hypersurfaces
\begin{equation}
\zeta_{\alpha}
\equiv - \psi - \H \frac{\delta\rho_{\alpha}}{\rho_\alpha'} \, .
\end{equation}
The perturbed energy conservation equation for each fluid can then
be written as
\begin{eqnarray}
 \label{dotzetaalpha}
 \zeta_\alpha' &=&
 3\frac{\H^2}{\rho_\alpha'}\delta P_{{\rm intr},\alpha} - \Sigma_\alpha
 -\frac{\nabla^2}{3\H}\left[
 \frac{aQ_\alpha}{\rho_\alpha'}\R_\alpha\right]
\nonumber \\
&&
-\left(\frac{\H}{a}\right)^\prime\frac{aQ_\alpha}{\rho_\alpha'}
 \left(\frac{\delta\rho_{\alpha}}{\rho_\alpha'} -
 \frac{\delta\rho}{\rho'}\right) - \frac{\H}{\rho_\alpha'}\left(
\delta Q_{\alpha}
-\frac{Q_\alpha'}{\rho_\alpha'}\delta\rho_{\alpha}\right) \,,
 \end{eqnarray}
where the intrinsic non-adiabatic pressure perturbation of each
fluid is given by
\be
\label{deltaPintralpha}
\delta P_{\rm{intr},\alpha} \equiv
\delta P_{\alpha} - c^2_{\alpha}\delta\rho_{\alpha} \,,
\ee
the scalar shear along worldlines comoving with the $\alpha$-fluid
is
\begin{equation}
\label{Sigmaalpha}
\Sigma_\alpha \equiv \frac{1}{3}\nabla^2 \left( \sigma + V_\alpha
\right) \,,
\end{equation}
and, extending \eq{defpsi1com}, the curvature perturbation comoving
with the $\alpha$-fluid is
\begin{equation}
\R_\alpha \equiv \psi + \H(v_\alpha+B) \,.
\end{equation}
Thus we see that $\zeta_\alpha$ is constant on large scales for
adiabatic perturbations of any perfect fluid, with $\delta
P_{\rm{intr},\alpha}=0$, whose energy is conserved, $Q_\alpha=0$
\cite{Wands00}. In fact we shall show later that $\zeta_\alpha$ is
constant even in the presence of energy transfer, $Q_\alpha\neq0$,
so long as that energy transfer is adiabatic \cite{MW2004}.

\subsubsection{Multiple fields}
\label{multi_field_sec}

If we consider $N$ scalar fields  with  Lagrangian density
\begin{equation}
{{\cal L}} = -U(\varphi_1,\cdots,\varphi_N)
 -\frac{1}{2} \sum_{I=1}^{N} g^{\mu\nu}
 \varphi_{I,\mu}\varphi_{I,\nu}\,,
\label{lag}
\end{equation}
and minimal coupling to gravity, then the total energy, pressure
and momentum perturbations are given by
\begin{eqnarray}
\delta\rho &=& \sum_I\left[
a^{-2} \varphi_I' \left( {\delta\varphi}_I' - \varphi_I' \A1 \right)
 + U_{I}\delta\varphi_I \right] \,,
\label{eq:densityphi} \\
\delta P &=& \sum_I\left[
a^{-2} \varphi_I' \left( {\delta\varphi}_I' -\varphi_I' \A1 \right)
  - U_{I}\delta\varphi_I \right] \,,
\label{eq:pressurephi} \\
(\rho+P)(v+B)_{,i}
 &=& - \sum_I a^{-1} {\varphi}_I' \delta\varphi_{I,i} \,,
\label{eq:mtmphi}
\end{eqnarray}
where $U_I \equiv \partial U/\partial \varphi_I$.
These then give the gauge-invariant comoving density perturbation
(\ref{defrho1com})
\begin{eqnarray}
\label{def:rhom}
 \delta\rho_m =
 a^{-2} \sum_I \left[ \varphi_I' \left(
{\delta\varphi}_I' - \varphi_I' \A1 \right) - (\varphi_I''-\H\varphi_I')
\delta\varphi_I \right]\,.
\end{eqnarray}
The comoving density is sometimes used to represent the total
matter perturbation, but for a single scalar field it is
proportional to the non-adiabatic pressure (\ref{defPnad}):
\begin{equation}
 \label{dPnadsingle}
\delta P_{\rm nad} = - \frac{2a^2U_{,\varphi}}{3\H\varphi'}
\delta\rho_\com \,.
\end{equation}
{}From the Einstein constraint equation (\ref{eq:rhomcon}) this
will vanish on large scales ($k/aH\to0$) if $\Psi$ remains finite,
and hence single scalar field perturbations become adiabatic in
this large-scale limit, even without assuming slow-roll
\cite{Gordon2000,M2001}.

The anisotropic stress, $\pi_{ij}$, vanishes to linear order for
any number of scalar fields minimally coupled to gravity.

The first-order scalar field perturbations obey the wave equation
\begin{eqnarray}
\label{eq:pertKG} \label{eq:scalareom}
{\delta\varphi}_I'' + 2\H{\delta\varphi}_I'
 - {\nabla^2} \delta\varphi_I + a^2 \sum_J U_{IJ}
\delta\varphi_J
=
-2a^2U_{I}\A1 + \varphi_I' \left[ {\A1}' + 3{\psi}' -
 {\nabla^2} \sh \right]. \label{eq:perturbation}\nonumber\\
\end{eqnarray}
where the terms on the right-hand-side represent the effect of
metric perturbations at first-order (sometimes called the
gravitational back-reaction).

The scalar field wave equation in a perturbed FRW cosmology can most
easily be written in closed form in terms of the field perturbations
in the spatially flat gauge, defined in \eq{defvpflat1}, which in the
multi-field case have the gauge invariant definitions
\begin{equation}
\label{eq:defdphipsi}
{\delta\varphi}_{\fg I} \equiv \delta\varphi_I +
\frac{\varphi_I'}{\H} \psi \,.
\end{equation}
Note that since the scalar field can be thought of as a potential
for the 4-velocity, this variable is a rescaling of the curvature
perturbation on the comoving-orthogonal hypersurface for each field
\begin{equation}
{\delta\varphi}_{\fg I} = \frac{\varphi_I'}{\H} \R_I \,.
\end{equation}
Using the Einstein equations to eliminate the metric perturbations
on the right-hand-side of Eq.~(\ref{eq:perturbation}) yields
\cite{Sasaki:1995aw,Hwang9608,Nambu:1996gf,Taruya:1997iv}
\begin{equation}
{\delta\varphi}_{\fg I}^{\prime\prime} + 2\H
{\delta\varphi}_{\fg I}^{\prime}
 - {\nabla^2} {\delta\varphi}_{\fg I} + a^2 \left[ \sum_J U_{IJ}
 - \frac{8\pi G}{a^2} \left( \frac{a^2\varphi_I'\varphi_J'}{\H}\right)^\prime \right]
{\delta\varphi}_{\fg J} = 0\,.
\end{equation}
The effect of gravitational coupling is now evident due to the terms
proportional to Newton's gravitational constant. It is also evident
that this gravitational coupling vanishes at first order for fields
whose time derivative vanishes in the background solution, which is
why at lowest order in a slow-roll approximation during inflation
one can neglect the gravitational coupling.

In the next section we will discuss how the coupled equations for
multiple fields may be partially decoupled by identifying the
adiabatic and isocurvature field perturbations on large scales.

\subsection{First order vector perturbations}

The divergence-free part of the 3-momentum [see
Eqs.~(\ref{decompBi}), (\ref{defv}) and~(\ref{defTmtm})]
 \be
 {\delta q}_i = (\rho+P)(v_{{\rm vec}i}-S_i) \,,
 \ee
obeys the momentum conservation equation
\begin{equation}
\label{eq:evolvec}
{\delta q}_i' + 4\H \delta q_i = - \nabla^2 \Pi_i \,,
\end{equation}
where the vector part of the anisotropic stress, \eq{def:Pi_ij}, is
given by $a^2\partial_{(i}\Pi_{j)}$. The gauge-invariant vector
metric perturbation is then directly related to the divergence-free
part of the momentum via the constraint equation
\begin{equation}
 \label{eq:convec}
\nabla^2 \left( {F}_i' + S_i \right) = - 16\pi G a^2 \delta q_i \,.
\end{equation}
Thus the Einstein equations constrain the gauge-invariant vector
metric perturbation, $F_i'+S_i$, to vanish in the presence of only scalar
fields, for which the divergence-free momentum necessarily
vanishes.

Equation~(\ref{eq:convec}) shows that vector metric perturbations
can be supported only by divergence-free momenta, but even then
equation (\ref{eq:evolvec}) shows that the vector perturbations
are redshifted away by the Hubble expansion on large scales unless
they are driven by an anisotropic stress.

\subsection{First order tensor perturbations}

There is no constraint equation for the tensor perturbations as
these are the free gravitational degrees of freedom (gravitational
waves). The spatial part of the Einstein equations yields a wave
equation
\begin{equation}
\label{teneq}
{h}_{ij}'' + 2\H{h}_{ij}' - {\nabla^2} h_{ij} = 8\pi G a^2
\Pi_{ij} \,,
\end{equation}
where $\Pi_{ij}^{(TT)}$ is the transverse and tracefree part of the
anisotropic stress (\ref{def:Pi_ij}).

We can decompose arbitrary tensor perturbations into eigenmodes of
the spatial Laplacian, $\nabla^2e_{ij}=-(k^2/a^2)e_{ij}$, with
comoving wavenumber $k$, and scalar amplitude $h(t)$:
\begin{equation}
\label{eq:defh} h_{ij} = h(t) e_{ij}^{(+,\times)}(x)\,,
\end{equation}
with two possible polarisation states, $+$ and $\times$.
In the absence of any such anisotropic stress, e.g., in the presence
of scalar fields and perfect fluids, the amplitude, defined in
Eq.~(\ref{eq:defh}), of the tensor metric perturbation with comoving
wavenumber, $k$, obeys the wave equation for a massless scalar field
(\ref{eq:pertKG}) in an unperturbed FRW metric.
\begin{equation}
\label{heq}
{h}'' + 2\H h' + k^2 h = 0
\,.
\end{equation}

\section{Adiabatic and entropy perturbations}
\setcounter{equation}{0}

\subsection{Multiple fluids}

We will refer to primordial perturbations as the perturbations at
the epoch of primordial nucleosynthesis. The abundances of the light
elements provide constraints on the matter content and expansion
rate of the universe at this epoch, so we will assume that the
material content of the universe is known (photons, neutrinos,
baryonic matter and cold dark matter) and the gravitational laws are
described by general relativity. This is expected to be some time
after an early inflationary epoch when the perturbations originated
as vacuum fluctuations on much smaller scales. By the time of
primordial nucleosynthesis, the scales responsible for the large
scale structure of our observable universe today were far outside
the Hubble scale and well described by large scale limit.

In the standard Hot Big Bang the entropy density of the universe is
dominated by the number of relativistic particles and for most of
the history of the universe it is proportional to the number of
photons, $s=1.8g_sn_\gamma$, where $g_s$ is the effective number of
light species \cite{KolbTurner}. In particular the perturbed
baryon-entropy ratio $n_B/s$ (assuming $g_s$ remains constant) is
given by \cite{Mollerach:1989hu}
\begin{equation}
 \label{def:SnB}
S_B \equiv \frac{\delta(n_B/n_\gamma)}{n_B/n_\gamma} = \frac{\delta
  n_B}{n_B} - \frac{\delta n_\gamma}{n_\gamma} \,.
\end{equation}
Written in terms of the energy density of photons,
$\rho_\gamma\propto n_\gamma^{4/3}$, and baryons, $\rho_B\propto
n_B$, at the time of primordial nucleosynthesis this becomes
\begin{equation}
 \label{SB}
S_B = \frac{\delta\rho_B}{\rho_B} -
 \frac{3}{4} \frac{\delta\rho_\gamma}{\rho_\gamma} \,.
\end{equation}

More generally, density perturbations in an $n$-component system can
be decomposed into an overall density perturbation and $n-1$
relative density perturbations between the different components. The
overall density perturbation is naturally gauge-dependent, however
the gauge transformation rule for the linear density perturbation
(\ref{rhotransform1})
suggests a natural gauge-invariant definition of the relative density
perturbation at first-order,
\begin{equation}
\delta\rho_{IJ} \propto \delta\rho_I - \frac{\rho_I'}{\rho_J'}
 \delta\rho_J \,,
\end{equation}
corresponding to the density perturbation of fluid $I$ on surfaces of
uniform density of the fluid $J$.
Comparing this expression with the conventional definition of the
primordial baryon-entropy perturbation (\ref{SB}) suggests a
gauge-invariant definition of the relative perturbation between {\em
any} two fluids \cite{MWU,MW2004}
\begin{equation}
 \label{defSIJ}
 S_{IJ} \equiv  3\H \left( \frac{\delta\rho_J}{\rho_J'} -
 \frac{\delta\rho_I}{\rho_I'} \right)
 \equiv 3\left( \zeta_I - \zeta_J \right)
 \,,
\end{equation}
which reduces to baryon-entropy perturbation~(\ref{SB}) for
$S_B\equiv S_{B\gamma}$. Hence we refer to $S_{IJ}$ as the relative
entropy perturbation for two fluids. It is the correct
generalisation of the the entropy perturbation defined in
Ref.~\cite{KS} to the case of non-interacting fluids.

The non-adiabatic pressure perturbation is given by
\begin{equation}
\label{defPnad}
\delta P_{\rm nad} \equiv \delta P - \frac{P'}{\rho'}\delta\rho \,.
\end{equation}
For a detailed recent discussion of the non-adiabatic pressure see
Ref.~\cite{Christopherson:2008ry}.

In a system of more than one fluids the total non-adiabatic pressure
perturbation, $\delta P_{\rm nad}$, may be further split into two
parts \cite{KS},
\be \label{deltaPnad} \delta P_{\rm nad}\equiv \delta P_{\rm
intr}+\delta P_{\rm rel} \,. \ee
The first part is due to the intrinsic entropy perturbation of each
fluid
\be
 \label{deltaPintr} \delta P_{\rm intr}=\sum_\alpha \delta
P_{\rm{nad},\alpha} \,,
 \ee
where the intrinsic non-adiabatic pressure perturbation of each
fluid was given in Eq.~(\ref{deltaPintralpha}). The second part of
the non-adiabatic pressure perturbation (\ref{deltaPnad}) is due to
the {\em relative entropy perturbation} $\cal{S}_{\alpha\beta}$
between different fluids (\ref{defSIJ})
\be
 \label{deltaPrel}
  \delta P_{\rm rel} \equiv -\frac{1}{6\H\rho'}
\sum_{\alpha,\beta}\rho_\alpha'\rho_\beta'
\left(c^2_\alpha-c^2_\beta\right)\S_{\alpha\beta} \,.
 \ee

In analogy with the non-adiabatic pressure perturbation for each
fluid (\ref{deltaPintralpha}), we can identify an intrinsic
non-adiabatic part of the energy transfer perturbation
\cite{MWU,MW2004} that appears in the perturbed energy conservation
equation for each fluid (\ref{pertenergyexact})
\be \label{deltaQintralpha}
 \delta Q_{{\rm intr},\alpha} \equiv
 \delta Q_\alpha - \frac{Q_\alpha'}{\rho_\alpha'}
 \delta\rho_\alpha \,.
 \ee
This is automatically zero if the local energy transfer $Q_\alpha$
is a function of the local density $\rho_\alpha$ so that $\delta
Q_\alpha = (d{Q}_\alpha/d\rho_\alpha)\delta\rho_\alpha$, just as the
intrinsic non-adiabatic pressure perturbation
(\ref{deltaPintralpha}) vanishes when $\delta P_\alpha =
(d{P}_\alpha/d\rho_\alpha)\delta\rho_\alpha$.

We can also identify in Eq.~(\ref{pertenergyexact}) a relative
non-adiabatic energy transfer
\begin{eqnarray}
 \label{deltaQrelalpha}
\delta Q_{{\rm rel},\alpha}
 &=& Q_\alpha\frac{\H'-\H^2}{\H^2} \left(
\frac{\delta\rho_\alpha}{\rho_\alpha'} -\frac{\delta\rho}{\rho'}
\right)
 \nonumber\\
&=& - \frac{Q_\alpha}{6\H\rho} \sum_\beta \rho_\beta'
\S_{\alpha\beta} \,,
 \end{eqnarray}
due to the presence of relative entropy perturbations whenever the
background energy transfer is non-zero, $Q_\alpha\neq0$.

The perturbed energy conservation Eq.~(\ref{dotzetaalpha}) for each
fluid can then be written as
\begin{eqnarray}
 \label{dotzetaalphafinal}
 \zeta_\alpha' &=&
 3\frac{\H^2}{\rho_\alpha'}\delta P_{{\rm intr},\alpha}
 -\frac{\H}{\rho_\alpha'} \left( \delta Q_{{\rm intr},\alpha}
  + \delta Q_{{\rm rel},\alpha} \right)
  - \Sigma_\alpha
 -\frac{\nabla^2}{3\H}\left[
 \frac{aQ_\alpha}{\rho_\alpha'}\R_\alpha\right]
 \end{eqnarray}
where the non-adiabatic pressure perturbation of each fluid is given
by Eq.~(\ref{deltaPintralpha}) and the non-adiabatic energy transfer
is given by Eqs.~(\ref{deltaQintralpha}) and (\ref{deltaQrelalpha}).
We thus see that $\zeta_\alpha$ is constant for adiabatic
perturbations in the large scale limit where the shear of comoving
worldlines, $\Sigma_\alpha$ defined in Eq.~(\ref{Sigmaalpha}),
vanishes \cite{MW2004}.

More generally \cite{Lyth:2003im} one finds a conserved perturbation
whenever there is a local conservation equation of the form
$dy/d\tau=-\theta f(y)$, where $\theta$ is the local expansion rate
and $\tau$ the proper time along comoving worldlines. When one
integrates this conservation equation one finds the local
logarithmic expansion as a function of $y$:
\begin{equation}
\tilde{N} = \int \theta d\tau = \int \frac{dy}{f(y)} \equiv F(y)\,.
\end{equation}
Thus the difference $\delta F = F(y_B) - F(y_A)$ evaluated along
different worldlines remains a fixed constant of integration if one
evaluates $\delta F$ on constant time hypersurfaces separated by a
fixed expansion, and spatially flat hypersurfaces provide a suitable
time-slicing on large scales. The classic example of such a conserved
quantity is the curvature perturbation on uniform density
hypersurfaces, $\zeta$ defined in Eq.~(\ref{defzeta}), which is
conserved on large scales when $P=P(\rho)$ and thus the perturbations
are adiabatic. But as we have seen the same result holds for the
curvature perturbation on uniform-density hypersurfaces for any fluid
whose pressure perturbation and energy transfer is adiabatic.

Thus the matter isocurvature perturbation (\ref{SB}) is constant on
large scales if there is negligible energy transfer between
non-relativistic matter and radiation. However at even higher
energies we can still define a conserved perturbation associated
with conserved baryon number density \cite{Lyth:2003ip}
 \begin{equation}
\tilde\zeta_B = \frac{\delta n_B}{n_B} - \psi
 \end{equation}
so long as we have a conserved quantum number associated with baryon
number. The observed stability of the proton implies that baryon
number is conserved up to very high energy, possibly the GUT scale,
and thus it should be possible to relate any primordial baryon
isocurvature perturbation to physics at very high energies.

\subsection{Multiple fields}
 \label{twofield}

In the background FRW cosmology driven by multiple scalar fields it
is possible to identify an adiabatic direction along the background
trajectory in field space
\begin{equation}
 \label{defhatr}
\hat{r}_I = \frac{\varphi_I'}{\sqrt{\varphi_I^{\prime2}}} \,.
\end{equation}
The background solution, even in the presence of multiple fields,
can then be described in terms of an effective single field, $r$,
obeying the usual Klein-Gordon equation
\begin{equation}
r''+2\H r' + U_{r} = 0 \,.
\end{equation}
where ${r'}=\sum_I \hat{r}_I\varphi_I'$ and $U_{r} = \sum_I
\hat{r}_I U_I$. However field perturbations need not follow this
background trajectory and we encounter qualitatively different
behaviour from that in the single field case when we consider
inhomogeneous perturbations about the background trajectory.

In analogy with our treatment of fluid perturbations one can
identify adiabatic and isocurvature field perturbations in a
cosmology with more than one scalar field. Indeed even a single
scalar field, $\varphi$, can support a non-adiabatic pressure
perturbation, given in Eq.~(\ref{dPnadsingle}). We refer to this as
the intrinsic pressure perturbation for the field. However this
intrinsic non-adiabatic perturbation for a single field is
proportional to the comoving density perturbation and thus vanishes
in a scalar field dominated universe on large scales according to
Eq.~(\ref{eq:rhomcon}), leaving effectively only adiabatic
perturbations in this large scale limit.

For multiple fields we can perform a local rotation in field space
to identify the adiabatic part of arbitrary perturbations along the
background trajectory \cite{MW1998,Gordon2000}
\begin{equation}
\delta r \equiv \sum_I \hat{r}_I \delta\varphi_I
 \,.
\end{equation}
The generalisation to non-canonical fields with arbitrary metric in
field space is given in
Refs.~\cite{GrootNibbelink:2001qt,DiMarco:2002eb}.

Field perturbations orthogonal to the adiabatic field are
isocurvature field perturbations, or entropy perturbations in
analogy with the fluid density perturbations (\ref{SB}),
\begin{equation}
 \delta s_{IJ} \propto \frac{\delta\varphi_I}{\varphi_I'} -
 \frac{\delta\varphi_J}{\varphi_J'} \,.
\end{equation}

Note that the adiabatic field perturbation is naturally
gauge-dependent, whereas the isocurvature field perturbations are
gauge-independent at first order. For simplicity we will consider
the case of two scalar fields where the direction in field space is
given by the angle $\theta$, see Figure~1, so that
$\hat{r}=(\cos\theta,\sin\theta)$. We then have
\begin{eqnarray}
 \label{defdeltar}
 \delta r = \cos\theta \delta\varphi_1 + \sin\theta \delta\varphi_2
\,,\\
 \label{defdeltas}
 \delta s = -\sin\theta \delta\varphi_1 + \cos\theta \delta\varphi_2
\end{eqnarray}
and we will work with the adiabatic field perturbation in the
spatially flat gauge
\begin{equation}
\delta r_\fg = \delta r + \frac{r'}{\H}\psi \,,
\end{equation}
Note that $\delta r_\fg$ is thus proportional to the total
comoving curvature perturbation
\begin{equation}
 \label{Radiabatic}
\R = \psi + \frac{\H}{r'} \delta r = \frac{\H}{r'} \delta r_\fg
 \,.
\end{equation}
since the adiabatic field, $r$, is the potential for the total
velocity, $u_\mu\propto dr/dx^\mu$.

\begin{figure}
 \label{fig:decompose}
\begin{center}
{ \includegraphics*[height=5cm]{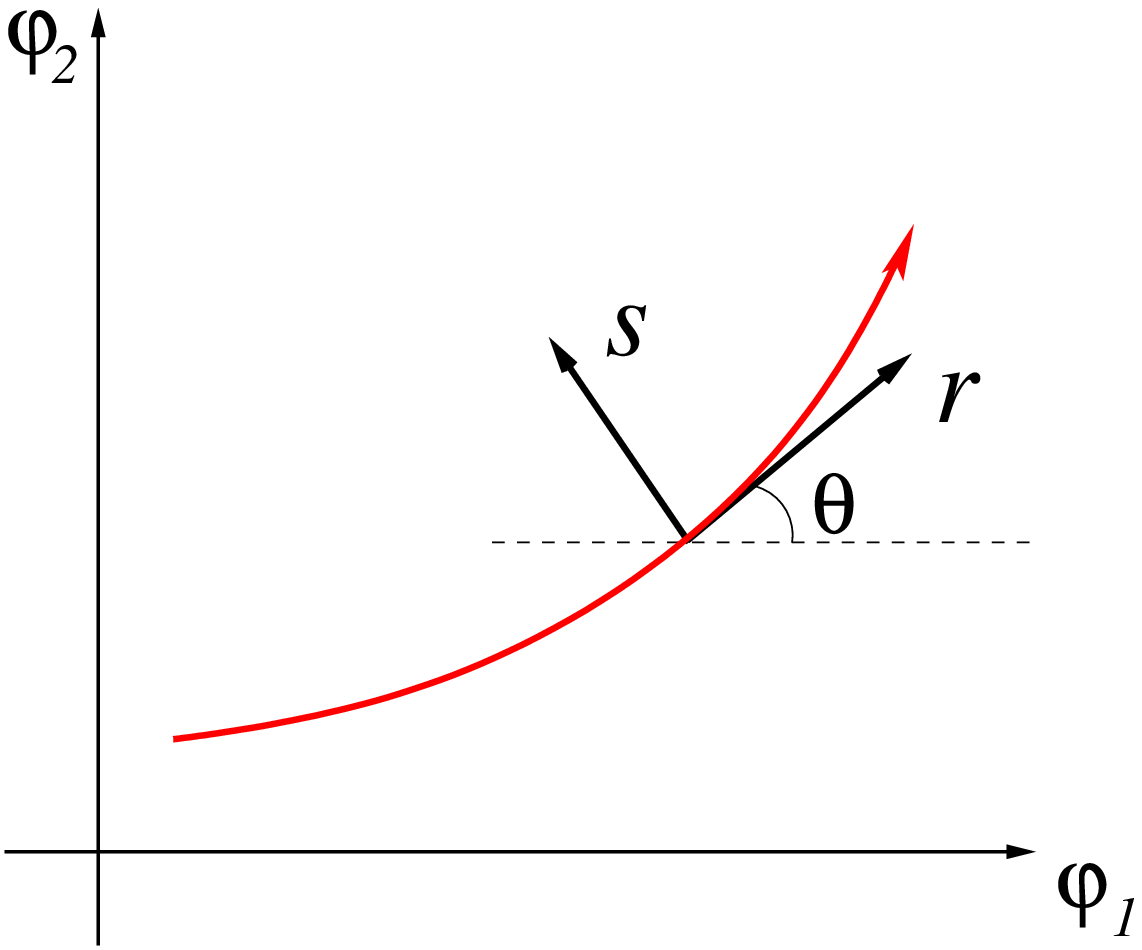} } \caption{$r$ and
$s$ are used to denote the instantaneous adiabatic and entropy
fields respectively, along and orthogonal to the curved background
solution in field space.}
\end{center}
\end{figure}

The adiabatic and isocurvature field perturbations obey the coupled
evolution equations \cite{Gordon2000}
\begin{eqnarray}
\label{eq:twoscalareom}
 {\delta r}^{\prime\prime}_\fg
+ 2\H{\delta r}^{\prime}_\fg
 + \left[ {k^2} + a^2U_{rr}
  - {\theta}^{\prime2}
  - \frac{8\pi G}{a^2} \left( \frac{a^2 r^{\prime2}}{\H}
  \right)^\prime
 \right] \delta r_\fg \nonumber\\
= 2(\theta'\delta s)' - 2\left( \frac{a^2U_r}{r'} +
\frac{\H'}{\H} \right) \theta' \delta s\,,\\
\label{eq:entropyeom}
 {\delta s}'' + 2\H{\delta s}' + \left({k^2}
  + a^2U_{ss} + 3{\theta}^{\prime2} \right) \delta s =
\frac{\theta'}{r'} \frac{k^2}{2\pi G} \Psi\,,
\end{eqnarray}
where
\begin{eqnarray}
 U_{rr} &\equiv& (\cos^2 \theta) U_{11} +(\sin
2\theta)U_{12}+(\sin^2 \theta) U_{22},\\
 U_{ss} &\equiv& (\sin^2 \theta) U_{11} -(\sin
2\theta)U_{12}+(\cos^2 \theta) U_{22}. \label{Vdd}
\end{eqnarray}

We can identify a purely adiabatic mode where $\delta s=0$ and
remains zero on large scales. However a non-zero isocurvature
perturbation appears as a source term in the perturbed inflaton
equation (\ref{eq:twoscalareom}) whenever ${\theta}' \ne 0$ and the
inflaton trajectory is curved in field space. Note that ${\theta}'$
is given by \cite{Gordon2000}
\begin{equation}
\theta' = - \frac{a^2U_s}{r'} \,,
\end{equation}
where $U_s=(\cos\theta) U_2-(\sin\theta) U_1$ is the potential
gradient orthogonal to the inflaton trajectory in field space. In
the slow-roll approximation the background field always follows the
potential gradient so the adiabatic-isocurvature coupling is
suppressed in this slow-roll limit the integrated effect of
isocurvature field perturbations on the adiabatic field perturbation
cannot in general be neglected.

Equation~(\ref{eq:twoscalareom}) shows that the isocurvature
perturbation $\delta s$ works as a source term for the adiabatic
curvature perturbation. This is in fact clearly seen if we take the
time derivative of the comoving curvature perturbation
(\ref{Radiabatic}):
\begin{equation}
 \label{dotR}
 \R' = \frac{\H}{\H'-\H^2}{k^2}\Psi
+\frac{2\H}{r'}{\theta'}\delta s\,.
\end{equation}
Therefore ${\cal R}$ (or equivalently $\zeta$) is not conserved even
in the large-scale limit in the presence of an isocurvature field
perturbation, $\delta s$, with a non-straight trajectory in field
space (${\theta}' \ne 0$).

By contrast, the solution for the isocurvature field perturbation is
independent of any initial adiabatic perturbation on large-scales.
The adiabatic perturbation provides a source term for the
isocurvature field only through the spatial gradient of the
longitudinal gauge metric potential, $\Psi$, which rapidly becomes
negligible on super-Hubble scales during slow-roll inflation.

\section{Perturbations from inflation}
\setcounter{equation}{0}

The standard hot big bang model of cosmology has a major shortcoming
in that there is no causal explanation for the existence of
primordial density perturbations on super-Hubble scales during the
radiation dominated era. The CMB, and in particular the acoustic
peaks in the temperature and polarisation anisotropies seen by the
WMAP satellite \cite{Komatsu:2008hk} provide strong evidence that
these primordial density perturbations do exist on scales much
larger than the causal horizon at early times. The detailed
distribution of primordial inhomogeneities is left as an unexplained
initial condition in the standard hot big bang.

The primary success of inflation \cite{Starobinsky:1980te,Guth:1980zm}
is to give a model for the origin of the primordial density
perturbations from vacuum fluctuations during a period of accelerated
expansion at very early times. This relies on speculative and
uncertain physics - in particular it requires some form of energy
density with negative pressure in general relativity - but the
unexpected discovery that the universe is accelerating today appears
to show that cosmological inflation does happen. Zero-point vacuum
fluctuations of any light, weakly-coupled scalar field will be
stretched up to super-Hubble scales during inflation and leave an
approximately scale-invariant and Gaussian distribution of
perturbations on large scales
\cite{Mukhanov:1981xt,Hawking:1982cz,Starobinsky:1982ee,Guth:1982ec,Bardeen83,Mukhanov:1985rz,Sasaki1986,Mukhanov1988}.

The simplest model for inflation is that it is driven by a vacuum
(potential) energy density which is a function of one or more scalar
fields. We can describe the homogeneous FRW solution using the
inflaton field, $r$, which describes the evolution along
the trajectory, $\hat{r}$ defined in Eq.~(\ref{defhatr}), in a
possibly multi-dimensional field space. For a sufficiently flat
potential the evolution can be well-described by the slow-roll
approximation which assumes that the energy density is potential
dominated
\begin{equation}
\H^2 \simeq \frac{8\pi Ga^2}{3} U(r) \,.
\end{equation}
This is equivalent to requiring that the first slow-roll parameter
is small:
\begin{equation}
\epsilon \equiv - \frac{\dot{H}}{H^2}
 = - \frac{\H'-\H^2}{\H^2} \ll 1 \,.
\end{equation}
Note that the condition for accelerated expansion requires
$\epsilon<1$. In terms of the potential we have
\begin{equation}
\epsilon \simeq \frac{1}{16\pi G} \left( \frac{U_r}{U} \right)^2 \,,
\end{equation}

We also assume the universe is overdamped, such that
\begin{equation}
3\H r' \simeq -a^2 U_r \,.
\end{equation}
This implies that we can neglect the decaying mode of the overdamped
system and we have a unique trajectory in field space for a single
field. This is a self-consistent approximation when the inflaton
field is light compared with the Hubble scale, which requires that
the second slow-roll parameter is small:
\begin{equation}
|\eta_{rr}|\ll 1 \quad {\rm where}\ \eta_{rr} \equiv
\frac{U_{rr}}{3H^2} \,.
\end{equation}

In the two-field model described in section~\ref{twofield}, allowing
for non-adiabatic perturbations, we can define three parameters
\cite{Wands:2002bn}, $\eta_{rr}$, $\eta_{rs}$ and $\eta_{ss}$,
describing the curvature of the potential, where in general we have
\begin{equation}
\eta_{IJ} \equiv \frac{1}{8\pi G} \frac{U_{IJ}}{U} \,.
\end{equation}

The background slow-roll solution is described in terms of the
slow-roll parameters by
\begin{equation}
 \label{twofieldbackground}
\dot{r}^2 \simeq \frac23 \epsilon U \, , \quad H^{-1} \dot\theta
\simeq -\eta_{r s} \,,
\end{equation}
while the perturbations on large scales (comoving wavenumber $k\ll
\H$) obey
\begin{eqnarray}
\label{twofieldperturbations} H^{-1} \dot{\delta r}_\fg &\simeq&
 \left( 2\epsilon -\eta_{rr} \right) \delta r_\fg
 - 2\eta_{rs}\delta s \,,
  \nonumber\\
H^{-1} \dot{\delta s} &\simeq& -\eta_{ss}\delta s \,,
\end{eqnarray}
where we neglect spatial gradients. Although $U_s\simeq 0$ at lowest
order in slow-roll, this does not mean that the inflaton and entropy
perturbations decouple. $\dot\theta$ given by
Eq.~(\ref{twofieldbackground}) is in general non-zero at first-order
in slow-roll and large-scale entropy perturbations do affect the
evolution of the adiabatic perturbations when $\eta_{\sigma s}\neq
0$.

While the general solution to the two second-order perturbation
equations (\ref{eq:twoscalareom}) and (\ref{eq:entropyeom}) has four
independent modes, the two first-order slow-roll equations
(\ref{twofieldperturbations}) give the approximate form of the
squeezed state on large scales. This has only two modes which we can
describe in terms of the dimensionless comoving curvature and
isocurvature perturbations:
\begin{equation}
\label{eq:defRS} \R \equiv \frac{\H}{r'} \delta r_\fg \,, \quad \S
\equiv \frac{\H}{r'} \delta s \,.
\end{equation}
The normalisation of $\R$ coincides with the standard definition of
the comoving curvature perturbation, Eq.~(\ref{defR}). The
normalisation of the dimensionless entropy during inflation, $\S$,
is chosen here coincide with Ref.~\cite{Wands:2002bn}. It can be
related to the non-adiabatic pressure perturbation (\ref{defPnad})
on large scales
\begin{equation}
\delta P_{\rm nad} \simeq - \epsilon \eta_{rs} \frac{H^2}{2\pi G} \S
\,.
\end{equation}

The slow-roll approximation can provide a useful approximation to
the instantaneous evolution of the fields and their perturbations on
large scales during slow-roll inflation, but is not expected to
remain accurate when integrated over many Hubble times, where
inaccuracies can accumulate. In single-field inflation the constancy
of the comoving curvature perturbation after Hubble exit, which does
not rely on the slow-roll approximation, is crucial in order to make
accurate predictions of the primordial perturbations using the
slow-roll approximation only around Hubble crossing. In a two-field
model we must describe the evolution after Hubble exit in terms of a
general transfer matrix:
\begin{equation}
\label{defTransfer} \left(
\begin{array}{c}
{\R} \\ {\S}
\end{array}
\right) = \left(
\begin{array}{cc}
1 & {T}_{\R\S} \\ 0 & {T}_{\S\S}
\end{array}
\right) \left(
\begin{array}{c}
\R \\\S
\end{array}
\right)_* \,.
\end{equation}
On large scales the comoving curvature perturbation still remains
constant for the purely adiabatic mode, corresponding to $\S=0$, and
adiabatic perturbations remain adiabatic. These general results are
enough to fix two of the coefficients in the transfer matrix, but
${T}_{\R\S}$ and ${T}_{\S\S}$ remain to be determined either within
a given theoretical model, or from observations, or ideally by both.
The scale-dependence of the transfer functions depends upon the
inflaton-entropy coupling at Hubble exit during inflation and can be
given in terms of the slow-roll parameters as \cite{Wands:2002bn}
\begin{eqnarray}
\label{dTdlnk} \frac{\partial}{\partial\ln k} {T}_{\R\S} &=&
2\eta_{rs} +
 (2\epsilon - \eta_{rr} + \eta_{ss}) {T}_{\R\S}
\,,\nonumber\\
\frac{\partial}{\partial\ln k} {T}_{\S\S} &=&
 (2\epsilon - \eta_{rr} + \eta_{ss}) {T}_{\S\S} \,.
\end{eqnarray}

\subsection{Initial power spectra}

The expectation value of the fluctuations of a homogeneous field are
given by
\begin{equation}
\langle \delta\varphi_I ({\bf k}_1) \delta\varphi_I ({\bf k}_2)
\rangle = (2\pi)^3 \delta^3({\bf k}_1+{\bf k}_2)
P_{\delta\varphi_I}(k_1) \,,
\end{equation}
where angle brackets denote the ensemble average (i.e., the average
over infinitely many realisations of the field, which is equivalent
to taking the spatial average in an infinite space).
The dimensionless power spectrum of the field (equivalently
the variance of the field per logarithmic range of $k$) is given by
\begin{equation}
{\cal P}_{\delta\varphi_I}(k) = \frac{4\pi k^3}{(2\pi)^3}
P_{\delta\varphi_I}(k) \,.
\end{equation}

For weakly-coupled, light fields we can neglect interactions on
wavelengths below the Hubble scale, so that vacuum fluctuations
correspond to \cite{Lidsey:1995np}
\begin{equation}
P_{\delta\varphi_I}(k) = \frac{1}{2ka^2} \,.
\end{equation}
This gives rise to the power spectrum of field fluctuations on the
Hubble scale ($k=\H=aH$) during inflation given by
\begin{equation}
 \label{H2pi}
{\cal P}_{\delta\varphi_I}
 \simeq \left( \frac{H}{2\pi} \right)_*^2 \,,
\end{equation}
where we use a $*$ to denote quantities evaluated at Hubble-exit.
If a field has a mass comparable to the Hubble scale or larger then
the vacuum fluctuations on wavelengths greater than the effective
Compton wavelength are suppressed. In addition fluctuations in
strongly interacting fields may develop correlations before Hubble
exit. But during slow-roll inflation the correlation between vacuum
fluctuations in weakly coupled, light fields at Hubble-exit is
suppressed by slow-roll parameters. This remains true under a local
rotation in fields space to another orthogonal basis such as the
instantaneous inflaton and entropy directions (\ref{defdeltar}) and
(\ref{defdeltas}) in field space.

The curvature and isocurvature power spectra at Hubble-exit are
given by
\begin{equation}
 \label{Rstar}
\left. {\cal P}_{\R} \right|_* \simeq \left. {\cal P}_{\S} \right|_*
\simeq
 \left( \frac{H^2}{2\pi\dot\sigma} \right)_*^2
 \simeq \frac{8}{3} \left( \frac{U}{\epsilon M_{\rm Pl}^4} \right)_*
 \,,
\end{equation}
while the cross-correlation is first-order in slow-roll
\cite{GrootNibbelink:2001qt,Byrnes:2006fr},
\begin{equation}
\left. {\cal C}_{\R\S} \right|_* \simeq -2C\eta_{rs} \left. {\cal
P}_{\R} \right|_* \,,
\end{equation}
where $C=2-\ln2-\gamma\approx0.73$ and $\gamma$ is the Euler number.
The normalisation chosen for the dimensionless entropy perturbation
in Eq.~(\ref{eq:defRS}) ensures that the curvature and isocurvature
fluctuations have the same power at horizon exit
\cite{Wands:2002bn}.
The spectral tilts at horizon-exit are also the same and are given
by
\begin{equation}
\label{nR*} \Delta n_\R|_* \simeq \Delta n_\S|_* \simeq -6\epsilon +
2\eta_{rr} \,.
\end{equation}
where $\Delta n_X \equiv d \ln{\cal P}_X/d \ln k$.

The tensor perturbations (\ref{teneq}) are decoupled from scalar
metric perturbations at first-order and hence the power spectrum has
the same form as in single field inflation. Thus the power spectrum
of gravitational waves on super-Hubble scales during inflation is
given by
\begin{equation}
 \label{Tstar}
\left. {\cal P}_{\rm T} \right|_*
 \simeq \frac{16H^2}{\pi M_{\rm Pl}^2}
 \simeq \frac{128}{3} \frac{U_*}{M_{\rm Pl}^4} \,,
\end{equation}
and the spectral tilt is
\begin{equation}
 \label{nT}
\Delta n_{\rm T}|_* \simeq -2 \epsilon \,.
\end{equation}

\subsection{Primordial power spectra}

The resulting primordial power spectra on large scales can be
obtained simply by applying the general transfer matrix
(\ref{defTransfer}) to the initial scalar perturbations. The scalar
power spectra probed by astronomical observations are thus given by
\cite{Wands:2002bn}
\begin{eqnarray}
\label{PR}
{\cal P}_\R &=& (1+T_{\R\S}^2) {\cal P}_\R|_* \\
{\cal P}_\S &=& T_{\S\S}^2 {\cal P}_\R|_* \\
\label{CRS} {\cal C}_{\R\S} &=& T_{\R\S} T_{\S\S} {\cal P}_\R|_* \,.
\end{eqnarray}
Note that the primordial curvature and isocurvature perturbations from
inflation are in general correlated~\cite{Langlois:1999dw} (see also
\cite{Bucher:1999re,Trotta:2001yw,Amendola:2001ni}).
The cross-correlation can be given in terms of a dimensionless
correlation angle:
\begin{equation}
\label{defDelta} \cos\Theta \equiv \frac{{\cal
C}_{\R\S}}{\sqrt{{\cal P}_\R{\cal
      P}_\S}} = \frac{T_{\R\S}}{\sqrt{1+T_{\R\S}^2}} \,.
\end{equation}

We see that if we can determine the dimensionless correlation angle,
$\Theta$, from observations, then this determines the off-diagonal
term in the transfer matrix
\begin{equation}
\label{defTRS} T_{\R\S} = \cot\Theta \,,
\end{equation}
and we can in effect measure the contribution of the entropy
perturbation during two-field inflation to the resultant curvature
primordial perturbation. In particular this allows us in principle
to deduce from observations the power spectrum of the curvature
perturbation at Hubble-exit during two-field slow-roll inflation
\cite{Wands:2002bn}:
\begin{equation}
{\cal P}_\R|_* = {\cal P}_\R  \sin^2\Theta \,.
\end{equation}

The scale-dependence of the resulting scalar power spectra depends
both upon the scale-dependence of the initial power spectra and of
the transfer coefficients. The spectral tilts are given from
Eqs.~(\ref{PR}--\ref{CRS}) by
\begin{eqnarray}
 \label{gentilt}
\Delta n_\R &=& \Delta n_\R|_* + H_*^{-1}
(\partial{T}_{\R\S}/\partial t_*) \sin 2\Theta \,,
 \nonumber\\
\Delta n_\S &=& \Delta n_\R|_* + 2 H_*^{-1}
(\partial\ln{T}_{\S\S}/\partial t_*)
\,,\\
\Delta n_{\cal C} &=& \Delta n_\R|_* + H_*^{-1} \left[
(\partial{T}_{\R\S}/\partial t_*) \tan\Theta +
(\partial\ln{T}_{\S\S}/\partial t_*) \right] \,,
 \nonumber
\end{eqnarray}
where we have used Eq.~(\ref{defTRS}) to eliminate $T_{\R\S}$ in
favour of the observable correlation angle $\Theta$.
Substituting Eq.~(\ref{nR*}) for the tilt at Hubble-exit, and
Eqs.~(\ref{dTdlnk}) for the scale-dependence of the transfer
functions, we obtain \cite{Wands:2002bn}
\begin{eqnarray}
\label{srtilts}
\Delta n_{\R} &\simeq& -(6-4\cos^2\Theta) \epsilon \nonumber\\
&&  + 2\left( \eta_{rr}\sin^2\Theta + 2\eta_{rs}
\sin\Theta\cos\Theta + \eta_{ss}\cos^2\Theta \right)
\,,\nonumber\\
\Delta n_\S &\simeq& -2\epsilon + 2\eta_{ss} \,,\\
\Delta n_{\cal C} &\simeq& -2\epsilon + 2\eta_{ss} +
2\eta_{rs}\tan\Theta
 \nonumber\,.
\end{eqnarray}

Although the overall amplitude of the transfer functions are
dependent upon the evolution after Hubble-exit and through reheating
into the radiation era, the spectral tilts can be expressed solely
in terms of the slow-roll parameters at Hubble-exit during inflation
and the correlation angle, $\Theta$, which can in principle be
observed.

If the primordial curvature perturbation results solely from the
adiabatic inflaton field fluctuations during inflation then we have
$T_{\R\S}=0$ in Eq.~(\ref{PR}) and hence $\cos\Theta=0$ in
Eqs.~(\ref{srtilts}), which yields the standard single field result
\begin{equation}
 \Delta n_\R \simeq -6\epsilon +2\eta_{rr} \,.
\end{equation}
Any residual isocurvature perturbations must be uncorrelated with
the adiabatic curvature perturbation (at first-order in slow-roll)
with spectral index
\begin{equation}
\Delta n_\S \simeq - 2\epsilon + 2\eta_{ss} \,.
\end{equation}

On the other hand, if the observed primordial curvature perturbation
is produced due to some entropy field fluctuations during inflation,
we have $T_{\R\S}\gg1$ and $\sin\Theta\simeq0$. In a two-field
inflation model any residual primordial isocurvature perturbations
will then be completely correlated (or anti-correlated) with the
primordial curvature perturbation and we have
\begin{equation}
\Delta n_\R \simeq \Delta n_{\cal C} \simeq \Delta n_\S \simeq
 - 2\epsilon + 2\eta_{ss} \,.
\end{equation}

The gravitational wave power spectrum is frozen-in on large scales,
independent of the scalar perturbations, and hence
\begin{equation}
{\cal P}_{\rm T} = {\cal P}_{\rm T}|_* \,.
\end{equation}
Thus we can derive a modified consistency relation
\cite{Lidsey:1995np} between observables applicable in the case of
two-field slow-roll inflation:
\begin{equation}
\frac{{\cal P}_{\rm T}} {{\cal P}_\R} \simeq -8 \Delta n_{\rm T}
\sin^2\Theta  \,.
\end{equation}
This relation was first obtained in Ref.~\cite{Bartolo:2001rt} at
the end of two-field inflation, and verified in Ref.~\cite{Tsuji03}
for slow-roll models. But it was realised in
Ref.~\cite{Wands:2002bn} that this relation also applies to the
primordial perturbation spectra in the radiation era long after
two-field slow-roll inflation has ended and hence may be tested
observationally.

More generally, if there is any additional source of the scalar
curvature perturbation, such as additional scalar fields during
inflation, then this could give an additional contribution to the
primordial scalar curvature spectrum without affecting the
gravitational waves, and hence the more general result is the
inequality \cite{Starobinsky:1994mh,Sasaki:1995aw}:
\begin{equation}
\frac{{\cal P}_{\rm T}} {{\cal P}_\R} \leq -8 \Delta n_{\rm T}
\sin^2\Theta \,.
\end{equation}
This leads to a fundamental difference when interpreting the
observational constraints on the amplitude of primordial tensor
perturbations in multiple inflation models.  In single field
inflation, observations directly constrain $r=[{\cal P}_{\rm
T}/{\cal P}_\R]_*$ and hence, from Eqs.~(\ref{Rstar}) and
(\ref{Tstar}), the slow-roll parameter $\epsilon$. However in
multiple field inflation, non-adiabatic perturbations can enhance
the power of scalar perturbations after Hubble exit and hence
observational constraints on the amplitude of primordial tensor
perturbations do not directly constrain the slow-roll parameter
$\epsilon$.

Current CMB data alone require $r<0.55$ (assuming power-law
primordial spectra) \cite{Komatsu:2008hk} which in single-field
models is interpreted as requiring $\epsilon<0.04$. But in multiple
field models $\epsilon$ could be larger if the primordial density
perturbation comes from non-adiabatic perturbations during
inflation.


\section{Non-linear evolution and non-Gaussianity}
\label{non_linear_sec}
\setcounter{equation}{0}

A powerful technique to calculate the primordial curvature
perturbation resulting from many inflation models, including
multi-field models, is to note that the curvature perturbation
$\zeta$ defined in Eq.~(\ref{defzeta}) can be interpreted as a
perturbation in the local expansion
\cite{Starobinsky:1982ee,Starobinsky:1986fx,Sasaki:1995aw,LMS}
\begin{equation}
 \label{deltaN}
\zeta = \delta N \,,
\end{equation}
where $\delta N$ is the perturbed expansion to uniform-density
hypersurfaces with respect to spatially flat hypersurfaces, which is
given to first-order by
\begin{equation}
 \label{linearzetarho}
\zeta = - H \frac{\delta\rho_\fg}{\dot\rho} \,,
\end{equation}
where $\delta\rho_\fg$ must be evaluated on spatially flat ($\psi=0$)
hypersurfaces (see Sections \ref{geo_quant_sec} and \ref{udg_sec}).

An important simplification arises on large scales where anisotropy
and spatial gradients can be neglected, and the local density,
expansion, etc., obeys the same evolution equations as in a homogeneous
FRW universe
\cite{Salopek:1990jq,Sasaki:1995aw,Sasaki98,Wands00,Lyth:2003im,Rigopoulos03,LMS}.
Thus we can use the homogeneous FRW solutions to describe the local
evolution, which is known as the ``separate universe'' approach
\cite{Salopek:1990jq,Sasaki:1995aw,Sasaki98,Wands00,Rigopoulos03}. In
particular we can evaluate the perturbed expansion in different parts
of the universe resulting from different initial values for the fields
during inflation using the homogeneous background solutions
\cite{Sasaki:1995aw}.

In the slow-roll approximation the integrated expansion on
super-Hubble scales from some initial spatially flat hypersurface up
to a subsequent fixed density hypersurface (say at the epoch of
primordial nucleosynthesis) is some function of the local field
values on the initial hypersurface, $N(\varphi_I|_\psi)$. More
generally we expect this to hold whenever we can neglect the
decaying mode for the field perturbations on super-Hubble scales.
The resulting primordial curvature perturbation on the
uniform-density hypersurface is then
\begin{equation}
 \label{lineardeltaN}
\zeta = \sum_I N_{I} \delta\varphi_{I\fg} \,,
\end{equation}
where $N_I\equiv \partial N/\partial \varphi_I$ and
$\delta\varphi_{I\fg}$ is the field perturbation on some initial
spatially-flat hypersurfaces during inflation (\ref{eq:defdphipsi}).
In particular the power spectrum for the primordial density
perturbation in a multi-field inflation can be written (at leading
order) in terms of the field perturbations after Hubble-exit as
\begin{equation}
{\cal P}_\zeta = \sum_I N_I^2 {\cal P}_{\delta\varphi_{I\fg}} \,.
\end{equation}

This approach is readily extended to estimate the non-linear effect
of field perturbations on the metric perturbations
\cite{Sasaki98,Lyth:2003im,Lyth:2005fi,LMS}.
We can take Eq.~(\ref{deltaN}) as our definition of the non-linear
primordial curvature perturbation, $\zeta$, so that in the radiation
dominated era the non-linear extension of Eq.~(\ref{linearzetarho})
is given by \cite{LMS}
\begin{equation}
 \label{defzetarho}
 \zeta = \frac14 \ln \left( \frac{\rho_\fg}{\rho_0} \right) \,,
\end{equation}
where $\rho_\fg(t,{\bf x})$ is the perturbed (inhomogeneous) density
evaluated on a spatially flat hypersurface and $\rho_0(t)$ is the
background (homogeneous) density.
See
Refs.~\cite{Rigopoulos03,Rigopoulos:2004ba,Rigopoulos:2005xx,Langlois:2005ii,Langlois:2005qp,Langlois:2006iq}
for alternative approaches to define the non-linear extension of
$\zeta$.

This non-linear curvature perturbation as a function of the initial
field fluctuations can simply be expanded as a Taylor expansion
\cite{Lyth:2005fi,Seery05b,Byrnes06,Seery06b}
\begin{equation}
 \label{Taylor}
\zeta \simeq \sum_I N_{I} \delta\varphi_{I\fg} + \frac12
\sum_{I,J} N_{IJ} \delta\varphi_{I\fg} \delta\varphi_{J\fg}
+\frac16\sum_{I,J,K} N_{IJK} \delta\varphi_{I\fg}
\delta\varphi_{J\fg} \delta\varphi_{K\fg} + \ldots \,.
\end{equation}
where we now identify (\ref{lineardeltaN}) as the leading-order
term.

We expect the field perturbations at Hubble-exit to be close to
Gaussian for weakly coupled scalar fields during inflation
\cite{Maldacena02,Seery05a,Rigopoulos:2005bq,Seery05b,Seery06a}. In
this case the bispectrum of the primordial curvature perturbation at
leading (fourth) order, can be written as
\cite{Gangui:1993tt,Komatsu:2001rj}
\bea
\label{fNLmultifielddefn}
B_\zeta(\bkone,\bktwo,\bkthree) &=&
 \frac65 \fNL \left[ P_\zeta(k_1) P_\zeta(k_2) + P_\zeta(k_2)
P_\zeta(k_3) + P_\zeta(k_3) P_\zeta(k_1) \right] \,.
\eea
where $P_\zeta(k)=2\pi^2{\cal P}_\zeta(k)/k^3$, and the
dimensionless non-linearity parameter is given, using the $\delta
N$-formalism,
by\footnote{Note, that the factor ``$5/6$'' in \eq{fNLmultifield} is
a historical convention, due to the original definition
\cite{Komatsu:2001rj} which was given in terms of the Newtonian
potential, which on large scales in the matter era is given by
$\Phi=-(3/5)\zeta$.}
 \cite{Lyth:2005fi}
\begin{equation}
 \label{fNLmultifield}
 \fNL = \frac{5}{6} \frac{N_A N_B N^{AB}}{\left(N_C N^C\right)^2} \,.
\end{equation}
Similarly to the bispectrum, the connected part of the trispectrum in
this case can be written as \cite{Seery06b,Byrnes06}
\bea
\label{tauNLgNLdefn}
 T_\zeta (\bkone,\bktwo,\bkthree,\bkfour) &=&
\tau_{NL}\left[P_\zeta(|{\bf k}_1+{\bf k}_3|)P_\zeta(k_3)P_\zeta(k_4)
+(11\,\,\rm{perms})\right] \nonumber \\
&&+\frac{54}{25}g_{NL}\left[P_\zeta(k_2)P_\zeta(k_3)P_\zeta(k_4)
+(3\,\,\rm{perms})\right]\,.
\eea
where
\bea
 \label{tauNLmultifield}
  \tau_{NL}&=&\frac{N_{AB}N^{AC}N^BN_C}{(N_DN^D)^3}\,, \\
 \label{gNLmultifield}
  g_{NL}&=&\frac{25}{54}\frac{N_{ABC}N^AN^BN^C}{(N_DN^D)^3}\,.
 \eea
The expression for $\tau_{NL}$ was first given in
\cite{Alabidi:2005qi}. Note that we have factored out products in
the trispectrum with different $k$ dependence in order to define the
two $k$ independent non-linearity parameters $\tau_{NL}$ and
$g_{NL}$. This gives the possibility that observations may be able
to distinguish between the two parameters \cite{Okamoto:2002ik}.

In many cases there is single direction in field-space, $\chi$,
which is responsible for perturbing the local expansion, $N(\chi)$,
and hence generating the primordial curvature perturbation
(\ref{Taylor}). For example this would be the inflaton field in
single field models of inflation, or it could be a late-decaying
scalar field \cite{Mollerach:1989hu,Linde:1996gt,Enqvist02} as in
the curvaton scenario \cite{Lyth02,Moroi01}.
In this case the curvature perturbation (\ref{Taylor}) is given by
\begin{equation}
 \label{localN}
\zeta \simeq
 N' \delta\chi_\fg
 + \frac12 N'' \delta\chi^{2}_\fg
 + \frac16 N''' \delta\chi^{3}_\fg
+ \ldots \,,
\end{equation}
and the non-Gaussianity of the primordial perturbation has the
simplest ``local'' form
\begin{equation}
 \label{localzeta}
\zeta = \zeta_1 + \frac35f_{NL}\zeta_1^2 +
\frac{9}{25}g_{NL}\zeta_1^3 +\ldots
\end{equation}
where $\zeta_1=N'\delta\chi_\fg$ is the leading-order Gaussian
curvature perturbation and the non-linearity parameters $f_{NL}$ and
$g_{NL}$, are given by \cite{Lyth:2005fi,Sasaki:2006kq}
\bea
 \label{fNL1field}
  f_{NL}&=&\frac56\frac{N''}{(N')^2}\,, \\
 \label{gNL1field}
  g_{NL}&=&\frac{25}{54}\frac{N'''}{(N')^3}\,,
\eea The primordial bispectrum and trispectrum are then given by
Eqs.~(\ref{fNLmultifielddefn}) and~(\ref{tauNLgNLdefn}), where the
non-linearity parameters $f_{NL}$ and $g_{NL}$, given in
Eqs.~(\ref{fNLmultifield}) and~(\ref{gNLmultifield}), reduce to
Eqs.~(\ref{fNL1field}) and~(\ref{gNL1field}) respectively, and
$\tau_{NL}$ given in Eq.(\ref{tauNLmultifield}) reduces to
\bea
\label{tauNL1field}
 \tau_{NL}&=&\frac{(N'')^2}{(N')^4}=\frac{36}{25}f_{NL}^2\,.
\eea
Thus $\tau_{NL}$ is proportional to $f_{NL}^2$ (first shown in
\cite{Okamoto:2002ik} using the Bardeen potential, and in
\cite{Lyth:2005fi} using this notation). However the trispectrum
could be large even when the bispectrum is small because of the
$g_{NL}$ term \cite{Okamoto:2002ik,Sasaki:2006kq}.

In the case of where the primordial curvature perturbation is
generated solely by adiabatic fluctuations in the inflaton field,
$r$, the curvature perturbation is non-linearly conserved on large
scales \cite{Lyth:2003im,LMS,Langlois:2005qp} and we can calculate
$N'$, $N''$, $N'''$, etc, at Hubble-exit. In terms of the slow-roll
parameters, we find
\bea
 N'&=& \frac{{H}}{\dot{{\varphi}}}
\simeq \frac{1}{\sqrt{2}}\frac{1}{\Mp}\frac{1}{\sqrt{\epsilon}} \sim {\mathcal{O}} \left(\epsilon^{-1/2}\right)\,,  \\
N''&\simeq&-\frac12\frac{1}{\Mp^2}\frac{1}{\epsilon}(\eta_{rr}-2\epsilon) \sim {\mathcal{O}} \left( 1 \right)\,,  \\
N'''&\simeq&\frac{1}{\sqrt{2}}\frac{1}{\Mp^3}\frac{1}{\epsilon\sqrt{\epsilon}}\left(\epsilon\eta_{rr}-\eta_{rr}^2+\frac12\xi_r^2\right)
\sim {\mathcal{O}} (\epsilon^{1/2})\,,
 \eea
where we have used the reduced Planck mass $\Mp^2=(8\pi G)^{-1}$ and
introduced the second-order slow-roll parameter
$\xi_\sigma^2=\Mp^4U_r U_{rrr}/U^2$.
Hence the non-linearity parameters for single field inflation,
(\ref{fNL1field}) and~(\ref{gNL1field}), are given by
\bea
 f_{NL}&=&\frac56(\eta_{rr}-2\epsilon)\,, \\
 g_{NL}&=&\frac{25}{54}\left(2\epsilon\eta_{rr}-2\eta_{rr}^2+
 \xi_r^2\right)\,. \eea
with $\tau_{NL}$ given by Eq.~(\ref{tauNL1field}). Although there
are additional contributions to the primordial bispectrum and
trispectrum coming from the intrinsic non-Gaussianity of the field
perturbations at Hubble-exit, these are also suppressed by slow-roll
parameters in slow-roll inflation. Thus the primordial
non-Gaussianity is likely to be too small to ever be observed in the
conventional inflaton scenario of single-field slow-roll
inflation~\cite{Acquaviva:2002ud,Maldacena02}. Indeed any detection
of primordial non-Gaussianity $f_{NL}> 1$ would appear to rule out
this inflaton scenario.

However significant non-Gaussianity can be generated due to
non-adiabatic field fluctuations. Thus far it has proved difficult
to generate detectable non-Gaussianity in the curvature perturbation
during conventional slow-roll inflation, even in multiple field
models \cite{Rigopoulos:2005ae,Vernizzi:2006ve},
though see Refs.~\cite{Byrnes:2008wi,Cogollo:2008bi}.
But detectable non-Gaussianity might be produced in non-slow-roll
\cite{Chen:2008wn} or non-canonical scalar field inflation
\cite{Alishahiha:2004eh,Langlois:2008qf}, or when the curvature
perturbation is generated from isocurvature field perturbations at
the end of inflation \cite{BernardeauUzan,LRreview}, during
inhomogeneous reheating
\cite{Dvali:2003em,Kofman:2003nx,Zaldarriaga,Dvali:2003ar,Kolb,Byrnes05},
or some time after inflation has ended in the curvaton model
\cite{Lyth02,LUW,Bartolo:2003jx,Malik:2006pm,Sasaki:2006kq}.

\section{Summary and outlook}

Linear perturbations have become part of the standard toolbox of
modern cosmology. Earlier confusion surrounding apparently different
behaviour found in different coordinate bases has largely been
resolved through the use of variables which have gauge-invariant
definitions. In Section \ref{gi_var_sec} we have emphasised the
physical interpretation of these gauge-invariant variables.

The power spectrum of primordial perturbations revealed by the
cosmic microwave background and large scale galaxy surveys is a
powerful probe of inflationary models of the very early universe,
and a challenge for alternative theories. Linear theory enables us
to relate the primordial spectra to quantum fluctuations in the
metric and matter fields at much higher energies. In the simplest
single field inflation models, it is possible to equate the
primordial density perturbation with the curvature perturbation,
$\zeta$ defined in \eq{defzeta}, during inflation which remains
constant on large scales for adiabatic density perturbations. More
generally, if one allows for non-adiabatic perturbations then it
becomes necessary to allow for variation in the curvature
perturbation, even on super-Hubble scales. Nonetheless it is still
possible in many cases to identify the primordial density
perturbation with a perturbed expansion in the $\delta N$ approach
described in Section~\ref{non_linear_sec} where the integrated
expansion, $N$, is a function of the local field values on spatially
flat slices during inflation. As one goes beyond the textbook
examples, it becomes necessary to have a clear physical definition
of the perturbation variables to consistently extend the background
FRW equations to the inhomogeneous perturbations.

The new frontier in the study of cosmological perturbations is the
study of non-linear primordial perturbations, at second-order and
beyond. Many of the familiar certainties of linear perturbation theory
no longer apply. We have shown in Section \ref{gaugesec} that
quantities that were automatically gauge-independent at first order
(including the non-adiabatic pressure perturbation, anisotropic
stress, and the tensor metric perturbation) become gauge dependent at
second order. We have shown in Section \ref{gi_var_sec} that it is
possible to use the same methodology to construct gauge-invariant
variables at second (and higher) order. A variable with an unambiguous
physical meaning will have a gauge invariant definition.  The
resulting gauge invariant definitions inevitably become more
complicated than those at first order and we have only presented
explicit definitions at second order for a few cases. Likewise we have
not attempted to present the second order dynamical equations in a
comprehensive manner as was done at first order in
Section~\ref{sect:dynamics}. Our aim has been to provide an
introduction to some of the issues that arise at higher order.
Early works on non-linear and second order perturbation theory include
Refs.~\cite{Tomita1967,Peebles80}, more recently see also
Refs.~\cite{Pyne:1995bs,Tomita:2005nu,Noh2004,Nakamura:2004wr,Nakamura:2004rm,Nakamura:2006rk}.

Non-linearities allow additional information to be gleaned from the
primordial perturbations. Much effort is currently being devoted to
the study of higher-order correlations. Non-Gaussianity in the
distribution of primordial density perturbations would reveal
interactions beyond the linear theory. Such interactions are minimal
(suppressed by slow-roll parameters) in the simplest single field
inflation model, so any detection of primordial non-Gaussianity would
cause a major upheaval in our thinking about the very early
universe. In principle the $\delta N$ approach can be easily extended
to higher-orders, simply by extending the Taylor series for the
integrated expansion as a function of the field perturbations beyond
linear order. This enables one to compute higher-order correlation
functions for $\zeta$ as shown in Section \ref{non_linear_sec}. The
challenge is then to develop transfer functions to relate the
primordial $\zeta$ to observables beyond linear
order~\cite{Bartolo:2004if,Bartolo:2005kv,Bartolo:2006cu,Bartolo:2006fj},
although large primordial non-Gaussianity (e.g., $\fNL\gg1$) is
expected to dominate over non-linearity in the transfer functions.

The Klein-Gordon equation in closed form at second order shows that
at second order scalar perturbations will also be sourced by terms
quadratic in first order field perturbations
\cite{M2006,Malik:2007nd}. This and second order perturbation theory
in general, provides an alternative to using the $\delta N$
formalism in calculating the primordial non-gaussianity $\fNL$
\cite{Seery:2008qj}.
The main advantage of perturbation theory is that it can also be
extended to sub-horizon scales, whereas the $\delta N$ formalism is
only valid on super-horizon scales, and in some cases it has been
shown to be numerically more efficient \cite{Malik:2006pm}.

Typically non-linear effects are going to be small, given that
scalar metric perturbations are only of order $10^{-5}$. Primordial
tensor modes are smaller, and vector modes are effectively zero
during scalar field driven inflation. But the additional information
available from higher-order correlations and the use of optimised
filters for specific forms of non-Gaussianity means that there are
already impressive constraints on the degree of primordial
Gaussianity.

Qualitatively new effects appear beyond linear order. The
non-linearity of the field equations inevitably leads to mixing
between scalar, vector and tensor modes and the existence of
primordial density perturbations then inevitably generates vector
and tensor modes
\cite{Mollerach:2003nq,Nakamura:2004rm,Ananda:2006af,Baumann:2007zm,Lu:2007cj}.
As shown in Section \ref{gaugesec},
if we continue studies of scalar
perturbations to higher order then the distinction between the
different types of perturbations becomes gauge dependent and
consistent computation will require careful (gauge invariant)
definition of the variables being used.

\section*{Acknowledgement}

DW is supported by STFC.
The authors are grateful to all their collaborators for the
collaborative work upon which much of this review is based, including
Kishore Ananda, Marco Bruni, Ed Copeland, Juan Garcia-Bellido, Chris
Gordon, Kazuya Koyama, David Langlois, Andrew Liddle, Jim Lidsey,
David Lyth, David Matravers, Misao Sasaki, David Seery, and Filippo
Vernizzi, and we thank Hooshyar Assadullahi, Chris Byrnes, Anne Green,
and Roy Maartens for their careful reading of this manuscript.
Algebraic computations were assisted by the GRTensorII package for
Maple and by Cadabra \cite{Peeters}.

\newpage
\appendix
\section{Definitions and notation}
\label{defandnot_sect}

\subsection{Notation}

The sign convention is (+++) in the classification of Misner, Thorne,
and Wheeler \cite{MTW}. \\

Tensor indices:\\
Greek indices, such as $\alpha,\beta,\dots,\mu,\nu,\dots$, run from 0
to $3$. Latin indices, such as $a,b,\dots,i,j,\dots$, run from 1 to
$3$, that is only over spatial dimensions. \\

Spatial three-vectors are written in boldface,
i.e.~$\vect{v}\equiv v^i$, whenever convenient.\\

Throughout this work we use the units $c=\hbar=1$.

\subsection{Definitions}

The connection coefficient is defined as
\be \Gamma^{\gamma}_{\beta\mu}= \frac12 g^{\alpha\gamma} \left(
g_{\alpha\beta, \mu} + g_{\alpha\mu,\beta} -
 g_{\beta\mu,\alpha} \right) \,.
\ee
The Riemann tensor is defined as
\be
R^{\alpha}_{~\beta\mu\nu}=
\Gamma^{\alpha}_{\beta\nu,\mu}-\Gamma^{\alpha}_{\beta\mu,\nu}+
\Gamma^{\alpha}_{\lambda\mu}\Gamma^{\lambda}_{\beta\nu}-
\Gamma^{\alpha}_{\lambda\nu}\Gamma^{\lambda}_{\beta\mu} \,.
\ee
The Ricci tensor is a contraction of the Riemann tensor and given by
\be
R_{\mu\nu}=R^{\alpha}_{~\mu\alpha\nu} \,,
\ee
and the Ricci scalar is given by contracting the Ricci tensor
\be
R=R^{\mu}_{~\mu} \,.
\ee
The Einstein tensor is defined as
\be
G_{\mu\nu}=R_{\mu\nu}-\frac{1}{2}g_{\mu\nu}R \,.
\ee
The covariant derivatives are denoted by
\bea
&{}_{;\mu}&\equiv\nabla_{\mu}~~~ {\rm{Covariant~differentiation~with~
respect~to}}~g_{\mu\nu}
 \,.  \nonumber
\eea
Partial derivatives are denoted by
\be X_{,i}\equiv\frac{\p X}{\p x^i}\,. \ee
Symmetrisation and anti-symmetrisation are, as usual, defined as
\be
V_{(i,j)}\equiv\frac{1}{2}\left(V_{i,j}+V_{j,i}\right)\,,
\qquad
V_{[i,j]}\equiv\frac{1}{2}\left(V_{i,j}-V_{j,i}\right)\,.
\ee

\subsection{Lie derivatives}

The Lie derivatives with respect to a vector field $\xi^\mu$ of a
scalar $\vp$, a covariant vector $v_\mu$, and a covariant tensor
$t_{\mu\nu}$ are given by (see e.g. Ref.~\cite{Stephani2004})
\bea
\label{lie_scalar} \pounds_{\xi}\vp&=&
\xi^\lambda\vp_{,\lambda}\,,\\
\label{lie_vector} \pounds_{\xi} v_\mu
&=& v_{\mu,\alpha}\xi^\alpha+v_\alpha \xi^\alpha_{,~\mu}\,,\\
\label{lie_tensor} \pounds_{\xi}t_{\mu\nu}
&=&t_{\mu\nu,\lambda}\xi^\lambda
+t_{\mu\lambda}\xi^\lambda_{,~\nu}+t_{\lambda\nu}\xi^\lambda_{,~\mu}\,.
\eea
%

\subsection{Covariant derivatives}

The covariant derivatives of a scalar $\vp$, a covariant vector
$V_\mu$, and a covariant tensor $t_{\mu\nu}$ are given by
(e.g.~\cite{Stephani2004})
\bea
\vp_{;\mu}&=&\vp_{,\mu}\,,\nonumber\\
V_{\mu;\nu}&=&V_{\mu,\nu}-\Gamma^\alpha_{\mu\nu}V_\alpha\,, \nonumber\\
t_{\mu\nu;\lambda}&=&t_{\mu\nu,\lambda}
-\Gamma^\alpha_{\mu\lambda}t_{\alpha\nu}
-\Gamma^\beta_{\nu\lambda}t_{\beta\mu}\,.
\eea

\section{Components of connection coefficients and tensors}

In the following no gauge is specified, and we leave quantities
undecomposed whenever convenient (the decomposition rules are given in
\eqs{vectordecomp}, (\ref{decompBi}), and (\ref{decompCij})).

\subsection{Connection coefficients}

%
The connection coefficients up to and including second order are
\bea
\Gamma^0_{00} &=&
\frac{a'}{a}+\phi_1'+\frac{1}{2}\phi_2'-2\phi_1\phi_1'
+\frac{a'}{a}B_{1k}B^{1k}+B_1^{k}B_{1k}'+B_1^k\phi_{1,k} \,, \\
\Gamma^0_{0i} &=&
\phi_{1,i}+\frac{1}{2}\phi_{2,i}
+\frac{a'}{a}\left(B_{1i}+\frac{1}{2}B_{2i}\right)
-2\phi_1\phi_{1,i}-2\frac{a'}{a}\phi_1 B_{1i}
+B_1^k C_{1ki}' \nonumber \\
&&
+\frac{1}{2} B_1^k \left( B_{1k,i} - B_{1i,k} \right)\,, \\
\Gamma^i_{00} &=&
\frac{a'}{a}\left(B_{1}^{~i}+\frac{1}{2}B_{2}^{~i}\right)
+\left(B_{1}^{~i\prime}+\frac{1}{2}B_{2}^{~i\prime}\right)
+\phi_{1,}^{~i}+\frac{1}{2}\phi_{2,}^{~i}-\phi_1'B_1^{~i}   \nonumber \\
&&
-2\frac{a'}{a}C_1^{ik}B_{1k}
-2C_1^{ik}B'_{1k}-2C_1^{ik}\phi_{1,k} \,, \\
\Gamma^i_{j0} &=&
\frac{a'}{a}\delta^i_j+{C^i_{1~j}}'+\frac{1}{2}{C^i_{2_j}}'
+\frac{1}{2}\left( B_{1~,j}^{~i} - B_{1j,}^{~~i}\right)
+\frac{1}{4}\left( B_{2~,j}^{~i} - B_{2j,}^{~~i}\right) \nonumber \\
&&
-2C_1^{ik}C_{1jk}'-B_1^{~i}\left(\frac{a'}{a}B_{1j}+\phi_{1,j}\right)
+C_1^{ik}\left( B_{1j,k} - B_{1k,j}\right)\,, \\
\Gamma^0_{ij} &=&
\left[\frac{a'}{a}-2\frac{a'}{a}\left(\phi_1+\frac{1}{2}\phi_{2}\right)
\right]\delta_{ij} +C_{1ij}'+\frac{1}{2}C_{2ij}'
+2\frac{a'}{a}C_{ij}+\frac{a'}{a}C_{2ij}
\nonumber \\
&&
-\frac{1}{2}\left(B_{1j,i} + B_{1i,j} \right)
-\frac{1}{4}\left(B_{2j,i} + B_{2i,j} \right)
+B_1^{~k}\left(C_{1jk,i}+C_{1ik,j}-C_{1ij,k}\right)\nonumber \\
&&
+\phi_1\left[B_{1j,i} + B_{1i,j}
-4\frac{a'}{a}C_{1ij}-2C_{1ij}'\right]
+\delta_{ij}\frac{a'}{a}\left[4\phi_1^2-B_1^{~k} B_{1k} \right] \,, \\
\Gamma^i_{jk} &=&
C^{~i}_{1k,j}+C^{~i}_{1j,k}-C^{~~~~i}_{1jk,}
+\frac{1}{2}\left(C^{~i}_{2k,j}+C^{~i}_{2j,k}-C^{~~~~i}_{2jk,}\right)
-\frac{a'}{a}\delta_{kj} \left(B_{1}^{~i}+\frac{1}{2}B_{2}^{~i}\right)
\nonumber \\
&&
+\frac{1}{2}B_1^{~i}\left( B_{1k,j} + B_{1j,k} \right)
+B_1^{~i}\left[2\frac{a'}{a}\phi_1\delta_{kj}
-C_{1jk}'-2\frac{a'}{a}C_{1jk}\right]
+2\frac{a'}{a}\delta_{kj}B_{1l} C_1^{li} \nonumber \\
&&-2C_1^{il}
\left(C_{1kl,j}+C_{1jl,k}-C_{1jk,l}
\right)\,,
\eea
including scalar, vector, and tensor perturbations.

\subsection{Energy-momentum tensor for $N$ scalar fields}
\label{scalarTmunu_sect}

The energy-momentum tensor for $N$ scalar fields with potential
$U(\vp_I)$ is then split into background, first, and second order
perturbations, using \eq{tensor_split1}, as
\be
T^\mu_{~\nu}\equiv T^\mu_{(0)\nu} +\delta T^\mu_{(1)\nu}
+\frac{1}{2} \delta T^\mu_{(2)\nu}\,,
\ee
and we get for the components, from \eq{multiTmunu}, at zeroth order
\bea
T^{0}_{(0)0} &=& -\left(\sum_K\frac{1}{2a^2}{\vp'_{0K}}^2+U_0\right)\,,
\qquad
T^{i}_{(0)j} =
\left(\frac{1}{2a^2}{\sum_K\vp'_{0K}}^2-U_0\right)\delta^i_{~j}\,,
\eea
at first order
\bea
\label{deltaT100}
\delta T^{0}_{(1)0} &=& -\frac{1}{a^2}\sum_K\left(
\vp_{0K}'\dvpK1'-{\vp'_{0K}}^2\A1
\right)-\dU1   \,,\nonumber \\
\label{deltaT10i}
\delta T^{0}_{(1)i}
&=& -\frac{1}{a^2}\sum_K\left(\vp_{0K}'\dvpK1_{,i}\right)\,, \\
\delta T^{i}_{(1)j} &=& \frac{1}{a^2}\left[
\sum_K\left(\vp_{0K}'\dvpK1'-{\vp'_{0K}}^2\A1\right)-a^2\dU1
\right]\delta^i_{~j}\,,\nonumber
\eea
and at second order in the perturbations
\bea
\delta T^{0}_{(2)0} &=& -\frac{1}{a^2}\sum_K\Big[
\vp_{0K}'\dvpK2'-4\vp_{0K}'\phi_1\dvpK1'-{\vp'_{0K}}^2\phi_2
+4{\vp'_{0K}}^2\A1^2+\dvpK1'^2+a^2\dU2 \nonumber\\
&&\qquad\qquad+\dvpdvpKll-{\vp_{0K}'}^2 B_{1k}B_1^{~k}
\Big]
\,,\nonumber \\
\label{deltaT20i}
\delta T^{0}_{(2)i}
&=& -\frac{1}{a^2}\sum_K\left(\vp_{0K}'\dvpK2_{,i}
-4 \A1\vp_{0K}'\dvpK1_{,i}+2\dvpK1'\dvpK1_{,i}
\right)\,, \\
\delta T^{i}_{(2)j}
&=& \frac{1}{a^2}\sum_K\Big[
\vp_{0K}'\dvpK2'-4\vp_{0K}'\A1\dvpK1'-{\vp'_{0K}}^2\A2
+4{\vp'_{0K}}^2\A1^2+\dvpK1'^2-\dvpdvpKll\nonumber\\
&&\qquad\qquad
-{\vp_{0K}'}^2 B_{1k}B_1^{~k}
-2\vp_{0K}'\delta\vp_{1K,l}B_{1}^{~~l}
-a^2\dU2\Big]\delta^i_{~j}\nonumber\\
&&
\qquad+ \frac{2}{a^2}\left(\vp_{0K}'B_{1}^{~i}+\dvpK1_{,}^{~i}
\right)\delta\vp_{1K,j}\,.\nonumber
\eea
%

\subsection{Energy-momentum tensor for fluid}
\label{fluidTmunu_sect}

We get for the components of the stress energy tensor, indices down,
in the background
\bea
T_{00} = -a^2\rho_0 \,, \qquad
T_{0i} = 0\,,  \qquad
T_{ij} = a^2 P_0 \,,
\eea
at first order,
\bea
{}^{(1)}\delta T_{00} &=& a^2\left(\delta\rho_1+2\rho_0\A1\right) \, , \\
{}^{(1)}\delta T_{0i} &=& a^2\left[
-\left(\rho_0+P_0\right)v_{1i}
+\rho_0B_{1i}\right] \,, \\
{}^{(1)}\delta T_{ij} &=& a^2\left[\delta
P_1\delta_{ij}+ 2P_0 C_{1ij} + \pi_{1ij}\right] \,, \eea
and at second order
\bea
{}^{(2)}\delta T_{00} &=&
a^2\left[\delta\rho_2+2\rho_0\A2+4\A1\delta\rho_1
+2\left(\rho_0+P_0\right)v_{1k}v_1^{~k}\right] \, , \\
{}^{(2)}\delta T_{0i} &=& -a^2\Big[
\left(\rho_0+P_0\right)
\left(v_{2i}+2\A1v_{1i}+4C_{1ik}v_1^{~k}\right)+\rho_0B_{2i}
+2\delta\rho_1 B_{1i}
\nonumber\\
&&\qquad
+2v_{1i}\left(\delta\rho_1+\delta P_1\right)
+4\pi_{1ik}v_1^{~k}\Big] \,, \\
{}^{(2)}\delta T_{ij} &=& a^2\Big[
\delta P_2\,\delta_{ij}+ 2P_0 C_{2ij}+4\delta P_1 C_{1ij}
+\left(\rho_0+P_0\right)\left(v_{1i}+B_{1i}\right)
\left(v_{1j}+B_{1j}\right)
\nonumber\\
&&\qquad+ \pi_{2ij} \Big]\,.
\eea
%

\section{Geometry of spatial hypersurfaces}
\label{App_n_components}

\subsection{Components at first and second order
of shear, expansion, and acceleration}

The calculation of the shear, defined above in \eq{defshear}, simplifies in
case of the unit normal vector field $n^\mu$, that is for
$n_i\equiv\bf{0}$,
\be
\sigma_{ij}=-n_0 \Gamma^0_{ij}-\frac{1}{3}\theta\ g_{ij}\,,
\ee
which gives (including vectors and tensors) at first order
\bea
{\delta}^{(1)}\sigma_{00}&=&0\,,
\qquad {\delta}^{(1)}\sigma_{0i}=0\,,\\
{\delta}^{(1)}\sigma_{ij}&=&a\Big[
C_{1ij}'-B_{1(i,j)}-\frac{1}{3}\delta_{ij}
\left(C_{1k}^{\prime~~k}-B_{1k,}^{~~~k}\right)\Big]\,,
\eea
and at second order
\bea
{\delta}^{(2)}\sigma_{00}&=&0\,,\\
{\delta}^{(2)}\sigma_{0i}&=&2a\Big[
B_1^k\left(C'_{1ik}-B_{1(1,k)}\right)
-\frac{1}{3}B_{1i}\left(C_{1k}^{\prime~~k}-B_{1k,}^{~~k}
\right)\Big]\,,\\
{\delta}^{(2)}\sigma_{ij}&=&a\Big[
C_{2ij}'-B_{2(i,j)}+2B_1^k\left(C_{1ki,j}+C_{1kj,i}-C_{1ij,k}\right)
+2\phi_1 \left(B_{1(i,j)}-C_{1ij}'\right)\nonumber \\
&&\qquad -\frac{4}{3}C_{1ij}\left(C_{1k}^{\prime~~k}-B_{1k,}^{~~k}
\right)
+\frac{1}{3}\delta_{ij}
\Big\{
-C_{2k}^{\prime~~k}+B_{2k,}^{~~~k}
+2\phi_1\left(C_{1k}^{\prime~~k}-B_{1k,}^{~~k}
\right) \nonumber \\
&&\qquad
+4C_{1}^{kl}\left(C_{1kl}'-B_{1k,l}\right)
-2B_{1}^l\left(2C_{1lk,}^{~~~k}-C_{1~~k,l}^{~k}\right)
\Big\}\Big]\,.
\eea

The expansion is given from \eq{theta}
in the background as
\be
\theta_0=\frac{3a'}{a^2} \,,
\ee
at first order
\be
\delta\theta_1=\frac{1}{a}\left[
-3\frac{a'}{a}\A1+{C_{1k}^{~~k}}'-B_{1k,}^{~~k}\right]\,,
\ee
and at second order
\bea
\delta\theta_2 =\frac{1}{a}\Bigg[
&&-{3}\frac{a'}{a}\left(\A2-3\A1^2\right)
+\left({C_{2k}^{~~k}}'-B_{2k,}^{~~k}\right)
+2\phi_1\left(B_{1k,}^{~~k}-{C_{1k}^{~~k}}'\right)\nonumber\\
&&-{3}\frac{a'}{a}B_{1k}B_{1}^k
-4C_1^{kl}C_{1kl}'+4C_1^{kl}B_{1l,k}
+4B_1^lC_{1lk,}^{~~~k}-2B_1^k C^l_{1~l,k} \Bigg]\,.\nonumber\\
\eea
The acceleration is given from \eq{defacc} at first order as
\be
a_{(1)0}=0\,, \qquad a_{(1)i}=\A{}_{1,i}\,,
\ee
and at second order as
\bea
a_{(2)0}&=&2B_{1}^k\A{}_{1,k}\,, \qquad
a_{(2)i}=\left[
\A{}_{2,i}+\left(B_{1k}B_1^k-2\A1^2\right)_{,i}\right]\,.
\eea

\subsection{Curvature of spatial three-hypersurfaces}
\label{3Rsec}

The intrinsic curvature of spatial three-hypersurfaces is given at
first and second order, respectively, by
\bea
\delta{}^{(3)}R_1&=&\frac{4}{a^2}\nabla^2\psi_1\,,\\
\delta{}^{(3)}R_2&=&\frac{1}{a^2}\Big[
4\nabla^2\psi_2-4C_{1km,}^{~~~~m}C_{1~~,n}^{kn}
+3C_{1mn,}^{~~~~k}C^{mn}_{1~~,k} -C^k_{1~k,n}C^{m~~n}_{1~m,}
\nonumber\\
&& \qquad+4C_1^{mn}\left(
C_{1mn,~k}^{~~~~~k}+C^k_{1~k,mn}-C_{1mk,n}^{~~~~~~k}-C_{1kn,m}^{~~~~~~k}
\right)\nonumber\\
&& \qquad +2\left(C^k_{1~k,j} C^{jn}_{1~,n}
+C_{1jk,}^{~~~j}C^{m~~~k}_{1~m,} -C^k_{1~n,m}C^{mn}_{1~~,k}
\right)\,,
\eea
where we used
\be 2\left(C^{mn}_{~~~,mn}-C^{m~~k}_{~m,~~k}\right)=4\nabla^2\psi\,.
\ee
%

\section{Governing equations}

It is often convenient to have all relevant equations available
``at a single glance''. We therefore reproduce all governing equations given in
previous sections in this appendix together.
No gauge is specified, i.e.~without choosing a particular hypersurface
or gauge restrictions, and we leave quantities undecomposed whenever
possible (the decomposition rules are given in \eqs{vectordecomp},
(\ref{decompBi}), and (\ref{decompCij})).

\subsection{Background}

Energy conservation for the $\alpha$-fluid in the background is given
from \eq{nablaTalpha} as
\be
\rho_{0\alpha}'=-3\H\left(\rho_{0\alpha}+P_{0\alpha}\right)+aQ_{0\alpha}\,,
\ee
and the total energy conservation is then given by summing over the
individual fluids and using \eq{Qconstraint} as
\be
\rho_{0}'=-3\H\left(\rho_{0}+P_{0}\right)\,.
\ee

The Friedmann constraint is given from the $0-0$ component \eq{Einstein} as
\be
H^2 = \frac{8\pi G}{3}\rho_0 \qquad \H^2 = \frac{8\pi G}{3}a^2\rho_0 \,.
\ee
The trace gives
\be
\frac{a'^2}{a^2}-2\frac{a''}{a}=8\pi G a^2 P_0\,.
\ee

\subsection{First order}

In this subsection we give the governing equations on large scales in
the general case without any gauge restrictions, i.e.~without choosing
a particular hypersurface.

\subsubsection{Field equations}
\label{field1}

The $0-0$ Einstein equation is given from \eq{Einstein} as
\be
3\H\left( \H\phi_1+ \psi_1'\right) -\nabla^2\left(\psi_1+\H
\sigma_1\right)= -4\pi G a^2\delta\rho_1\,.
\ee
The $0-i$ Einstein equation is
\be
\H\phi_1+ \psi_1'=
-4\pi Ga^2\left(\rho_0+P_0\right)\left(v_1+B_1\right)\,,
\ee
the off-trace is
\be
\sigma_1'+2\H\sigma_1+\psi_1-\phi_1=8\pi G a^2\Pi\,,
\ee
and the trace is
\be
\psi_1''+2\H\psi_1'+\H\phi_1'
+\left(2\frac{a''}{a}-\frac{a'^2}{a^2}\right)\phi_1
=4\pi G a^2 \left(\delta P_1+\frac{2}{3}\nabla^2\Pi\right)\,.
\ee

\subsubsection{Energy-momentum conservation}
\label{energy1}

Energy and momentum conservation of the $\alpha$-fluid is given from
\eq{nablaTalpha} at first order as
\be
\delta\rho_{\alpha}' +3\H(\delta\rho_{\alpha}+\delta P_{\alpha})
- 3\left(\rho_{\alpha}+P_{\alpha}\right)\psi'
+ a^{-1}(\rho_{\alpha}+P_{\alpha}) \nabla^2\left(\Va+\sh\right)
= aQ_{\alpha}\phi + a\delta Q_{\alpha} \,,
\ee
The momentum conservation equation of the $\alpha$-fluid is
\be
{V}_\alpha'
+\left[\frac{aQ_\alpha}{\rho_\alpha+P_\alpha}(1+c^2_\alpha)
-3\H c^2_\alpha\right]\Va +a\phi
+\frac{a}{\rho_\alpha+P_\alpha}\left[
\delta P_\alpha+\frac{2}{3}\nabla^2\Pi_\alpha
- Q_\alpha V -f_\alpha\right] =0\,.
\ee
Total energy and momentum conservation follows from the above, by
summing over all individual fluids and using \eqs{Qconstraint}, and is
given by
\bea
\delta\rho_{1}'
+3\H\left( {\delta\rho_{1}}+{\delta P_{1}}\right)
&+&\left(\rho_{0}+P_{0}\right)\left[
\nabla^2\left(\sigma_1+v_1+B_1\right)-3\psi_1'\right]
=0\,, \\
\left[\left(\rho_{0}+P_{0}\right)\left(v_1+B_1\right)
\right]'
&+&\left(\rho_{0}+P_{0}\right)
\left[4\H\left(v_1+B_1\right)+\phi_1\right]
+\delta P_1+\frac{2}{3}\nabla^2\Pi=0\,.\nonumber \\
\eea
%

\subsection{Second order}

\subsubsection{Energy-momentum conservation}
\label{energy2}

In the multi-fluid case, energy conservation of the $\alpha$-fluid is
given from \eq{nablaTalpha} at second order as
\bea
&&\delta\rho_{2\alpha}'+3\H\left( {\delta\rho_{2\alpha}}
+{\delta P_{2\alpha}}\right)
+\left(\rho_{0\alpha}+P_{0\alpha}\right)
\left(-3\psi_2'+\nabla^2 E_2'+\nabla^2v_{2\alpha}\right)\nonumber\\
&&+2\left( {\delta\rho_{1\alpha}}+{\delta P_{1\alpha}}\right)
\left(-3\psi_1'+\nabla^2 E_1'+\nabla^2v_{1\alpha}\right)\nonumber\\
&&
+2\left( {\delta\rho_{1\alpha}}+{\delta P_{1\alpha}}\right)_{,k}
v_{1\alpha}^k
+2\left(\rho_{0\alpha}'+P_{0\alpha}'\right)v_{1\alpha}^k
\left(v_{1(\alpha)k}+B_{1k}\right)
\nonumber\\
&&
+2\left(\rho_{0\alpha}+P_{0\alpha}\right)\left[
\left(v_{1(\alpha)k}'+B_{1k}'\right)\left(2v_{1(\alpha)k}+B_{1k}\right)
+4\H v_{1(\alpha)}^k\left(v_{1(\alpha)k}+B_{1k}\right)\right]\nonumber\\
&&
-4\left(\rho_{0\alpha}+P_{0\alpha}\right)C_{1ij}'C_1^{ij}
+2\left(\rho_{0\alpha}+P_{0\alpha}\right)
\left(2v_{1\alpha}^k\A1_{,k}+\A1\nabla^2v_{1\alpha}\right)\nonumber\\
&&
+2\pi_{1(\alpha)ij}\left(C_1^{\prime ij}-2\H C_1^{ij}\right)
+2\pi_{1(\alpha)}^{kl}v_{1(\alpha)k,l}
+2v_{1\alpha}^k\pi_{1(\alpha)kl,}^{~~~l}\nonumber\\
&&
+2\left(\rho_{0\alpha}+P_{0\alpha}\right)
\left(-3\psi_{1,l}+\nabla^2 E_{1,l}\right)v_{1\alpha}^l\nonumber\\
&&=a\Big\{
\delta Q_{2\alpha}+2\A1\delta Q_{1\alpha}
+Q_{0\alpha}\left(\A2-\A1^2+v_{1k}v_1^k\right)
+\frac{4}{a}f_{1(\alpha)k}v_1^k\Big\}
\,.
\eea

For a single fluid we find by summing over the individual fluids and
using \eq{Qconstraint}
\bea
&&\delta\rho_2'+3\H\left( {\delta\rho_{2}}+{\delta P_{2}}\right)
+\left(\rho_0+P_0\right)
\left(-3\psi_2'+\nabla^2 E_2'+\nabla^2v_2\right)\nonumber\\
&&+2\left( {\delta\rho_{1}}+{\delta P_{1}}\right)
\left(-3\psi_1'+\nabla^2 E_1'+\nabla^2v_1\right)
+2\left( {\delta\rho_{1}}+{\delta P_{1}}\right)_{,k}v_1^k
+2\left(\rho_0'+P_0'\right)v_1^k\left(v_{1k}+B_{1k}\right)
\nonumber\\
&&
+2\left(\rho_0+P_0\right)\left[
\left(v_{1k}'+B_{1k}'\right)\left(2v_{1k}+B_{1k}\right)
+4\H v_1^k\left(v_{1k}+B_{1k}\right)\right]
-4\left(\rho_0+P_0\right)C_{1ij}'C_1^{ij}
\nonumber\\
&&
+2\left(\rho_0+P_0\right)\left(2v_1^k\A1_{,k}+\A1\nabla^2v_1\right)
+2\pi_{1ij}\left(C_1^{\prime ij}-2\H C_1^{ij}\right)
+2\pi_{1}^{kl}v_{1k,l}+2v_1^k\pi_{1kl,}^{~~~l}\nonumber\\
&&
+2\left(\rho_0+P_0\right)
\left(-3\psi_{1,l}+\nabla^2 E_{1,l}\right)v_1^l
=0\,.
\eea

\newpage
\begin{center}
\subsection*{List of Symbols}

\begin{tabular}{l l}
$a$ & Scale factor \\

$a_{\mu}$   &   Acceleration \\

$c_{\rm{s}}$    &   Adiabatic sound speed \\

$c_\alpha$  &   Adiabatic sound speed of the $\alpha$ component\\

$f_{(J)}$   &   Momentum transfer perturbation of Jth component\\

$g_{\mu\nu}$    &   Metric tensor \\

$h_{ij}$    &   Tensor metric perturbation \\

$\hp_{\mu\nu}$ & Projection tensor,
$\hp_{\mu\nu}\equiv g_{\mu\nu}+n_\mu n_\nu$\\

$k^i$   &   Comoving wave-vector \\

$k$ &   Comoving wavenumber $k^2\equiv k^ik_i$\\

$\Mpl$  &   Planck mass $=G^{-1}$\\

$\Mp$  &   Reduced Planck mass $=(8\pi G)^{-1}$\\

$n^{\mu}$   &   Unit time-like vector field \\

$n_{\rm{s}}$    &   Spectral index of curvature perturbations \\

$\Delta n_X$    &   Scale dependence of perturbation spectrum
of a quantity $X$\\

$q^\mu$ &   momentum current density (in rest frame)\\

$q^i$   &   Wave vector\\

$ds$    &   Infinitesimal line element \\

$t$ &   Coordinate time \\

$u^{\mu}$   &   4-velocity \\

$v$ &   Scalar velocity perturbation \\

$v_\alpha$  &   Scalar velocity perturbation of $\alpha$-component\\

$\bar v_J$  &   Rescaled perturbation in Jth orthogonal field,
        $a\delta\bar\sigma_J$ \\

$v^i$   &   Vector velocity perturbation \\

$x^i$   &   Spatial coordinate \\

\end{tabular}

\newpage
\begin{tabular}{l l}

$B$ &   Shift (scalar metric perturbation) \\

$E$ &   Scalar spatial metric perturbation \\

$F_i$   &   Vector spatial metric perturbation \\

$G$ &   Newton's constant \\

$G_{\mu\nu}$    &   Einstein tensor \\

$H$ &   Hubble parameter, $H\equiv\frac{\dot a}{a}$\\

$H^{(1)}_{\nu}$ &   Hankel function of the first kind of degree $\nu$ \\

$K_{\mu\nu}$    &   Extrinsic curvature \\

$L$ &   Lagrangian \\

${\cal{L}}$ &   Lagrangian density \\

$N$ &   Number of e-folds (integrated expansion)\\

$\hp_{\mu\nu}$  &   Projection tensor,
$\hp_{\mu\nu}\equiv g_{\mu\nu}+n_{\mu}n_{\nu}$ \\

$P$ &   Pressure \\

$P_\alpha$  &   Pressure of the $\alpha$-component\\

$Q_{(J)}$   &   Energy transfer parameter of the Jth component \\

$Q^{\mu}_{(J)}$ &   Energy momentum four vector of the Jth component\\

${\cal{P}}_X$   &   Power spectrum of a quantity $X$ \\

$R$ &   Ricci scalar \\

$R_{\mu\nu}$    &   Ricci tensor \\

${}^{(3)}R$ &   Intrinsic spatial curvature of three-hypersurface \\

${\cal{R}}$ &   Curvature perturbation in comoving gauge \\

$S$ &   Action \\

$S_i$   &   Vector metric perturbation \\

$S_{IJ}$    &   Entropy perturbation \\

$T_{\mu\nu}$    &   Energy momentum tensor \\

$U(\varphi)$    &   Potential of scalar field \\

\end{tabular}


\newpage
\begin{tabular}{l l}

$\alpha$ &  Arbitrary scalar function (temporal gauge function)\\

$\beta$ &   Arbitrary scalar function (spatial gauge function)\\

$\delta^{\mu}_{~\nu}$   &   Kronecker delta \\

$\eta$  &   Conformal time, $a d\eta\equiv d t$\\

$\zeta$ &   Curvature perturbation on uniform density hypersurface \\

$\gamma^{i}$ & Arbitrary divergence-free vector function (spatial
  gauge function)\\

$\gamma_{ik}$   &   Metric tensor on spatial 3-hypersurface\\

$\kappa$    &   Curvature of background spacetime \\

$\lambda$       &       comoving wave length $k=\frac{2\pi}{k}$\\

$\lambda_{\rm{phys}}$       &
physical wave length $\lambda_{\rm{phys}}=a\lambda$\\

$\omega_{\mu\nu}$   &   Vorticity tensor\\

$\pi^i$ &   Anisotropic stress vector \\

$\pi^{\mu}_{~\nu}$  &   Anisotropic stress tensor \\

${}^{\rm{tensor}}\pi^{i}_{j}$   &   Tensorial anisotropic
stress tensor \\

$\varphi$   &   Scalar field \\

$\varphi_I$ &   one of multiple scalar fields \\

$\phi$  &   Lapse function (scalar metric perturbation) \\

$\psi$  &   Curvature perturbation (scalar metric perturbation) \\

$\rho$  &   Energy density \\

$\rho_\alpha$   &   Energy density of  $\alpha$-component \\

$\sigma$    &   Shear scalar \\

$\sigma_{\mu\nu}$   &   Shear tensor \\

$\tau$          & proper time, $d\tau^2=ds^2$ \\

$\theta$    &   Expansion \\

$\bar\xi^i$ &   Arbitrary vector valued function \\

$\Gamma(x)$ &   Gamma function \\

$\Pi$   &   Scalar anisotropic stress tensor \\

$\Phi$  &   Bardeen potential (lapse function in longitudinal gauge) \\

$\Psi$  &   Bardeen potential (curvature perturbation in
longitudinal gauge) \\

%
%
%
\end{tabular}
\end{center}

\newpage
\subsection*{Dimensions}

It can be useful as a quick check of the validity of an equation or
expression, particularly for the large expressions at second order, to
check that all terms have the correct dimensions.

\begin{center}
\begin{tabular}{l l}
Quantity \hspace{12mm} &  Dimension\\

$[a]$ & $1$ \\

$[\vp]$ & $T^{-1}$ \\

$[H]$ & $T^{-1}$ \\

$[U]$ & $T^{-4}$ \\

$[\dot\vp]$ & $T^{-2}$ \\

$[G]$ & $T^{2}$ \\

$[\eta]$ & $T$ \\

$[x^i]$ & $T$ \\

$[\xi^\mu]$ & $T$ \\

$[\phi]$ & $1$ \\

$[\psi]$ & $1$ \\

$[B]$ & $T$ \\

$[E]$ & $T^2$ \\

$[C_{ij}]$ & $1$ \\

$[\R]$ & $1$ \\

$[\zeta]$ & $1$ \\

$[\rho]$ & $T^{-4}$ \\
\end{tabular}
\end{center}

Note, in geometric units $T\equiv L$.

\newpage


\end{document}